\documentclass[10pt,conference]{IEEEtran}

\usepackage{amsthm, amssymb}
\usepackage[cmex10]{amsmath}
\usepackage{xspace}
\usepackage{graphicx}
\usepackage{paralist}
\usepackage{mathtools}
\usepackage{calc}
\usepackage{cite}

\newtheorem{proposition}{Proposition}[section]
\newtheorem{definition}{Definition}[section]
\newtheorem{remark}{Remark}[section]

\newtheorem{lemma}{Lemma}[section]
\newtheorem{theorem}{Theorem}[section]
\newtheorem{corollary}{Corollary}[section]

\newcommand{\DefinedAs}{\,\stackrel{\text{def}}{=}\,}

\newcommand{\TLP}{\textsf{TLP}\xspace}
\newcommand{\LTL}{\textsf{LTL}\xspace}
\newcommand{\TEL}{\textsf{TEL}\xspace}
\newcommand{\THT}{\textsf{THT}}
\newcommand{\HT}{\textsf{HT}\xspace}
\newcommand{\NFA}{\textsf{NFA}\xspace}
\newcommand{\TAS}{\textsf{TAS}\xspace}

\newcommand{\Au}{\ensuremath{\mathcal{A}}}
\newcommand{\Lang}{\mathcal{L}}
\newcommand{\Nat}{\mathbb{N}}
\newcommand{\Left}{\textit{left}}
\newcommand{\Right}{\textit{right}}
\newcommand{\Down}{\textit{down}}
\newcommand{\Up}{\textit{up}}
\newcommand{\Instance}{\mathcal{I}}

\newcommand{\depthX}{\textit{d}_\Next}

\newcommand{\Fin}{\textit{Fin}}
\newcommand{\Inf}{\textit{Inf}}
\newcommand{\CON}{\textit{CON}}
\newcommand{\Init}{\textit{init}}
\newcommand{\Final}{\textit{final}}

\newcommand{\Model}{\textsf{M}}
\newcommand{\HModel}{\textsf{H}}
\newcommand{\TModel}{\textsf{T}}
\newcommand{\Main}{\textit{main}}
\newcommand{\pseudo}{\textit{pseudo}}
\newcommand{\Tag}{\textit{tag}}

\newcommand{\num}{\textit{num}}

\newcommand{\cell}{\textit{cell}}

\newcommand{\symSp}{u}

\newcommand{\Next}{\textsf{X}}
\newcommand{\Until}{\textsf{U}\,}
\newcommand{\Release}{\textsf{R}\,}
\newcommand{\Always}{\textsf{G}}
\newcommand{\Eventually}{\textsf{F}}

\newcommand{\NLOGSPACE}{\textsf{NLogspace}\xspace}
\newcommand{\PSPACE}{{\sc Pspace}\xspace}
\newcommand{\NP}{{\sc NP}\xspace}

\newcommand{\EXPSPACE}{{\sc Expspace}\xspace}

\newcommand{\NEXPTIME}{{\sc Nexptime}\xspace}
\newcommand{\CompEL}{\ensuremath{\Sigma_2}}

\newcommand{\DefORmini}{\ensuremath{\;\big|\;}}
\newcommand{\myparagraph}[1]{\paragraph*{#1}}

\newcommand{\tupleof}[1]{\langle#1\rangle}

\newcommand{\details}[1]{}

\begin{document}
\title{\Large \textbf{On the complexity of Temporal Equilibrium Logic}}

\author{\IEEEauthorblockN{Laura Bozzelli}
\IEEEauthorblockA{Technical University of Madrid (UPM), Madrid, Spain\\
Email: laura.bozzelli@fi.upm.es}
\and
\IEEEauthorblockN{David Pearce}
\IEEEauthorblockA{Technical University of Madrid (UPM), Madrid, Spain\\
Email: david.pearce@upm.es}}

\maketitle

\begin{abstract}Temporal Equilibrium Logic (\TEL)~\cite{CabalarV07} is a promising framework that extends the knowledge representation and reasoning capabilities of Answer Set Programming with temporal operators in the style of \LTL. To our knowledge it is the first nonmonotonic logic that accommodates fully the syntax of a standard temporal logic (specifically \LTL) without requiring further constructions. This paper provides a systematic complexity analysis for the (consistency) problem of checking the existence of a temporal equilibrium model of a \TEL formula. It was previously shown that this problem in the general case lies somewhere between \PSPACE and \EXPSPACE. Here we establish a lower bound matching the \EXPSPACE upper bound in~\cite{CabalarD11}. Additionally we analyse the complexity for various natural subclasses of \TEL formulas, identifying both tractable and intractable fragments. Finally the paper offers some new insights on the logic \LTL by addressing satisfiability for minimal \LTL models. The complexity results obtained highlight a substantial difference between interpreting \LTL over finite or infinite words.
\end{abstract}

\IEEEpeerreviewmaketitle

\section{Introduction}
In this paper we analyse the complexity of checking model existence in Temporal Equilibrium Logic (\TEL). \TEL was proposed by Cabalar and Vega~\cite{CabalarV07} as a nonmonotonic logic for temporal reasoning. In particular, \TEL provides an important extension of the language of answer set programming (ASP) by capturing temporal reasoning problems not representable in ASP. It is also apparently the only nonmonotonic extension of a standard modal temporal logic (viz. \LTL) that does not use additional operators or constructions.

 Answer Set Programming (ASP) is now well established as a successful paradigm for declarative programming, with its roots in the fields of knowledge
representation (KR), logic programming, and nonmonotonic reasoning (NMR)~\cite{DBLP:journals/cacm/BrewkaET11}.
Besides a fully declarative, modular
reading of problem descriptions, distinguishing features of ASP are its intrinsic handling of nondeterminism and the rich possibilities for knowledge representation, including the seamless handling of incomplete and defeasible knowledge, preferences at various levels,
as well as aggregates and other useful features.

An adequate logical foundation for ASP is provided by a formalism called
{\em Equilibrium Logic}~\cite{Pearce96,Pearce06}, a nonmonotonic extension of the superintuitionistic logic of {\em here-and-there} (\HT)~\cite{Heyting30}. This provides useful logical tools for the metatheory of ASP and a framework for defining extensions of the basic ASP language, for example to
arbitrary propositional and first-order theories, to languages with intensional functions, and to hybrid theories that combine classical and rule-based reasoning~\cite{DBLP:conf/jelia/PearceV04,DBLP:conf/jelia/CabalarCPV14,DBLP:conf/rr/BruijnPPV07,DBLP:conf/jelia/FinkP10}.

The nonmonotonic capability of ASP helps to solve typical representation issues in temporal reasoning such as the frame problem \cite{Mccarthy69} and the ramification problem \cite{Kautz86}. However, while ASP has been applied to a wide range of problems involving temporal reasoning, including prediction, planning, diagnosis and verification,  since it is not an intrinsically temporal formalism, it suffers some important limitations. Most ASP solvers deal with finite domains, a restriction that allows a grounding of the program into a finite set of propositional rules. This limitation means that time is usually represented by an extensional predicate with a finite domain fixed {\em a priori}, hampering the solution of  problems dealing with unbounded time.

Temporal scenarios dealing with unbounded time are typically best suited for modal temporal logics. 
However,
standard modal temporal logics, such as propositional linear-time temporal logic \LTL \cite{Pnueli77}, do not  accommodate default and nonmonotonic reasoning and are not designed to deal with many issues in knowledge representation. \TEL extends equilibrium logic and therefore includes KR features from ASP but is able to express concepts from modal temporal logic.  It shares the syntax of \LTL, but its semantics is an orthogonal combination of the \LTL semantics with the nonmonotonic semantics of Equilibrium Logic. As for Equilibrium Logic, \TEL models
(called \emph{temporal equilibrium models}) are the result of a kind of minimisation among models of the monotonic logic of Temporal Here-and-There (\THT), a combination of \LTL and \HT.
Considerable progress has already been made in the theoretical study of \TEL and its computational methods.
Key results include the use of \TEL to translate action languages \cite{CabalarV07}, an automata-theoretic approach for checking the existence of \TEL models \cite{CabalarD11},  
a decidable criterion for proving the strong equivalence of two \TEL theories \cite{CabalarD14}, and
a tool for computing models of temporal programs under \TEL semantics \cite{CabalarD11Tool}.

\myparagraph{Our contribution} We investigate the computational cost of the \TEL consistency  problem, that is checking for a given \THT\ formula
the existence of a temporal equilibrium model. This question was previously addressed in \cite{CabalarD11} by showing that the problem lies somewhere between \PSPACE\ and \EXPSPACE. Our first contribution consists in filling this computational gap by
providing a lower bound matching the \EXPSPACE\ upper bound in    \cite{CabalarD11}. 

As a second contribution, we give a systematic analysis, searching for natural subclasses of \THT\ formulas for which complexity decreases. In particular, we consider all the syntactical fragments of \THT\ obtained by restricting the set of allowed temporal modalities and/or by imposing a bound on the nesting depth of temporal modalities and/or the implication connective (including negation, expressed in terms of implication). The aim is to obtain a better understanding of what makes the initial problem \EXPSPACE-hard, and to identify interesting fragments with lower
complexity.  Overall, our results are rather negative. We show that the \TEL consistency problem remains \EXPSPACE-hard even in the following two simple cases: (1)
 the unique allowed temporal modality is $\Always$  (`always'), and (2)
 there is no nesting of implication.

The result for the first case is surprising since \LTL/\THT\ satisfiability for the fragment   where the unique allowed temporal modalities are $\Always$ and $\Eventually$ (`eventually') is just \NP-complete~\cite{SistlaC85,CabalarD11}.
On the other hand, the result for the second case highlights an important difference between propositional equilibrium logic and \TEL\. It is well-known that for logic programs without default negation (corresponding to \HT\ formulas where there is no nesting of implication\footnote{recall that in \HT/\THT\ negation is expressed in terms of implication}), the existence of classical models ensures the existence of stable models. This fails in the temporal extension, where as  pointed  in~\cite{CabalarD11}, the non-existence of equilibrium models may be also due to the lack of a finite justification for satisfying the criterion of minimal knowledge.

The \TEL consistency problem remains hard, and, precisely, \NEXPTIME-complete even for the simple case where  no nesting of temporal modalities is allowed. However, on the positive side, we identify many interesting \THT\ fragments with a lower  complexity.  For each of them, we  show that the \TEL consistency problem is complete
for some complexity class in  $\{$\NP, \CompEL,\,\PSPACE$\}$ (for an overview of the obtained results, see Subsection~\ref{sec:ResultSummary}). Some of these results also point out a peculiar difference between \LTL\ and \THT: due to the interpretation of the implication connective, in \THT, a temporal modality cannot expressed in terms of its `dual' modality. Thus, in \THT, dual temporal modalities, such as $\Eventually$ and $\Always$, need to be considered independently from one another. This is illustrated by one of our positive results: for the \THT\ fragment  whose allowed temporal modalities  are $\Eventually$ and $\Next$ (`next'), the complexity of the considered problem collapses to the second level
$\Sigma_2$ of the polynomial hierarchy. This also turns out to be  the unique case where, surprisingly,
\LTL/\THT\ satisfiability is harder than \TEL consistency.

As a third contribution, we provide new insights into the logic \LTL. We address minimal \LTL satisfiability, that is checking the existence of \LTL models which are minimal with respect to the partial order given by pointwise propositional containment. While for \LTL over finite words, the existence of \LTL models ensures the existence of minimal ones, for \LTL over infinite words, this is not true. In particular, we show the for the case of infinite words, minimal \LTL\ satisfiability is exponentially harder than \LTL satisfiability, and, precisely, \EXPSPACE-complete. To the best of our knowledge, there is no complexity result in the literature emphasizing the differences arising from interpreting \LTL over finite or infinite words.

\myparagraph{Related work} Several research areas of AI have combined modal temporal logics with formalisms from knowledge representation for reasoning about actions and planning (see e.g. \cite{fagin1995reasoning}). Combinations of NMR with modal logics designed for temporal reasoning are much more infrequent in the literature. The few exceptions are typically modal action languages with a nonmonotonic semantics defined under some syntactical restrictions. Recently, an alternative to \TEL has been introduced, namely, Temporal Answer Sets (\TAS), which relies on
 dynamic linear-time temporal logic \cite{GiordanoMD13}, a modal approach more expressive than \LTL. However, while the non-monotonic semantics of \TEL covers any arbitrary theory in the syntax of \LTL, \TAS uses a syntactic transformation that is only defined for theories with a rather restricted syntax. A framework unifying \TEL and \TAS has been proposed in \cite{AguadoPV13}.

\section{Temporal Equilibrium Logic}


We recall the framework of Temporal Equilibrium Logic (\TEL)~\cite{CabalarV07}. \TEL is defined by first introducing a monotonic and  intermediate version of standard linear temporal logic \LTL~\cite{Pnueli77}, the so-called logic of \emph{Temporal Here-and-There} (\THT)~\cite{CabalarV07}. The nonmonotonic semantics of \TEL is then defined by introducing a criterion for selecting models of \THT.

Let $\Nat$ be the set of natural numbers and for all $i,j\in\Nat$, let $[i,j] := \{h\in\Nat \mid i\leq h \leq j\}$.
  For an infinite word  $w$ over some
alphabet  and for all $i\geq 0$,
$w(i)$ is the $i^{th}$ symbol of $w$.

\myparagraph{Syntax and semantics of \THT} while the syntax of \THT\ coincides with that of   \LTL, the semantics of \THT\ is instead  an orthogonal combination of the superintuitionistic propositional logic of Here-and-There (\HT)
\cite{Heyting30} and  \LTL.  Fix a finite set $P$ of atomic propositions. The set of \THT\ formulas $\varphi$ over $P$ is defined by the following abstract syntax.
\[
  \varphi:=
p \DefORmini
\bot      \DefORmini
\varphi \vee \varphi \DefORmini
\varphi \wedge \varphi \DefORmini
\varphi \rightarrow \varphi \DefORmini
\Next \,\varphi     \DefORmini
\varphi\,\Until\varphi  \DefORmini
\varphi\,\Release \varphi
\]
\noindent where $p\in P$ and $\Next$,
$\Until$, and $\Release$, are the standard `next',
`until', and `release' temporal modalities.
Negation is defined as $\neg \varphi \DefinedAs \varphi \rightarrow \bot$ while $\top \DefinedAs \neg \bot$. As usual
$\varphi_1 \leftrightarrow \varphi_2 $ stands for
$(\varphi_1 \rightarrow \varphi_2) \wedge  (\varphi_1 \leftarrow \varphi_2 )$. The classical temporal operators $\Always$ (`always') and $\Eventually$
(`eventually') can be defined in terms of $\Until$ and $\Release$ as follows: $\Eventually \varphi \DefinedAs \top \Until \varphi$ and
$\Always \varphi \DefinedAs \bot \Release \varphi$. The size $|\varphi|$ of a formula $\varphi$ is the number of distinct subformulas of $\varphi$.
The \emph{temporal height} (resp. \emph{implication height}) of $\varphi$ is the maximum number of nested temporal modalities (resp. nested implications) in $\varphi$. Notice that  negation is counted as an additional implication. Thus, for example, formula $\neg p  \rightarrow p$ has implication height equal to $2$.

 Recall that \LTL over $P$ is interpreted on infinite words over $2^{P}$, called in the following \emph{\LTL interpretations}. By contrast, the semantics of \THT\ is defined in terms of infinite words over $2^{P}\times 2^{P}$, which can also be  viewed as pairs of \LTL-interpretations. 
Formally, a \emph{\THT\ interpretation} is a pair $\Model= (\HModel,\TModel)$ consisting of two \LTL interpretations: $\HModel$ (the `here' interpretation) and $\TModel$ (the `there' interpretation) such that
 \[
 \text{for all } i\geq 0,\text{ } \HModel(i)\subseteq \TModel(i)
\]
  Intuitively, $\HModel(i)$ represents the set of propositions which are true at position $i$, while $\TModel(i)$ is the set of propositions which \emph{may} be true (i.e. which are not falsified in an intuitionistic sense). A \THT\ interpretation $\Model= (\HModel,\TModel)$ is said to be \emph{total} whenever $\HModel=\TModel$.
In the following, for \emph{interpretation}, we mean a \THT\ interpretation.
Given an interpretation  $\Model= (\HModel,\TModel)$, a position $i\geq 0$, and a \THT\ formula $\varphi$, the satisfaction relation $\Model,i\models \varphi$ is inductively defined as follows:
\[ \begin{array}{lll}
  \Model,i\not\models \bot &&\\
  \Model,i\models p & \Leftrightarrow & \,p\in \HModel(i)\\
  \Model,i\models \varphi\vee \psi & \Leftrightarrow &  \text{either } \Model,i\models \varphi \text{ or } \Model,i\models \psi \\
  \Model,i\models \varphi\wedge \psi & \Leftrightarrow &   \Model,i\models \varphi \text{ and } \Model,i\models \psi \\
  \Model,i\models \varphi \rightarrow \psi & \Leftrightarrow &  \text{for all }
    \HModel'\in\{\HModel,\TModel\},\\
 &&  \text{either } (\HModel',\TModel),i\not\models \varphi
    \text{ or } (\HModel',\TModel),i\models \psi\\
  \Model,i\models \Next\varphi &
               \Leftrightarrow & \Model,i+1\models \varphi\\
  \Model,i\models  \varphi \,\Until\psi &
               \Leftrightarrow &  \text{there is } j\geq i \text{ such that } \Model,j\models \psi \text{ and }\\
               &&       \text{for all } k\in [i,j-1],\,  \Model,k\models \varphi\\
  \Model,i\models  \varphi \,\Release\psi &
               \Leftrightarrow &  \text{for all } j\geq i,  \text{ either } \Model,j\models \psi \text{ or there}\\
        &&      \text{is } k\in [i,j-1] \text{ such that } \Model,k\models \varphi
\end{array}
\]
We say that $\Model$ is a  ($\THT$)  model of $\varphi$, written $\Model\models \varphi$, whenever $\Model,0\models \varphi$.
A $\THT$ formula $\varphi$ is $\THT$ \emph{satisfiable} if it admits a $\THT$ model. A formula $\varphi$ is $\THT$ \emph{valid} if every
interpretation $\Model$ is a $\THT$ model of $\varphi$. 
Note that the semantics of \THT\ is defined similarly to that of \LTL except for the clause for the implication connective $\rightarrow$ which must be checked in both the components $\HModel$ and $\TModel$ of $\Model$.
As a  consequence  $\Model,i \not\models \varphi$ does not correspond to $\Model,i \models \neg\varphi$ (i.e.,
$\Model,i \models \neg\varphi$ implies that $\Model,i \not\models \varphi$, but the converse direction does not hold in general).
However,  if we restrict the semantics to total interpretations, $(\TModel,\TModel)\models \varphi$ corresponds to the satisfaction relation
$\TModel\models \varphi$ in \LTL. More precisely, the \LTL models $\TModel$ of $\varphi$ correspond to the total interpretations $(\TModel,\TModel)$
which are $\THT$ models of $\varphi$. As shown in \cite{CabalarD11}, $\THT$ satisfiability can be reduced in linear-time to $\LTL$ satisfiability. 
With regard to $\THT$ validity, a $\THT$ valid formula is also an $\LTL$ valid formula, but the converse in general does not hold. For example, the \emph{excluded middle} axiom $\varphi\vee \neg\varphi$ is not a valid $\THT$ formula since, as highlighted above, for an interpretation $\Model=(\HModel,\TModel)$,  $\Model  \not\models \varphi$ does not imply that $\Model \models \neg\varphi$. Similarly, the temporal formulas $\Eventually\varphi \leftrightarrow \neg\Always \neg\varphi$ and
$\varphi_1\Until\varphi_2 \leftrightarrow \neg\varphi_1\Release\neg\varphi_2$, which are well-known valid \LTL formulas (and allow to express, in \LTL, a temporal modality  in terms of its dual modality), are not $\THT$ valid formulas. Thus, in $\THT$, dual temporal modalities, like $\Eventually$ and $\Always$, or $ \Until$ and $\Release$,  need to be considered independently one from the other one.
We summarize some observations made above and some additional observations (which easily follows from the semantics of \THT\ and \LTL) in the following proposition, where for clarity, we use $\models_\LTL$ to denote the satisfaction relation in \LTL.

\begin{proposition}\label{prop:fundamentalsTHT} Let $(\HModel,\TModel)$ be an interpretation and $\varphi$ be a \THT\ formula.
\begin{compactenum}
  \item If $(\HModel,\TModel),i\models \varphi$, then $(\TModel,\TModel),i\models\varphi$ (for all $i\geq 0$).
  \item $(\HModel,\TModel),i\models \neg\varphi$ iff $(\TModel,\TModel),i\models \neg\varphi$ (for all $i\geq 0$).
  \item $(\TModel,\TModel)\models\varphi$ iff $\TModel \models_\LTL \varphi$.
\item If $\varphi$ has implication height at most $1$, then $(\HModel,\TModel)\models\varphi$ implies $\HModel \models_\LTL \varphi$.
\end{compactenum}
\end{proposition}

\myparagraph{The non-monotonic logic \TEL} this logic is obtained from \THT\ by restricting the semantics to a subclass of models, called \emph{temporal equilibrium models}.
For two \LTL interpretations $\HModel$ and $\TModel$, we write $\HModel \sqsubseteq \TModel$ to mean that
$\HModel(i) \subseteq \TModel(i)$ for all $i\geq 0$. We write $\HModel \sqsubset \TModel$ to mean that $\HModel \sqsubseteq \TModel$ and
$\HModel \neq \TModel$.

\begin{definition}[Temporal equilibrium model] Given a \THT\ formula $\varphi$,  a \emph{(temporal) equilibrium model} of $\varphi$ is a total model $(\TModel,\TModel)$ of $\varphi$
satisfying the following minimality requirement:  whenever $\HModel \sqsubset \TModel$, then
$(\HModel,\TModel)\not\models\varphi$.
\end{definition}

If we restrict the syntax to \HT formulas (i.e., $\THT$ formulas where no temporal modality is allowed) and the semantics to
 \HT interpretations $(\HModel(0),\TModel(0))$, then (non-temporal) equilibrium models coincide with stable models of answer set programs in their
 most general form~\cite{Ferraris05a}. In particular, the  interpretation of negation is that of default negation in logic programming: formula $\neg\varphi$ holds ($\varphi$ is false by default) if there is no evidence regarding $\varphi$, i.e., $\varphi$ cannot be derived by the rules of the logic program. As an example, let us consider the \THT\ formula $\varphi$ given by $\varphi=\Always(\neg p \rightarrow \Next p)$. Its intuitive meaning corresponds to the first-order logic program consisting of rules of the form $p(s(X)) \leftarrow \textit{not}\, p(X)$, where time has been reified as an extra parameter $X=0,s(0),s(s(0)),\ldots$. Thus, at any time instant, if there is no evidence regarding $p$, then $p$ will become true at the next instant. Initially, we have no evidence regarding $p$, so this will imply $\Next p$. To derive $\Next\Next p$, the only possibility would be the rule $\neg \Next  p \rightarrow \Next\Next p$, an instance of $\varphi$. As the body of this rule is false, $\Next\Next p$ becomes false by default, and so on. It is easy to see that the unique equilibrium model of $\varphi$ is $((\emptyset \{p\})^{\omega},(\emptyset \{p\})^{\omega})$, corresponding to the unique \LTL model of formula $\neg p\wedge \Always(\neg p \leftrightarrow \Next  p)$.

Note that an \LTL satisfiable formula may have no temporal stable model.
 A familiar example from non-temporal ASP is the logic program rule $\neg p \rightarrow p$, whose unique classical model is $\{p\}$ and whose $\HT$ models are $(\emptyset, \{ p \})$ and $(\{ p \}, \{ p \})$.
 As a second example, consider the temporal formula $\varphi$ given
 by $\varphi=\Always(\neg\Next p \rightarrow p) \wedge \Always(\Next p \rightarrow p)$. This formula is \LTL-equivalent to $\Always p$. Thus, the unique \LTL model is $\TModel=\{p\}^{\omega}$. However, $(\TModel,\TModel)$
 is not an equilibrium model of $\varphi$, since the interpretation  $(\HModel,\TModel)$, where $\HModel=(\emptyset)^{\omega}$ is a \THT\ model of $\varphi$.

 In general, for $\HT$ formulas, the non-existence of equilibrium models is due to the unrestricted use of nested implication (recall that negation is expressed in terms of implication).
 For the temporal case, as pointed  in~\cite{CabalarD11}, the non-existence of equilibrium models
 may be also due to the lack of a finite justification which ensures the minimal fulfilment of the given formula.
 For example, for the formula $\varphi=\Always\Eventually \, p$, any $\LTL$ model $\TModel$ must contain infinite occurrences of $p$ (hence, no prefix of $\TModel$ can justifies the fulfilment of $\varphi$).
 Even if $\varphi$ is $\THT$/$\LTL$ satisfiable, one can easily check that there is no equilibrium model of $\varphi$.


\subsection{\textbf{Summary of the results}}\label{sec:ResultSummary}

We are interested in the following decision problem.

\myparagraph{The \TEL consistency problem:} let $\Lang$ be $\THT$ or a fragment of $\THT$. The $\TEL$ consistency decision problem for $\Lang$, written $\CON(\Lang)$, is the set of all $\Lang$-formulas  for which there exists  an equilibrium model.\vspace{0.2cm}

In particular, we consider the syntactical fragments of \THT\ obtained by restricting the set of allowed temporal modalities and/or by bounding the temporal/implication height. Formally, given   $O_1,O_2,\ldots\in \{\Next,\Eventually,\Always,\Until,\Release\}$, we denote by $\THT(O_1,O_2,\ldots)$ the fragment of $\THT$ for which only the temporal modalities $O_1,O_2,\ldots$ are allowed.  For  $k\geq 0$ and $m\geq 0$,  $\THT_m^{k}(O_1,O_2,\ldots)$ denotes  the fragment  of
$\THT(O_1,O_2,\ldots)$ where the temporal height is at most $m$ and the implication height is at most $k$. We write nothing for $m$ and/or $k$ when no bound is imposed. For instance, $\THT_2(\Always)$ denotes the fragment where the unique allowed temporal modality is $\Always$ and the temporal height is at most $2$.
The results obtained in this paper are illustrated in Fig.~\ref{table}. Notice that $\THT_0=\HT$ and checking the existence of equilibrium models for $\HT$ formulas is a well-known $\CompEL$-complete problem~\cite{EiterG95,Pearce06}. Moreover, membership in \EXPSPACE for the  \TEL consistency problem of full \THT\ has been established
in~\cite{CabalarD11} by a generalisation of the standard automata-theoretic approach for solving \LTL satisfiability.

Additionally, in Section~\ref{sec:MinimalityLTL}, we investigate the complexity of checking for a given
\THT\ formula $\varphi$, the existence of a \emph{minimal \LTL model}, i.e.
an \LTL model $\TModel$ of $\varphi$ such that   for all $\HModel \sqsubset \TModel$,
$\HModel \not\models_\LTL \varphi$. Notice that in general \LTL satisfiability does not ensure the existence of minimal \LTL models. An example is given by the formula $\Always\Eventually p$ which is \LTL satisfiable but does not admit minimal \LTL models.

\begin{table}[!t]
\renewcommand{\arraystretch}{1.3}
\caption{Computational cost of the \TEL consistency problem }
\label{table}
\centering
\begin{tabular}{|c|c|}
\hline   $m\geq 1$, $k\geq 1$ \,\, &
        {\bfseries \TEL consistency problem} \\
\hline \hline
\phantom{\large{$I^{I}$}}\THT,  $\THT_{m+1}^{1}(\Eventually,\Always)$,  \phantom{\large{$I^{I}$}} &    \EXPSPACE-complete  \\
  \phantom{\large{$I^{I}$}} $\THT_{m+1}^{k+1}(\Always)$,  $\THT_{m+1}^{k+1}(\Until)$  \phantom{\large{$I^{I}$}}   &    \text{ (Theorem~\ref{theorem:consistencyLowerBoundUntractable} and \cite{CabalarD11})}   \\
\hline
\phantom{\Large{$1$}}\THT(\Next),  \THT(\Eventually),  \THT(\Next,\Eventually) \phantom{\Large{$1$}}  &   $\CompEL$-complete \text{(Corollary~\ref{cor:MainResultXF})} \\
\hline
\phantom{\large{$I^{I}$}} $\THT_1$,    $\THT_{1}^{k+1}(\Eventually,\Always)$  \phantom{\large{$I^{I}$}} &   \NEXPTIME-complete   \\
\phantom{\large{$I^{I}$}}   $\THT_{1}^{k+1}(\Until)$,  $\THT_{1}^{k+1}(\Release)$ \phantom{\large{$I^{I}$}}  &    \text{(Theorems~\ref{theorem:consistencyLowerBoundTHTFGOneTemporalNesting} and~\ref{theo:MainUpperBoundTemporalOne})} \,\, \\
\hline
\phantom{\LARGE{$1$}} $\THT_1(\Next,\Always)$  \phantom{\LARGE{$1$}}  &    $\CompEL$-complete \text{(Theorem~\ref{theo:upperBoundFragmentOneGX})} \\
\hline
\phantom{\large{$I^{I}$}} $\THT^{1}(\Release)$,  $\THT^{1}(\Next,\Release)$ \phantom{\large{$I^{I}$}}  &    \PSPACE-complete  \\
\phantom{\large{$I$}} $\THT^{1}(\Until)$,  $\THT^{1}(\Next,\Until)$ \phantom{\large{$I$}}  &    \text{(Theorem~\ref{theorem:MainCONXROneImplication} and Cor.~\ref{cor:MainCONXU})} \\
\hline
\phantom{\LARGE{$1$}} $\THT^{1}_1$  \phantom{\LARGE{$1$}} &  \NP-complete    \text{(Theorem~\ref{theorem:MainCONXROneImplication})} \\
\hline
\phantom{\LARGE{$1$}} $\THT^{0}$  \phantom{\LARGE{$1$}}  &   \PSPACE-hard  \text{(Theorem~\ref{theorem:LowerBoundPositiveTHTMain})}\\
\hline
\phantom{\LARGE{$1$}} $\THT_{0}=\HT$  \phantom{\LARGE{$1$}}   &   $\CompEL$-complete \cite{EiterG95,Pearce06} \\
\hline
\end{tabular}
\end{table}

\section{Intractable fragments}\label{sec:UntractableFragments}

In this section we show that the \TEL consistency problem is in general \EXPSPACE-hard even for the fragments
$\THT_{2}^{1}(\Eventually,\Always)$, $\THT_{2}^{2}(\Always)$, and
$\THT_2^{2}(\Until)$. Moreover, the problem remains hard, and, precisely,  \NEXPTIME-complete when  no nesting of temporal modalities is allowed.
Notice that \EXPSPACE-hardness for   $\THT_{2}^{1}(\Eventually,\Always)$ is surprising since $\THT$ satisfiability for the fragment  $\THT(\Eventually,\Always)$ is just \NP-complete~\cite{CabalarD11,SistlaC85} and checking the existence of equilibrium models for $\HT^{1}$ formulas has the same complexity as satisfiability of classical propositional logic, i.e. \NP-complete.

\subsection{\textbf{\EXPSPACE-complete fragments}}

In this subsection, we establish the following result.

\begin{theorem}\label{theorem:consistencyLowerBoundUntractable} The \TEL consistency problems for $\THT_{2}^{\,2}(\Always)$, $\THT_{2}^{\,1}(\Eventually,\Always)$,
  and $\THT_2^{\,2}(\Until)$ are all \EXPSPACE-hard.
\end{theorem}

Theorem~\ref{theorem:consistencyLowerBoundUntractable} is proved by
polynomial-time reductions from a domino-tiling problem for grids with rows of singly exponential length~\cite{Harel92}. We fix
an instance $\Instance$ of such a problem, which  is a tuple $\Instance =\tupleof{C,\Delta,n,d_\Init,d_\Final}$, where $C$ is a finite set of colors, $\Delta \subseteq C^{4}$ is a set of tuples $\tupleof{c_\Down,c_\Left,c_\Up,c_\Right}$ of four colors, called \emph{domino-types}, $n>0$ is a  natural number (written in unary),
and $d_\Init,d_\Final\in\Delta$ are domino-types. A \emph{tiling of $\Instance$}  is a mapping $f:[0,k]\times [0,2^{n}-1] \rightarrow \Delta$ for some $k\geq 0$ satisfying the following:
\begin{itemize}
  \item two adjacent cells in a row have the same color on the shared edge: for all $(i,j)\in [0,k]\times [0,2^{n}-1]$ with $j< 2^{n}-1$,
   $[f(i,j)]_{\Right}=[f(i,j+1)]_{\Left}$;
  \item two adjacent cells in a column have the same color on the shared edge: for all $(i,j)\in [0,k]\times [0,2^{n}-1]$ with $i<k$,
   $[f(i,j)]_{\Up}=[f(i+1,j)]_{\Down}$;
  \item $f(0,0)=d_\Init$ (\emph{initialization});
  \item $f(k,2^{n}-1)=d_\Final$  (\emph{acceptance}).
\end{itemize}

\begin{remark}\label{remark:AssumptionTIling} Without loss of generality, we restrict ourselves to tilings $f:[0,k]\times [0,2^{n}-1] \rightarrow \Delta$
of $\Instance$ such that every cell except the last has content distinct from $d_\Final$, i.e. for all $(i,j)\neq (k,2^{n}-1)$, $f(i,j)\neq d_\Final$.
\end{remark}

 It is well-known that  checking the existence of a tiling for $\Instance$ is \EXPSPACE-complete~\cite{Harel92}.
 In the following, for each $\Lang\in \{\THT_{2}^{1}(\Eventually,\Always),\THT_2^{\,2}(\Always),\THT_2^{\,2}(\Until)\}$,  we construct in polynomial time an  $\Lang$-formula which admits an equilibrium model iff there exists a tiling of $\Instance$.
 Hence, Theorem~\ref{theorem:consistencyLowerBoundUntractable} follows.
We use the following set $P$ of atomic propositions for encoding tilings of $\Instance$:
%
\[
P = P_\Main \cup P_\Tag \cup \{\symSp\} \quad\quad P_{\Tag}=\{t_1,\ldots,t_9\}
\]
\[
P_\Main =\Delta\cup \{\$\} \cup P_\num \quad\quad P_\num = [1,n]\times \{0,1\}
\]
The atomic propositions in  $P_\num\subseteq P_\Main$ are used to encode the value of a $n$-bits counter numbering the cells of one row of a  tiling.
In particular, a cell with content $d\in\Delta$ and column number $j\in [0,2^{n}-1]$ is encoded by the words in
 $$\{(1,b_1)\}^{+}\ldots \{(n,b_n)\}^{+}\{d\}^{+}$$
 where $b_1 \ldots b_n$ is the binary encoding of the column number $j$. 
 Moreover, a row is encoded by words of the form $\{\$\}^{h}\cdot  \textit{cell}_0\ldots \textit{cell}_{2^{n}-1}$ for some $h\geq 1$, listing
 the encodings of cells from left to right.
Thus, a    tiling $f$ is encoded by  finite words  $w$  over $2^{P_\Main}$ where $w$
 corresponds to a sequence of row encodings, starting from the first row of $f$. Note that
  for all $i\geq 0$, $w(i)$ contains exactly one atomic proposition in $P_\Main$.
  The extra symbols in $P_\Tag$ and the additional proposition $u$ are used to mark segments of infinite words $\HModel$ in order to check that the projection of $\HModel$
  over $P_\Main$  has no prefix which encodes a tiling.

\myparagraph{Reductions} 
Here we focus on the fragment $\THT_{2}^{1}(\Eventually,\Always)$. The reductions  
for $\THT_2^{\,2}(\Always)$ and $\THT_2^{\,2}(\Until) $  are  in Appendix~\ref{APP:LowerBoundsUntilRelease}.
Our main tool is a notion of \emph{pseudo-tiling code}.

\begin{definition}[Pseudo-tiling codes for $\THT_{2}^{1}(\Eventually,\Always)$]\label{Def:PseudoTilingFGNested}  An interpretation $(\HModel,\TModel)$ $($over $P$$)$ is a \emph{pseudo-tiling code} for $\THT_{2}^{\,1}(\Eventually,\Always)$  if the following holds:\\
\emph{Unboundedness}: for infinitely many $i\geq 0$, $u\in \HModel(i)$.\\
\emph{Pseudo-tiling $\TModel$-requirement}: $\$\in \TModel(0)$ and:
  \begin{compactitem}
  \item $\TModel(i)\cap P_\Main$ is a singleton and $\TModel(i)\cap P_\Main=\HModel(i)\cap P_\Main$ for all $i\geq 0$;
  \item there is $i\geq 0$  such that $d_\Final\in \TModel(i)$.
   \item \emph{either} for all $i\geq 0$, $u\in \TModel(i)$ and  $\TModel(i)\cap P_\Tag=P_\Tag$ (\emph{full requirement}), \emph{or}
   $u\notin \TModel(0)$ and for all $i\geq 0$, $\TModel(i)\cap P_\Tag$ is a singleton.
  \end{compactitem}
\emph{$\HModel$-requirement:} if $\HModel\neq \TModel$ (i.e., $\HModel \sqsubset \TModel$), then $u\notin \HModel(0)$ and $\HModel(i)\cap P_\Tag$ is a singleton for all $i\geq 0$.
\end{definition}

A pseudo-tiling code $(\HModel,\TModel)$ for $\THT_{2}^{\,1}(\Eventually,\Always)$ is \emph{good} if it satisfies the full requirement. We observe the following fact.

\begin{remark}\label{remark:PseudoGoodForFGNested} Let $(\TModel,\TModel)$ be a total pseudo-tiling code for $\THT_{2}^{\,1}(\Eventually,\Always)$ which is not good. Then, $u\notin\TModel(0)$ and there exists  $\HModel$ such that $\HModel \sqsubset \TModel$, $(\HModel,\TModel)$ is a pseudo-tiling code and for all $i\geq 0$,
$\HModel(i)\cap (P\setminus \{u\})= \TModel(i)\cap (P\setminus \{u\})$.
\end{remark}

 We construct a $\THT_{2}^{1}(\Eventually,\Always)$ formula $\varphi_\Instance$ whose equilibrium models
 are the good pseudo-tiling codes $(\TModel,\TModel)$ such that  the projection of some prefix of $\TModel$ over $P_\Main$ encodes a tiling. In particular, we ensure that for a good pseudo-tiling code $(\TModel,\TModel)$, there exists $\HModel \sqsubset \TModel$ such that $(\HModel,\TModel)\models \varphi_\Instance$ iff $(\HModel,\TModel)$ is a pseudo-tiling code and   $\HModel$ is a ``slice" 
version of $\TModel$  witnessing that $\TModel$ ha  no prefix which  encodes a tiling.
 The construction of $\varphi_\Instance$ consists of three steps.
First, we define a $\THT_{2}^{1}(\Eventually,\Always)$ formula capturing the pseudo-tiling codes.

\begin{proposition}\label{Prop:FormulaPseudoFGNested} One can construct in polynomial time a $\THT_{2}^{\,1}(\Eventually,\Always)$ formula   $\varphi_\pseudo$  such that
 $(\HModel,\TModel)\models \varphi_\pseudo$ \emph{iff}
$(\HModel,\TModel)$ is a pseudo-tiling code for $\THT_{2}^{\,1}(\Eventually,\Always)$.
\end{proposition}
\begin{IEEEproof} The $\THT_{2}^{1}(\Eventually,\Always)$ formula   $\varphi_\pseudo$ is given by
\[
\begin{array}{l}
(\Always\Eventually \, u)\,\wedge \,  \$ \,\wedge \,    \displaystyle{\Always(\bigvee_{p\in P_\Main}(\,\, p\,\wedge \, \bigwedge_{p'\in P_\Main\setminus\{p\}}\neg p'))} \,\,\wedge \vspace{0.0cm}\\
 (\Eventually \, d_\Final) \,\wedge \,
 \Always(\displaystyle{\bigvee_{p\in P_\Tag}p}) \,\,\wedge  \vspace{0.0cm}\\
    \displaystyle{\Bigl(\bigl[u\vee \Eventually (\bigvee_{(p,p')\in P_\Tag: p\neq p'}(p\wedge p'))\bigr] \rightarrow \Always(\symSp\wedge\bigwedge_{p\in P_\Tag}p)\Bigr)}
\end{array}
\]
The first conjunct captures the unboundedness requirement, while the remaining conjuncts capture
 the pseudo-tiling $\TModel$-requirement and the $\HModel$-requirement.
\end{IEEEproof}

Second,   we use a family of $\THT_{2}^{1}(\Eventually,\Always)$ formulas    to mark by  propositions in $P_\Tag$ segments of infinite words on $2^{P_\Main}$.

\begin{proposition}\label{prop:FormualsForMarkFGNested} Let $t_{i_1},\ldots, t_{i_k}$ be  distinct propositions in $P_\Tag$, and
 $P_1,\ldots,P_k$ be non-empty subsets of $P_\Main$. Then,  one can construct in polynomial time a  $\THT_{2}^{\,1}(\Eventually,\Always)$ formula
$\theta(i_1|P_1,\ldots,i_k|P_k)$ over $P\setminus \{u\}$ such that: for all \emph{good pseudo-tiling codes} $(\HModel,\TModel)$ for $\THT_{2}^{\,1}(\Eventually,\Always)$ with $\HModel\neq \TModel$,
$$(\HModel,\TModel)\models\theta(i_1|P_1,\ldots,i_k|P_k) \text{ \emph{ iff} } $$
the projection of $\HModel$ over $P_\Tag$ is in $\{t_{i_1}\}^{+}\ldots \{t_{i_{k-1}}\}^{+}\{t_{i_k}\}^{\omega}$
 and for all $1\leq j\leq k$, all the main propositions which label the segment of $\HModel$ marked by $t_{i_j}$ are in $P_j$. Moreover,
 $$(\HModel,\TModel)\models\theta(i_1|P_1,\ldots,i_k|P_k) \text{ \emph{ iff} } \HModel\models_\LTL\theta(i_1|P_1,\ldots,i_k|P_k)$$
\end{proposition}
\begin{IEEEproof}
For a good pseudo-tiling code $(\HModel,\TModel)$ with $\HModel\neq \TModel$, $\TModel(i)\cap P_\Tag=P_\Tag$ and $\HModel(i)\cap P_\Tag$ is a singleton for all $i\geq 0$. 
%
%
Hence, $(\HModel,\TModel)\not\models \psi$ and
$(\TModel,\TModel)\models \psi$, where  $\psi = \displaystyle{\bigwedge_{p\in P_\Tag}\Always\,p}$. Then, 
$\theta(i_1|P_1,\ldots,i_k|P_k)$ is 
given by
 \[
\begin{array}{r}
 \underbrace{
  \displaystyle{\bigl(\bigvee_{t\in P_\Tag\setminus\{t_{i_1},\ldots,t_{i_k}\}} \Eventually \,t\bigr) \rightarrow \,\psi}}_{
  \text{$\HModel$ is only marked by tag propositions in $\{t_{i_1},\ldots,t_{i_k}\}$ }}\,\,\,\wedge \vspace{0.1cm}\\
   \underbrace{
  \displaystyle{\bigwedge_{j\in [1,k]} \Eventually t_{i_j}}}_{
  \text{every tag $t_{i_j}$  marks some position of $\HModel$ }}\,\,\,\wedge\vspace{0.1cm}\\
   \underbrace{
   \displaystyle{\bigl(\bigvee_{j\in [1,k]}\bigvee_{p\in P_\Main\setminus P_{j}} \Eventually(t_{i_j} \wedge p)\,\,\bigr) \rightarrow \,\psi}
   }_{
  \text{each $t_{i_j}$-position in $\HModel$  is labeled by a main proposition  in $P_j$ }}\,\,\,\wedge \vspace{0.1cm}\\
   \underbrace{
   \displaystyle{\bigl(\bigvee_{r,s\in [1,k]: s<r} \Eventually(t_{i_r} \wedge \Eventually \,t_{i_s})\,\,\bigr) \rightarrow \,\psi}}_{
  \text{the tags $t_{i_j}$ mark $\HModel$ according to the order $t_{i_1},\ldots,t_{i_k}$}}
\end{array}
\]
\end{IEEEproof}

\noindent The crucial step in the construction of $\varphi_\Instance$ is represented by the following result.

\newcounter{prop-FormulaBadFGNested}
\setcounter{prop-FormulaBadFGNested}{\value{proposition}}
\newcounter{sec-FormulaBadFGNested}
\setcounter{sec-FormulaBadFGNested}{\value{section}}

\begin{proposition}\label{Prop:FormulaBadFGNested} One can construct in polynomial time a $\THT_{2}^{\,1}(\Eventually,\Always)$ formula $\varphi_{\textit{bad}}$ over $P\setminus \{u\}$
such that for all  total interpretations $(\TModel,\TModel)$ which are good pseudo-tiling codes for $\THT_{2}^{\,1}(\Eventually,\Always)$,
there exists a good pseudo-tiling code for $\THT_{2}^{\,1}(\Eventually,\Always)$ of the form $(\HModel,\TModel)$ with $\HModel\neq \TModel$ and satisfying $\varphi_{\textit{bad}}$ \emph{iff}
there is \emph{no} prefix of $\TModel$  whose projection over $P_\Main$  encodes a tiling. Moreover,
for all good pseudo-tiling codes $(\HModel,\TModel)$ for $\THT_{2}^{\,1}(\Eventually,\Always)$ with
 $\HModel\neq \TModel$, $(\HModel,\TModel)\models \varphi_{\textit{bad}}$ \emph{iff}
 $\HModel\models_\LTL \varphi_{\textit{bad}}$.
\end{proposition}
\begin{IEEEproof}
the $\THT_{2}^{\,1}(\Eventually,\Always)$ formula $\varphi_{\textit{bad}}$ consists of various disjuncts
which capture all the possible conditions
such that  for a total good pseudo-tiling code $(\TModel,\TModel)$,  no prefix of $\TModel$ encodes a tiling iff some of these conditions is satisfied. These bad conditions can be summarized as follows, where for an \LTL interpretation $\TModel$ over $P$, a prefix of $\TModel$ is \emph{incomplete} if it has no position labeled by $d_\Final$:
\begin{itemize}
  \item The content of the first cell is not $d_\Init$.
  \item \emph{Either} some $\$$-position is preceded by an incomplete prefix and is followed by a $\Delta$-position,
\emph{or} some $P_\num$-position is preceded by an incomplete prefix and is followed by a $\$$-position.
  \item \emph{No} cell  preceded by an incomplete prefix has content $d_\Final$ and
is the last cell of a row. 
\item There are segments in $(\{\$\}\cup\Delta\setminus \{d_\Final\})\, P_\num^{+}\,\Delta^{+}$, preceded by incomplete prefixes, such that the suffix in
$P_\num^{+}\,\Delta^{+}$ is not a correct encoding of a cell.
\item There is a row  preceded by an incomplete prefix whose first (resp., last) cell has column number distinct
from $0$ (resp., $2^{n}-1$).
\item There are adjacent cells in a row, preceded by an incomplete prefix, whose column numbers are not consecutive.
\item \emph{Bad row (resp., column) condition}: there are two adjacent cells in a row (resp., column), preceded by an incomplete prefix,  which have different color on the shared edge. 
\end{itemize}

The above conditions are expressed in $\THT_{2}^{\,1}(\Eventually,\Always)$ by exploiting the formulas $\theta(i_1 | P_1,\ldots ,i_k| P_k)$ of  Proposition~\ref{prop:FormualsForMarkFGNested}. Here, we focus on the construction of the formula  expressing the bad column condition (a full proof of Proposition~\ref{Prop:FormulaBadFGNested} is in  Appendix~\ref{APP:FormulaBadFGNested}). Such a formula is
%
%
defined below, where  we use the following short hands:
$  R_m :=   P_\Main \setminus\{d_\Final\}$,
 $ R_c := P_\Main \setminus\{\$,d_\Final\}$, and
  $\Delta_R  := \Delta \setminus\{d_\Final\}$. Notice that we use the tag propositions
  $t_2$ and $t_7$ (resp., $t_3$ and $t_8$) to mark the cell-numbers (resp., the contents) of two cells.\vspace{0.2cm}

\noindent $\Bigl(\theta( 1 | R_m, 2| P_\num , 3| \Delta_R,  4 | R_c, 5 | \{\$\},  6 | R_c, 7 | P_\num, 8|\Delta,  9 | P_\Main) $\\
$ \vee\, \theta( 1 | R_m, 2| P_\num , 3| \Delta_R,  4 | R_c, 5 | \{\$\},   7 | P_\num, 8|\Delta,  9 | P_\Main)\,\,\vee$\\
$  \underbrace{\theta( 1 | R_\Main, 2| P_\num , 3| \Delta_R,   5 | \{\$\},  6 | R_c, 7 | P_\num, 8|\Delta,  9 | P_\Main)\phantom{
  }\Bigr)}_{
  \text{mark with $t_2$ and $t_7$ the cell-numbers of two cells $c$ and $c'$ of two adjacent rows}}$
%
 %
 \[
\begin{array}{l}
\text{\hspace{3.2cm}}\wedge\vspace{0.1cm}\\
\underbrace{\displaystyle{\bigwedge_{i\in [1,n]}\,\bigvee_{b\in \{0,1\}} \Bigl(\Eventually((i,b)\wedge t_2)\wedge \Eventually( (i,b)\wedge t_7)\Bigr)}}_{\text{the marked cells $c$ and $c'$ have the same column number}}\,\,\wedge\vspace{0.1cm}\\
  \underbrace{\displaystyle{\bigvee_{(d,d')\in \Delta\times \Delta: d_{\Up}\neq (d')_{\Down}}}\Bigl(\Eventually (d\wedge t_3)\wedge \Eventually ( d'\wedge t_8)\Bigr)}_{\text{the marked cells $c$ and $c'$ do \emph{not} have the same color on the shared edge}}
\end{array}
\]
\end{IEEEproof}

By using Propositions~\ref{Prop:FormulaPseudoFGNested} and~\ref{Prop:FormulaBadFGNested}, we deduce the following result from which
Theorem~\ref{theorem:consistencyLowerBoundUntractable} for the fragment
$\THT_2^{1}(\Eventually,\Always)$ directly follows.

\begin{lemma}\label{MainLemma:LowerBoundFGNested}
One can construct in polynomial time a $\THT_2^{\,1}(\Eventually,\Always)$ formula $\varphi_\Instance$  such that there is an equilibrium model of $\varphi_\Instance$ \emph{iff} there is a tiling of $\Instance$.
\end{lemma}
\begin{IEEEproof} Let $\varphi_\pseudo$ and $\varphi_{\textit{bad}}$ be the $\THT_2^{1}(\Eventually,\Always)$ formulas of Propositions~\ref{Prop:FormulaPseudoFGNested} and~\ref{Prop:FormulaBadFGNested}, respectively. Then:
\[
\varphi_\Instance =\varphi_\pseudo \wedge (\symSp \vee   \varphi_{\textit{bad}})
\]
We now prove that the construction is correct.
First, assume that there exists an equilibrium model $(\TModel,\TModel)$   of $\varphi_\Instance$. By construction of $\varphi_\Instance$ and Proposition~\ref{Prop:FormulaPseudoFGNested}, $(\TModel,\TModel)$ is a pseudo-tiling code.
We claim that $(\TModel,\TModel)$ is good as well. We assume the contrary and derive a contradiction.
By Remark~\ref{remark:PseudoGoodForFGNested}, $u\notin\TModel(0)$ and there exists $\HModel \sqsubset \TModel$ such that $(\HModel,\TModel)$ is a pseudo-tiling code and for all $i\geq 0$,
$\HModel(i)\cap (P\setminus \{u\})= \TModel(i)\cap (P\setminus \{u\})$. Since $u\notin\TModel(0)$ and $(\TModel,\TModel)\models \varphi_\Instance$,   $(\TModel,\TModel)\models \varphi_{\textit{bad}}$. Moreover, since  $\varphi_{\textit{bad}}$ is a formula over
$P\setminus \{u\}$ (Proposition~\ref{Prop:FormulaBadFGNested}), by Proposition~\ref{Prop:FormulaPseudoFGNested} we obtain that
$(\HModel,\TModel)$ satisfies $\varphi_\Instance$, which contradicts the hypothesis that $(\TModel,\TModel)$ is an equilibrium model of $\varphi_\Instance$. Thus,  $(\TModel,\TModel)$ is a good pseudo-tiling code.
If no prefix of $\TModel$  encodes a tiling, by Proposition~\ref{Prop:FormulaBadFGNested} there exists $\HModel \sqsubset \TModel$ such that $(\HModel,\TModel)\models \varphi_{\textit{bad}}$
and $(\HModel,\TModel)$ is a pseudo-tiling code; hence, by  Proposition~\ref{Prop:FormulaPseudoFGNested},
$(\HModel,\TModel)$ satisfies $\varphi_\Instance$, which contradicts the assumption that $(\TModel,\TModel)$ is an equilibrium model. Thus,
some prefix of $\TModel$ encodes a tiling. Hence, there exists a tiling  of $\Instance$.

Now, assume that there exists a tiling $f$ of $\Instance$. Let $(\TModel,\TModel)$ be any  \emph{good} pseudo-tiling code
such that   the projection of some prefix of $\TModel$ over $P_\Main$ is an encoding of $f$. Note that such a $(\TModel,\TModel)$ exists.
Since $\symSp\in\TModel(0)$, by construction and Proposition~\ref{Prop:FormulaPseudoFGNested},   $(\TModel,\TModel)$
satisfies $\varphi_\Instance$.
We assume that $(\TModel,\TModel)$ is not an equilibrium model and derive a contradiction, hence, the result follows. Thus, there is
$\HModel \sqsubset \TModel$ such that $(\HModel,\TModel)\models \varphi_\Instance$. By construction and  Proposition~\ref{Prop:FormulaPseudoFGNested},
$(\HModel,\TModel)$ is a pseudo-tiling code. Moreover, since $(\TModel,\TModel)$ is good, $(\HModel,\TModel)$ is good as well. Since $\HModel\neq \TModel$, $u\notin \HModel(0)$ (Definition~\ref{Def:PseudoTilingFGNested}).
 Hence, being
$(\HModel,\TModel)\models \varphi_\Instance$, by construction,  $(\HModel,\TModel)\models  \varphi_{\textit{bad}}$. By
Proposition~\ref{Prop:FormulaBadFGNested} there is no prefix of $\TModel$ which encodes a tiling. This contradicts the hypothesis, and we are done.

\end{IEEEproof}

\subsection{\textbf{The fragment $\THT_1$}}

We establish that  the \TEL consistency problem for the simple fragment  $\THT_1$, where no nesting of temporal modalities is allowed, is already \NEXPTIME-complete even for the smaller fragments $\THT_1(\Eventually,\Always)$, $\THT_1(\Until)$, and $\THT_1(\Release)$.

\subsubsection{Lower Bounds}

\begin{theorem}\label{theorem:consistencyLowerBoundTHTFGOneTemporalNesting} The
 \TEL consistency problems for   $\THT_1^{\,2}(\Eventually,\Always)$, $\THT_1^{\,2}(\Until)$, and $\THT_1^{\,2}(\Release)$ are \NEXPTIME-hard.
\end{theorem}

\noindent Theorem~\ref{theorem:consistencyLowerBoundTHTFGOneTemporalNesting} is proved by
 polynomial-time reductions from a domino-tiling problem for grids with rows and columns of
 exponential length~\cite{Boas97}.   An instance $\Instance =\tupleof{C,\Delta,n,d_\Init,d_\Final}$ of
  this problem is  as in the proof of
   Theorem~\ref{theorem:consistencyLowerBoundUntractable}.
  However, here, a  tiling of $\Instance$ is defined as a mapping $f:[0,2^{n}-1]\times [0,2^{n}-1] \rightarrow \Delta$,  i.e.,  the number of rows and the number of columns is $2^{n}$.   It is well-known that checking the existence of a tiling for
  $\Instance$ is \NEXPTIME-complete~\cite{Boas97}.
%
 We focus on the fragment
$\THT_1^{2}(\Eventually,\Always)$. The reductions for the fragments $\THT_1^{2}(\Until)$ and
$\THT_1^{2}(\Release)$ are given in Appendix~\ref{APP:consistencyLowerBoundTHTFGOneTemporalNesting}.

\myparagraph{Encoding of  tilings for $\THT_1^{\,2}(\Eventually,\Always)$} we use the following set $P$ of propositions:
\[
P = P_\Main \cup P_\Tag \cup \{\symSp\} \quad P_\Main =\Delta \cup P_\num^{r} \cup P_\num^{c}
\]
\[
P_\num^{r} =\{r\}\times [1,n]\times \{0,1\}   \quad   P_\num^{c} =\{c\}\times[1,n]\times \{0,1\}
\]
\[
P_{\Tag}=\{t_1,t_2,t_3\}\times \{\overline{p}\mid p\in P_\num^{r}\cup P_\num^{c}\}
\]
We use the atomic propositions in  $P_\num^{r}$ (resp., $P_\num^{c}$)  to encode the value of a $n$-bits counter numbering the $2^{n}$ rows (resp., columns) of a  tiling.
In particular, a cell with content $d\in\Delta$, row number $i\in [0,2^{n}-1]$, and column number $j\in [0,2^{n}-1]$  is encoded by
the subset of $P_\Main$ given by
 $$\{d,(r,1,b_1),\ldots,(r,n,b_n),(c,1,b'_1),\ldots,(c,n,b'_n)\}$$
 where $b_1 \ldots b_n$ (resp., $b'_1,\ldots,b'_n$) is the binary encoding of the row number $i$ (resp., column number $j$). 
 We call such subsets of $P_\Main$ \emph{cell-codes}. A  tiling $f$ is  then encoded by the infinite words $w$ over $2^{P_\Main}$ satisfying the following:
 \begin{itemize}
   \item for all $i,j\in [0,2^{n}-1]$, there is $h\geq 0$ such that $w(h)$ is the cell-code of the $(i,j)^{th}$ cell of $f$;
   \item for all $h\geq 0$, $w(h)$ encodes the $(i,j)^{th}$ cell of $f$ for some $i,j\in [0,2^{n}-1]$.
 \end{itemize}

 The extra symbols in $P_\Tag$ and the additional proposition $u$ are used to mark  infinite words $\HModel$ in order to check that the projection of $\HModel$ over $P_\Main$ does not encode a tiling. In particular,  a \emph{cell-number code}   is a subset of $P_\Tag$ of the form
 \[
 \{\overline{(r,1,b_1)},\ldots,\overline{(r,n,b_n)}, \overline{(c,1,b'_1)},\ldots,\overline{(c,n,b'_n)}\}
 \]

\myparagraph{Reduction for  $\THT_1^{\,2}(\Eventually,\Always)$} as in the proof of Theorem~\ref{theorem:consistencyLowerBoundUntractable}, we use a notion of \emph{pseudo-tiling code}.

\begin{definition}[Pseudo-tiling codes for $\THT_1^{2}(\Eventually,\Always)$]An interpretation $(\HModel,\TModel)$ $($over $P$$)$ is a pseudo-tiling code for  $\THT_1^{\,2}(\Eventually,\Always)$ if the following holds:\\
\emph{Pseudo-tiling $\TModel$-requirement}: for all $i\geq 0$, $\TModel(i)\cap P_\Main$ is a cell-code and $\HModel(i)\cap P_\Main=\TModel(i)\cap P_\Main$. Moreover,
      \begin{compactitem}
      \item there is $i\geq 0$ such that $\TModel(i)\cap P_\Main$ has row-number $0$, column-number $0$ and $d_\Init\in \TModel(i)$ (\emph{initialization});
        \item there is $i\geq 0$ such that $\TModel(i)\cap P_\Main$ has row-number $2^{n}-1$, column-number $2^{n}-1$, and $ d_\Final\in \TModel(i)$ (\emph{acceptance}).
      \end{compactitem}
\emph{Full $\TModel$-requirement}: for all $i$, $\TModel(i)\cap P_\Tag=P_\Tag$ and $\symSp\in \TModel(i)$;\\
\emph{$\HModel$-requirement}: if $\HModel \neq\TModel$, then $\symSp\notin \HModel(i)$ for all $i\geq 0$, and:
      \begin{compactitem}
      \item \emph{either} there is a cell-number code $P'\subseteq P_\Tag$  such that the projection of $\HModel$ over $P_\Tag$ is $(P')^{\omega}$;
        \item \emph{or} for all $i\geq 0$, $\HModel(i)\cap P_\Tag$ is a singleton contained in $\{t_1,t_2,t_3\}$.
      \end{compactitem}
\end{definition}

 We construct in polynomial time a $\THT_1^{2}(\Eventually,\Always)$ formula $\varphi_\Instance$ in such a way that (i) the total interpretations captured by
$\varphi_\Instance$ are the total interpretations   $(\TModel,\TModel)$ which are pseudo-tiling codes for $\THT_1^{2}(\Eventually,\Always)$, and (ii) there exists $\HModel \sqsubset \TModel$ such that $(\HModel,\TModel)\models \varphi_\Instance$ iff the projection of $\TModel$ over $P_\Main$  does not encode a tiling.
The construction of $\varphi_\Instance$ consists of two steps.
First, we define a formula capturing the pseudo-tiling codes.

\newcounter{prop-FormulaPseudoForFGOne}
\setcounter{prop-FormulaPseudoForFGOne}{\value{proposition}}
\newcounter{sec-FormulaPseudoForFGOne}
\setcounter{sec-FormulaPseudoForFGOne}{\value{section}}

\begin{proposition}\label{Prop:FormulaPseudoForFGOne} One can construct in polynomial time a $\THT_1^{\,2}(\Eventually,\Always)$ formula $\varphi_\pseudo$
 such that
 $(\HModel,\TModel)\models \varphi_\pseudo$ \emph{iff}
$(\HModel,\TModel)$ is a pseudo-tiling code for $\THT_1^{\,2}(\Eventually,\Always)$.
\end{proposition}

\noindent The proof of Proposition~\ref{Prop:FormulaPseudoForFGOne} is crucially based on the use of nested implication. In particular, we exploit the conjunct $\neg u \rightarrow u$ which is satisfied by an interpretation
 $(\HModel,\TModel)$ iff $u\in \TModel(0)$. For details, see Appendix~\ref{APP:FormulaPseudoForFGOne}. The second step in the construction of $\varphi_\Instance$ is given by the following result.

\begin{proposition}\label{Prop:FormulaBadFGOne} One can construct in polynomial time a $\THT_1^{\,1}(\Eventually,\Always)$  formula $\varphi_{\textit{bad}}$  such that for all  total interpretations $(\TModel,\TModel)$ which are pseudo-tiling codes for $\THT_1^{\,2}(\Eventually,\Always)$,
there exists a pseudo-tiling code for $\THT_1^{\,2}(\Eventually,\Always)$ of the form $(\HModel,\TModel)$ with $\HModel\neq \TModel$ and satisfying $\varphi_{\textit{bad}}$ \emph{iff}
the projection of $\TModel$ over $P_\Main$ does \emph{not} encode a tiling.
\end{proposition}
\begin{IEEEproof}
First, for all $t,t'\in \{t_1,t_2,t_3\}$ and $\tau\in\{r,c\}$, we
consider the   $\THT_1^{1}(\Eventually,\Always)$ formula $\varphi(t,t',\tau)$ given by
\[
\{\bigvee_{i\in [1,n]} [\Eventually\bigl( (t\vee t')\wedge (\tau,i,0)\bigr)\wedge \Eventually\bigl( (t\vee t') \wedge (\tau,i,1)\bigr)]\} \rightarrow \symSp
\]
Evidently, for each pseudo-tiling code $(\HModel,\TModel)$ for $\THT_1^{2}(\Eventually,\Always)$ with $\HModel\neq \TModel$, $(\HModel,\TModel)\models\varphi(t,t',r)$ (resp., $(\HModel,\TModel)\models\varphi(t,t',c)$) \emph{iff} for  all the positions of $\HModel$ marked by the propositions $t$ and $t'$, the associated cell-codes   have the same row-number (resp., cell-number). Then the $\THT_1^{1}(\Eventually,\Always)$ formula $\varphi_{\textit{bad}}$   consist of four disjuncts.
The first disjunct checks that there is a cell-number $(i,j)$ such that no cell-code has cell-number $(i,j)$.
\[
\begin{array}{l}
  \underbrace{\Bigl(\displaystyle{\bigvee_{p\in P_\Tag\setminus \{t_1,t_2,t_3\}}\Eventually \,p\Bigr)}}_{\text{ all the positions of $\HModel$ are marked by the same cell-number code $P'\subseteq P_\Tag$}} \,\,\,\wedge \vspace{0.1cm} \\
  \underbrace{\displaystyle{\Always\Bigl(\bigvee_{i\in[1,n]}\bigvee_{\tau\in\{r,c\}}\bigvee_{b\in\{0,1\}} [(\tau,i,b)\wedge \overline{(\tau,i,1-b)}]\Bigr)}}_{\text{at every position, the current cell-code has cell-number non-corresponding to $P'$}}
\end{array}
\]
The second disjunct 
checks that there are two cell-codes with the same cell-number but distinct content.\vspace{0.2cm}

 \noindent $(\Eventually t_1) \wedge \displaystyle{\bigwedge_{\tau\in \{r,c\}} \varphi(t_1,t_1,\tau)} \wedge
 \displaystyle{\bigvee_{d,d'\in \Delta: d\neq d'}[\Eventually(t_1\wedge d)\wedge \Eventually(t_1\wedge d')]}$\vspace{0.2cm}

Finally,   the third (resp., fourth) disjunct   checks that  there are two adjacent cells in a column (resp., row)  which do \emph{not} have the same color on the shared edge. We illustrate the construction of the fourth disjunct. %
\[
\begin{array}{r}
   \underbrace{(\Eventually t_1) \wedge (\Eventually t_2) \wedge \varphi(t_1,t_1,r) \wedge \varphi(t_2,t_2,r) \wedge \varphi(t_1,t_2,c)}_{\text{mark two cells $cl_1$ and $cl_2$ with the same column number}} \,\,\wedge \vspace{0.1cm}\\
   \displaystyle{
 \bigvee_{i\in [1,n]}}\Bigl[\,\, \Eventually((r,i,0)\wedge t_1)  \,\wedge \,  \Eventually((r,i,1)\wedge t_2)
 \, \wedge \text{\hspace{0.7cm}}\\ \displaystyle{\bigwedge_{j\in [1,i-1]}}\Bigl(\Eventually((r,j,1)\wedge t_1) \wedge  \Eventually((r,j,0)\wedge t_2)\Bigr)\, \wedge \text{\hspace{0.7cm}}\\
  \underbrace{\displaystyle{\bigwedge_{j\in [i+1,n]}\bigvee_{b\in\{0,1\}}\Bigl(\Eventually((r,j,b)\wedge t_1) \wedge  \Eventually((r,j,b)\wedge t_2)\Bigr)\,\,\Bigr]}}_{\text{ $cl_1$ and $cl_2$ have consecutive row-numbers}} \,\,
   \wedge\vspace{0.1cm}\\
   \underbrace{  \bigvee_{(d,d')\in \Delta\times \Delta: d_\Up \neq (d')_\Down}[\Eventually(t_1\wedge d)\wedge \Eventually(t_2\wedge d')] }_{\text{the  cells $cl_1$ and $cl_2$ do not have the same color on the shared edge}}
\end{array}
\]

By construction, for all pseudo-tiling codes $(\HModel,\TModel)$ for $\THT_1^{2}(\Eventually,\Always)$ such that
$\HModel\neq \TModel$,  if $(\HModel,\TModel)\models \varphi_{\textit{bad}}$ then $\TModel$ does not encode a tiling.
On the other hand, for each total pseudo-tiling code $(\TModel,\TModel)$ for $\THT_1^{2}(\Eventually,\Always)$ such that $\TModel$ does not encode a tiling, there exists
a pseudo-tiling code for $\THT_1^{2}(\Eventually,\Always)$ of the form $(\HModel,\TModel)$ such that $\HModel\neq \TModel$ and
$(\HModel,\TModel)$ satisfies $\varphi_{\textit{bad}}$. Hence, Proposition~\ref{Prop:FormulaBadFGOne} follows.
\end{IEEEproof}\vspace{0.1cm}

The  $\THT_1^{2}(\Eventually,\Always)$ formula $\varphi_\Instance$ is defined as follows:
\[
\varphi_\Instance =\varphi_\pseudo \wedge (\symSp \vee   \varphi_{\textit{bad}})
\]
where  $\varphi_\pseudo$ and $\varphi_{\textit{bad}}$ are the $\THT_1^{2}(\Eventually,\Always)$ formulas of Proposition~\ref{Prop:FormulaPseudoForFGOne} and~\ref{Prop:FormulaBadFGOne}, respectively.
By Propositions~\ref{Prop:FormulaPseudoForFGOne} and~\ref{Prop:FormulaBadFGOne}, we easily deduce the following result, hence,
 Theorem~\ref{theorem:consistencyLowerBoundTHTFGOneTemporalNesting} for the fragment $\THT_1^{2}(\Eventually,\Always)$  directly follows.

\begin{lemma}[Correctness of the construction] There is an equilibrium model of $\varphi_\Instance$ \emph{iff} there is a tiling of $\Instance$.
\end{lemma}

\subsubsection{Upper Bound for $\CON(\THT_1)$}

An interpretation $\Model$ is \emph{strongly ultimately periodic} if there is $i\geq 0$ such that
$\Model(k)=\Model(i)$ for all $k\geq i$. In such a case, the \emph{size}  of $\Model$ is defined as $j+1$, where  $j$ is the smallest $i$ satisfying the previous condition. In order to solve $\CON(\THT_1)$, we first  show that we can restrict ourselves to the equilibrium models which are strongly ultimately periodic and whose sizes are singly exponential in the size of the given formula.

\newcounter{lemma-SingleExpontialEMforTemporalOne}
\setcounter{lemma-SingleExpontialEMforTemporalOne}{\value{lemma}}
\newcounter{sec-SingleExpontialEMforTemporalOne}
\setcounter{sec-SingleExpontialEMforTemporalOne}{\value{section}}

\begin{lemma}\label{lemma:SingleExpontialEMforTemporalOne} Let $\varphi$ be a  $\THT_1$ formula having some equilibrium model.
Then, there exists a strongly ultimately periodic equilibrium model of $\varphi$ of size at most $2+2^{|\varphi|}$.
\end{lemma}

The proof of Lemma~\ref{lemma:SingleExpontialEMforTemporalOne}, which is detailed in Appendix~\ref{APP:SingleExpontialEMforTemporalOne}, exploits a notion of bisimilarity and contraction for interpretations. Bisimilar interpretations are indistinguishable from $\THT_1$ formulas, and the notion of contraction, which ensures bisimilarity, allows to `extract' from
a total interpretation   a strongly ultimately periodic interpretation
  of size singly exponential in $|P|$ by preserving the property of being an equilibrium model of a $\THT_1$ formula.

 Next, we show that for a $\THT_1$ formula $\varphi$ and a strongly ultimately periodic total interpretation
$\Model$ of size singly exponential in $|\varphi|$, checking that $\Model$ is an equilibrium model of $\varphi$ can be done in time singly exponential in $|\varphi|$. For this, we use a notion of extracted interpretation depending on $\varphi$, which generalizes a similar notion exploited in \cite{DemriS02} for solving \LTL satisfiability for  $\THT_1$ (considered as \LTL fragment).

\begin{definition}[Witness Extraction]\label{Def:WitnessExtraction} \emph{Given $\varphi\in \THT_1$ and an interpretation $\Model=(\HModel,\TModel)$,
a \emph{witness pattern of $\Model$ for $\varphi$} is an infinite sequence $n_0<n_1<\ldots$ of increasing natural numbers
such that there is $k\geq 0$ so that $\Model(n_i)=\Model(n_{k+1})$ for all $i\geq k+1$, and the finite set of positions
$W=\{n_0,\ldots,n_k\}$ \emph{minimally} satisfies the following conditions:
\begin{compactitem}
  \item $0\in W$ and if there is some subformula of $\varphi$ of the form $\Next\psi$, then $1\in W$;
  \item if $\Model$ is not total, then for some $i$, $\HModel(i)\subset \TModel(i)$ and $i\in W$;
  \item for each subformula $\varphi_1\Until\varphi_2$ of $\varphi$:
  \begin{compactitem}
    \item if $\Model\models \varphi_1\Until\varphi_2$, then the smallest position $i$ such that $\Model,i\models \varphi_2$ is in $W$.
    \item if $\Model\not\models \varphi_1\Until\varphi_2$ and $\Model\models \Eventually\varphi_2$, then the smallest position $i$ such that
    $\Model,i\not\models \varphi_1$ is in $W$. 
  \end{compactitem}
  \item for each subformula $\varphi_1\Release\varphi_2$ of $\varphi$:
  \begin{compactitem}
    \item if $\Model\not\models \varphi_1\Release\varphi_2$, then the smallest position $i$ such that $\Model,i\not\models \varphi_2$ is in $W$.
    \item if $\Model\models \varphi_1\Release\varphi_2$ and $\Model\not\models \Always\varphi_2$, then the smallest position $i$ such that
    $\Model,i\models \varphi_1\wedge\varphi_2$ is in $W$. 
  \end{compactitem}
\end{compactitem}
Note that witness patterns of $\Model$ for $\varphi$ exist. A \emph{witness extraction of $\Model$ for $\varphi$} is an interpretation $\Model_W$ of
the form $\Model_W=\Model(n_0),\Model(n_1),\ldots$, where $n_0<n_1<\ldots$ is a witness pattern of $\Model$ for $\varphi$.
Evidently, $\Model_W$ is strongly ultimately periodic with size at most
$|\varphi|+3$.}
\end{definition}

We establish the following result whose proof is in Appendix~\ref{APP:WitnessExtractionTemporalOne}.

\newcounter{lemma-WitnessExtractionTemporalOne}
\setcounter{lemma-WitnessExtractionTemporalOne}{\value{lemma}}
\newcounter{sec-WitnessExtractionTemporalOne}
\setcounter{sec-WitnessExtractionTemporalOne}{\value{section}}

\begin{lemma}\label{lemma:WitnessExtractionTemporalOne} Given  $\varphi\in \THT_1$, the following holds.
\begin{compactenum}
  \item Let $\Model$ and $\Model\,'$ be two interpretations such that $\Model\,'=\Model(n_0),\Model(n_1),\ldots$ where $n_0<n_1<\ldots$ is an infinite sequence of increasing natural numbers containing all the positions of some witness pattern of $\Model$ for $\varphi$. Then, for each subformula $\psi$ of $\varphi$, $\Model\models \psi$ \emph{iff} $\Model\,'\models \psi$.
  \item Let $\Model=(\TModel,\TModel)$ be a total strongly ultimately periodic interpretation satisfying $\varphi$ of size $m$. Then
   $\Model$ is an equilibrium model of $\varphi$ \emph{iff} for each $\HModel \sqsubset \TModel$ such that
   $(\HModel, \TModel)$ is a strongly ultimately periodic interpretation of size at most $m+|\varphi|+3$, $(\HModel, \TModel)\not\models\varphi$.
\end{compactenum}
\end{lemma}

By Lemmata~\ref{lemma:SingleExpontialEMforTemporalOne} and~\ref{lemma:WitnessExtractionTemporalOne}, we obtain the desired result.

\begin{theorem}\label{theo:MainUpperBoundTemporalOne}
 $\CON(\THT_1)$ is in \NEXPTIME.
\end{theorem}
\begin{IEEEproof}
Let $\varphi$ be a $\THT_1$  formula.
By Lemma~\ref{lemma:SingleExpontialEMforTemporalOne}, if $\varphi$
  has an equilibrium model, then there is some equilibrium model $(\TModel,\TModel)$ of $\varphi$ which is strongly ultimately periodic and whose size is at most $2+2^{|\varphi|}$.
Nondeterministically guessing such a $(\TModel,\TModel)$ and checking that $(\TModel,\TModel)$ satisfies $\varphi$ can be done in singly exponential time. Moreover, by Lemma~\ref{lemma:WitnessExtractionTemporalOne}, for verifying that $(\TModel,\TModel)$ is an equilibrium model, it suffices to check that for every strongly ultimately periodic interpretation $(\HModel_W,\TModel_W)$  of size at most $|\varphi|+3$, it holds that
  $(\HModel_W,\TModel_W)\not\models \varphi$ whenever $(\HModel_W,\TModel_W)$ satisfies the following condition.
  \vspace{0.1cm}

\noindent  \emph{Downward condition:} there is $\HModel \sqsubset \TModel$ such that $(\HModel,\TModel)$
    is strongly ultimately periodic with size  at most $5+2^{|\varphi|}+ |\varphi|$, and $(\HModel_W,\TModel_W)$ is a witness extraction of
    $(\HModel,\TModel)$ for $\varphi$.\vspace{0.1cm}

  By Definition~\ref{Def:WitnessExtraction}, one can deduce that checking whether $(\HModel_W,\TModel_W)$ satisfies the downward condition can be done in singly exponential (deterministic) time. Thus, since the number of strongly ultimately periodic interpretations  of size at most $|\varphi|+3$ is singly exponential in the  size of $|\varphi|$, membership in \NEXPTIME\ for $\CON(\THT_1)$ follows.
\end{IEEEproof}

\section{Tractable fragments}\label{sec:TractableFragments}

We now turn to the syntactical fragments of $\THT$, as defined in Subsection~\ref{sec:ResultSummary}, which are not captured by the results of
Section~\ref{sec:UntractableFragments}. For each of these fragments, except  the fragment $\THT^{0}$, we will show that the \TEL consistency problem is complete
for some complexity class in  $\{$\NP, \CompEL, \PSPACE \!$\}$.
For the fragment $\THT^{0}$, where no use of implication (and negation) is allowed, we are only able to provide a \PSPACE\ lower bound, as established by the following theorem. Notice that Theorem~\ref{theorem:LowerBoundPositiveTHTMain}, whose proof is given in Appendix~\ref{APP:LowerBoundPositiveTHT},  is, in fact,  surprising since a $\THT^{0}$ formula is always satisfiable.

\begin{theorem}\label{theorem:LowerBoundPositiveTHTMain} $\CON(\THT^{\,0})$ is \PSPACE-hard.
\end{theorem}

\subsection{\textbf{The fragment $\THT_1(\Next,\Always)$}}

The proposed approach for the fragment $\THT_1(\Next,\Always)$  is based on the  notion of witness extraction of  Definition~\ref{Def:WitnessExtraction}. The main result is as follows.

\begin{lemma}\label{lemma:smallSizePropertyXGOne} Let   $\varphi\in \THT_1(\Next,\Always)$ and
$\Model$ be an equilibrium model of $\varphi$. Then, every   witness extraction of $\Model$ for $\varphi$ is still an equilibrium model of $\varphi$. \end{lemma}
\begin{IEEEproof}
 let $\Model=(\TModel,\TModel)$ be an equilibrium model of $\varphi$ and $\Model_W =(\TModel_W,\TModel_W)$ be a witness extraction of $\Model$ for $\varphi$. We show that $\Model_W$ is an equilibrium model of $ \varphi$.
 By Lemma~\ref{lemma:WitnessExtractionTemporalOne}(1), $\Model_W$ satisfies $\varphi$.
Fix $\HModel_W \sqsubset \TModel_W$. It remains to prove that $(\HModel_W,\TModel_W)\not\models \varphi$. Let $n_0 < n_1 < \ldots$ be the witness pattern of $\Model$ for $\varphi$ such that $\TModel_W=\TModel(n_0), \TModel(n_1),\ldots$.
Define $\HModel$ as the \LTL interpretation where: for all $i\geq 0$, if $i=n_j$ for some $j$, then $\HModel(i)=\HModel_W(j)$; otherwise, $\HModel(i)=\TModel(i)$. Evidently,
$\HModel \sqsubset \TModel$. Let $\Model'=(\HModel,\TModel)$ and $\Model'_W=(\HModel_W,\TModel_W)$. Note that
 $\Model'_W=\Model'(n_0), \Model'(n_1),\ldots$.
 We prove that for each subformula $\psi$ of $\varphi$, $\Model'\models \psi$ iff $\Model'_W\models \psi$. Hence, since $\Model'\not\models \varphi$ ($(\TModel,\TModel)$ is an equilibrium model of $\varphi$), the result follows.

The unique non-trivial case is  when $\psi=\Always\psi'$. The implication  $\Model'_W\not\models \Always\psi'$ $\Rightarrow$ $\Model'\not\models \Always\psi'$ easily follows from the construction and the fact that $\psi'$ has no temporal modalities. Now, assume that $\Model'_W\models \Always\psi'$. We need to prove that for all $i\geq 0$, $\Model',i\models \psi'$. If $i=n_j$ for some $j\geq 0$, then $\Model'(i)=\Model'_W(j)$. Thus, since $\psi'$ has no temporal modalities, by hypothesis, the result follows. Otherwise, by construction, $\Model'(i)=(\TModel(i),\TModel(i))$. 
We assume that $\Model',i\not\models \psi'$ and derive a contradiction. Since $\psi'$ has no temporal modalities, we obtain that  $(\TModel,\TModel)\not\models \Always\psi'$. By Lemma~\ref{lemma:WitnessExtractionTemporalOne}(1), $(\TModel_W,\TModel_W)\not\models \Always\psi'$, hence, $\Model'_W=(\HModel_W,\TModel_W)\not\models \Always\psi'$ as well
(Proposition~\ref{prop:fundamentalsTHT}(1)), which contradicts the hypothesis, and we are done.
\end{IEEEproof}

By applying Lemmata~\ref{lemma:WitnessExtractionTemporalOne} and \ref{lemma:smallSizePropertyXGOne}, we obtain: 

\begin{theorem}\label{theo:upperBoundFragmentOneGX}
 $\CON(\THT_1(\Next,\Always))$ is $\CompEL$-complete.
\end{theorem}
\begin{IEEEproof}
The lower bound  directly follows from $\CompEL$-completeness of $\CON(\HT)$~\cite{EiterG95,Pearce06}. For the matching upper bound, let
$\varphi$ be a $\THT_1(\Next,\Always)$  formula.
By Lemma~\ref{lemma:smallSizePropertyXGOne} and Definition~\ref{Def:WitnessExtraction}, if $\varphi$ has an equilibrium model, then there is some equilibrium model $(\TModel,\TModel)$ of $\varphi$ which is strongly ultimately periodic and whose size is at most $|\varphi|+3$. Nondeterministically guessing such a $(\TModel,\TModel)$ and checking that $(\TModel,\TModel)$ satisfies $\varphi$ can be done in polynomial time. Moreover, by Lemma~\ref{lemma:WitnessExtractionTemporalOne}(2), to verify that $(\TModel,\TModel)$ is an equilibrium model, it suffices to check that each
strongly ultimately periodic interpretation of size at most $2(|\varphi|+3)$   and  of the form $(\HModel,\TModel)$ such that $\HModel \sqsubset \TModel$,  does not satisfy  $\varphi$. Universally guessing such a $(\HModel,\TModel)$ and checking that it does not satisfy $\varphi$ can be done in polynomial time. Hence, the result follows.
\end{IEEEproof}

\subsection{\textbf{The fragments $\THT^{\,1}(\Next,\Release)$, $\THT^{\,1}(\Next,\Until)$, and $\THT_1^{\,1}$}}

By Theorem~\ref{theorem:consistencyLowerBoundUntractable}, the \TEL consistency problem for $\THT^{1}$ where there is no nesting of implication is already \EXPSPACE-complete. However, we now show that for the relevant fragments
$\THT^{1}(\Next,\Release)$ and $\THT^{1}(\Next,\Until)$ of $\THT^{1}$, where the combined use of modalities $\Until$ and $\Release$ is disallowed, the problem is instead \PSPACE-complete. Additionally, we establish that $\CON(\THT_1^{1})$ is \NP-complete.

\subsubsection{The fragments $\THT^{\,1}(\Next,\Release)$ and  $\THT_1^{\,1}$}
For these two fragments, we first show that  \LTL satisfiability always guarantees the existence of minimal \LTL models.

\begin{theorem}\label{theorem:minimalityRXOneImplication} Every \LTL satisfiable $\THT^{\,1}(\Next,\Release)$ $($resp.,
$\THT_1$$)$ formula admits a minimal \LTL  model.
\end{theorem}
\begin{IEEEproof} We focus on the fragment $\THT^{1}(\Next,\Release)$ (for the fragment $\THT_1$, details can be found in Appendix~\ref{APP:minimalityRXOneImplication}). The proof for
$\THT^{1}(\Next,\Release)$
is by contradiction. So, assume that there exists a $\THT^{1}(\Next,\Release)$ formula $\varphi$ such that $\varphi$ is \LTL
satisfiable but there is no minimal \LTL model of $\varphi$. Let $(\TModel_n)_{n\geq 0}$ be any infinite sequence of \LTL models of $\varphi$ satisfying the following:
\begin{itemize}
  \item $\TModel_0$ is any \LTL model of $\varphi$;
  \item for all $n\geq 0$, $\TModel_{n+1}$ is any \LTL model of $\varphi$ such that $\TModel_{n+1} \sqsubset \TModel_{n}$ and the following holds;\\
\noindent \emph{Finite minimal requirement for $n$:} there is no \LTL model $\HModel$ of $\varphi$ such that $\HModel \sqsubset \TModel_{n}$  and: (i) for all $i\in [0,n+1]$,
  $\HModel(i)\subseteq \TModel_{n+1}(i)$, and (ii) for some $i\in [0,n+1]$, $\HModel(i)\subset  \TModel_{n+1}(i)$.
\end{itemize}

By hypothesis, such a sequence $(\TModel_n)_{n\geq 0}$ exists. Let $\TModel$ be the \LTL interpretation defined as follows: for all $i\geq 0$,
\[
\TModel(i):= \displaystyle{\bigcap_{n\geq 0}\TModel_n(i)}
\]
We will show that $\TModel$ is a minimal model of $\varphi$, which contradicts the assumption. Hence, the result follows.
First, we observe the following.\vspace{0.2cm}

\noindent \emph{Claim~1:} 1) $\TModel_{n+1} \sqsubset \TModel_n$ and $\TModel \sqsubset \TModel_n$ for all $n\geq 0$;
\begin{compactenum}
\item[2)]  for all $i\geq 0$, there is $k\geq 0$ such that for all $n\geq k$, $\TModel_n(i)=\TModel(i)$;
\item[3)] for all $\HModel \sqsubset \TModel$, $\HModel\not\models_\LTL \varphi$.
  \end{compactenum}\vspace{0.1cm}

  \noindent \emph{Proof of Claim~1:} Properties~1 and~2 directly follow by construction. For Property~3, let
  $\HModel \sqsubset \TModel$, and $n$ be any natural number such that for some $i\in [0,n+1]$, $\HModel(i)\subset \TModel(i)$.
  By Property~1, $\TModel \sqsubset \TModel_{n+1}$ and $\TModel_{n+1} \sqsubset \TModel_n$. Hence, $\HModel \sqsubset \TModel_{n}$  and: (i) for all $i\in [0,n+1]$,
  $\HModel(i)\subseteq \TModel_{n+1}(i)$, and (ii) for some $i\in [0,n+1]$, $\HModel(i)\subset  \TModel_{n+1}(i)$. Thus,
   by the finite minimal requirement for $n$, $\HModel\not\models_\LTL \varphi$.\qed \vspace{0.1cm}

 Next, we prove the following.\vspace{0.1cm}

\noindent \emph{Claim~2:} Let $\phi$ be a $\THT^{1}(\Next,\Release)$ formula and $i\geq 0$ such that $\TModel,i\models_\LTL\neg\phi$. Then, there is $k\geq 0$ such that for all $n\geq k$,
$\TModel_n,i\models_\LTL\neg\phi$.
\vspace{0.1cm}

\noindent \emph{Proof of Claim~2:} first, we recall that for a \THT\ formula $\psi$ (considered as \LTL formula), the
\emph{\LTL normal form of $\psi$} is obtained by pushing inward negations to propositional literals using De Morgan's laws, the duality between
 $\Until$ and $\Release$, and the fact the in the classical interpretation of implication, formula $\xi_1 \rightarrow \xi_2$ can be
 rewritten as $\neg\xi_1 \vee \xi_2$. If $\psi'$ is the \LTL normal form of $\psi$, then $\psi$ and $\psi'$ are globally equivalent, i.e., for all
 \LTL interpretations $\TModel$ and positions $i\geq 0$, $\TModel,i\models_\LTL \psi$ iff $\TModel,i\models_\LTL \psi'$.

 Now, we prove  Claim~2.  Let $\phi$ be a $\THT^{1}(\Next,\Release)$ formula and $i\geq 0$ such that $\TModel,i\models\neg\phi$.
The proof is  by induction on the structure of the  normal form $\psi$ of $\neg\phi$. We crucially use the following fact:
since $\phi\in \THT^{1}(\Next,\Release)$, every subformula of $\psi$ of the form $\psi_1\Release \psi_2$ is \emph{positive}, i.e. $\psi_1\Release \psi_2\in \THT^{0}$.
\begin{itemize}
  \item $\psi=p$ or $\psi=\neg p$ for some $p\in P$: the result directly follows from Claim~1(2).
  \item $\psi=\psi_1\vee\psi_2$ or $\psi=\psi_1\wedge\psi_2$: the result easily follows from the induction hypothesis.
  \item $\psi=\Next\psi_1$: we apply the induction hypothesis on $\psi_1$ and position $i+1$.
  \item $\psi=\psi_1\Until\psi_2$: hence, there exists $j\geq i$ such that $\TModel,j\models_\LTL \psi_2$ and $\TModel,m\models_\LTL \psi_1$
  for all $m\in [i,j-1]$. By applying the induction hypothesis, there exist $k_i,\ldots,k_j$ such  that
  $\TModel_n,j\models_\LTL \psi_2$ for all $n\geq k_j$, and for all $m\in [i,j-1]$ and $n\geq k_m$,
  $\TModel_n,m\models_\LTL \psi_1$. Thus, by taking $k=\max(\{k_i,\ldots,k_j\})$, the result follows.
  \item $\psi=\psi_1\Release \psi_2$: hence, $\psi_1\Release \psi_2$ is a positive formula, i.e., $\psi_1\Release \psi_2\in \THT^{0}$. Evidently, for all $\LTL$ interpretations $\HModel$ and
  $\HModel'$ such that $\HModel\sqsubseteq\HModel'$ and for all positive formulas $\xi$, $\HModel,i \models_\LTL \xi$ implies $\HModel',i \models_\LTL \xi$. Thus, since $\TModel\sqsubseteq \TModel_n$ for all $n\geq 0$, the result follows.
\end{itemize}\qed

Since $\TModel_n$ is an \LTL model of $\varphi$ for all $n\geq 0$,  by Claim~2, we deduce that
$\TModel \models_\LTL \varphi$. Thus, by Claim~1(3), $\TModel$ is a minimal \LTL model of $\varphi$ which concludes.
\end{IEEEproof}

\noindent We establish now the main results for $\THT^{1}(\Next,\Release)$ and $\THT_1^{1}$.

\begin{theorem}\label{theorem:MainCONXROneImplication} A $\THT^{\,1}(\Next,\Release)$ $($resp., $\THT_1^{\,1}$$)$ formula $\varphi$ has an equilibrium model iff $\varphi$ is \LTL satisfiable.  Moreover, $\CON(\THT^{\,1}(\Next,\Release))$ amd $\CON(\THT^{\,1}(\Release))$
  are \PSPACE-complete, while $\CON(\THT_1^{\,1})$ is \NP-complete.
\end{theorem}
\begin{IEEEproof}
For the first part of
Theorem~\ref{theorem:MainCONXROneImplication}, if $\varphi$ has an equilibrium model, then by Proposition~\ref{prop:fundamentalsTHT}(3),
$\varphi$ is \LTL satisfiable. For the converse direction, assume that $\varphi$ is \LTL satisfiable. By Theorem~\ref{theorem:minimalityRXOneImplication},
$\varphi$ has a minimal \LTL model $\TModel$. Since $\varphi\in \THT^{1}$, by Proposition~\ref{prop:fundamentalsTHT}(3-4), $(\TModel,\TModel)$ is an equilibrium model of $\varphi$.

By well-known lower bounds for \LTL \cite{SistlaC85,DemriS02}, $\LTL$-satisfiability for the fragment
$\THT^{1}(\Release)$ is \PSPACE-hard. Thus, since \LTL satisfiability is \PSPACE-complete, and
\LTL satisfiability for the fragment $\THT_1^{1}$ is \NP-complete~\cite{DemriS02}, the second part of
Theorem~\ref{theorem:MainCONXROneImplication} follows as well.
\end{IEEEproof}\vspace{0.1cm}

\subsubsection{The fragment $\THT^{\,1}(\Next,\Until)$}
For this fragment, we show that the \TEL consistency problem can be reduced in linear-time to $\LTL$-satisfiability.

 Given an interpretation $(\HModel,\TModel)$ and a position $i\geq 0$, $i$ is an
\emph{empty position} of $(\HModel,\TModel)$ if $\HModel(i)=\emptyset$. A total interpretation having a finite number of non-empty positions is said to be \emph{almost-empty}. A $\THT$ formula $\varphi$ satisfies the \emph{almost-empty requirement} if every temporal equilibrium of $\varphi$ is almost-empty. We first observe the following.

\begin{lemma}\label{lemma:criteriumForRedToSAT} Let $\varphi\in \THT^{\,1}$ and satisfy the almost-empty requirement. Then, there exists an equilibrium model of $\varphi$ \emph{iff} the following formula is $\LTL$-satisfiable
\begin{equation}\label{eq:criteriumForRedToSAT}
\varphi \wedge \Eventually\Always \displaystyle{\bigwedge_{p\in P}\neg p}
\end{equation}
\end{lemma}
\begin{IEEEproof} Let $(\TModel,\TModel)$ be an  equilibrium model of $\varphi$. Since $\varphi$ satisfies the almost-empty requirement, by
Proposition~\ref{prop:fundamentalsTHT}(3), $\TModel$ is an $\LTL$ model of formula~(\ref{eq:criteriumForRedToSAT}).
Now, assume that formula~(\ref{eq:criteriumForRedToSAT})  has an $\LTL$-model. Hence, there is an almost-empty interpretation $(\TModel,\TModel)$ such that $\TModel\models_\LTL  \varphi$ and $(\TModel,\TModel)\models \varphi$. Since the number of non-empty positions of $\TModel$ is finite, we can also assume that for all $\HModel \sqsubset \TModel$, $\HModel\not\models_\LTL  \varphi$ (i.e., $\TModel$ is a minimal \LTL model of $\varphi$). Since $\varphi\in \THT^{1}$, by Proposition~\ref{prop:fundamentalsTHT}(4), there is no $\HModel \sqsubset \TModel$ such that $(\HModel,\TModel)\models \varphi$. Thus, $(\TModel,\TModel)$ is an equilibrium model of $\varphi$, which concludes.
\end{IEEEproof}

Next, we establish that the formulas in the fragment $\THT(\Next,\Until)$ satisfy the almost-empty requirement. For this, we need additional definitions.
For a $\THT$ formula $\varphi$, $\depthX(\varphi)$ denotes the nesting depth of modality $\Next$ in $\varphi$.

\begin{definition}[Set of witnesses for $\THT(\Next,\Until)$]\label{Def:WitnessesForUX} \emph{Let $\varphi$ be a $\THT(\Next,\Until)$ formula and
$\Model=(\TModel,\TModel)$ be a total interpretation. We denote by
$\Fin(\varphi,\Model)$ (resp., $\Inf(\varphi,\Model)$)  the
set of subformulas $\psi_1\,\Until\,\psi_2$ of $\varphi$ such that the number of positions $i$ so that
$\Model,i\models \psi_2$ is finite and non-empty (resp., infinite). Note that
$\Fin(\varphi,\Model)\cap \Inf(\varphi,\Model)=\emptyset$. For $\psi_1\,\Until\,\psi_2\in \Fin(\varphi,\Model)\cup \Inf(\varphi,\Model)$, a \emph{witness of $\Model$ for $\psi_1\,\Until\,\psi_2$} is a  position $j$ such that $\Model,j\models \psi_2$.}

\emph{Let $\Fin(\varphi,\Model)=\{\phi_1,\ldots,\phi_k\}$. Fix an ordering $\xi_1,\ldots,\xi_m$ of the subformulas in $\Inf(\varphi,\Model)$ such that for all $i,j\in [1,m]$, if  $i\neq j$
and $\xi_i$ is a subformula of $\xi_j$, then $i>j$.
A \emph{set of witnesses of $\Model$ for $\varphi$} is any set of the form
\[
\{(0,\varphi),(j_1,\phi_1),\ldots,(j_k,\phi_k)\} \cup \{(h_1,\xi_1),\ldots,(h_m,\xi_m)\}
\]
such that the following holds, where $\ell=\max(\{j_1,\ldots,j_k\})$:
\begin{compactitem}
\item $j_i$ is the the greatest witness of
$\Model$ for $\phi_i$ for all $i\in [1,k]$;
  \item $h_j$ is a witness of $\Model$ for $\xi_j$ for all $j\in [1,m]$;
  \item $h_1> \ell+\depthX(\varphi)$ and $h_{j+1}>h_{j}+\depthX(\varphi)$ for all $j\in [1,m-1]$.
\end{compactitem}
Note that by definition of $\Inf(\varphi,\Model)$, sets of witnesses of $\Model$ for $\varphi$ exist. Moreover, such sets have cardinality at most $|\varphi|+1$.}
\end{definition}

\newcounter{lemma-EquibibiumModelAlmostEmptyXUntil}
\setcounter{lemma-EquibibiumModelAlmostEmptyXUntil}{\value{lemma}}
\newcounter{sec-EquibibiumModelAlmostEmptyXUntil}
\setcounter{sec-EquibibiumModelAlmostEmptyXUntil}{\value{section}}

\begin{lemma}\label{lemma:EquibibiumModelAlmostEmptyXUntil}
Let $\varphi$ be a $\THT(\Next,\Until)$ formula and $\Model$ be an equilibrium model of $\varphi$.
Then, $\Model$ is almost-empty.
\end{lemma}
\begin{IEEEproof} 
 Fix a set of witnesses $W$ of $\Model=(\TModel,\TModel)$ for $\varphi$. Let $\ell$ be the greatest position occurring in $W$.
We define an $\LTL$ interpretation $\HModel_W \sqsubseteq \TModel$ as follows:
\begin{itemize}
  \item for all $i\geq 0$, $\HModel_W(i)=\TModel(i)$ if $i\leq \ell + \depthX(\varphi)$, and
  $\HModel_W(i)=\emptyset$ otherwise.
\end{itemize}

 We show that $\HModel_W=\TModel$, hence, $\Model=(\TModel,\TModel)$ is almost empty, and  the result follows.
 For this, since  $\Model=(\TModel,\TModel)$ is an equilibrium model of $\varphi$, it suffices to prove that
 $(\HModel_W,\TModel),0\models \varphi$. Since $(0,\varphi)\in W$, the result directly follows from the following claim, which can be proved by structural induction on $\psi$ by using Definition~\ref{Def:WitnessesForUX} and
 Proposition~\ref{prop:fundamentalsTHT}(1). For details, see Appendix~\ref{APP:EquibibiumModelAlmostEmptyXUntil}.  \vspace{0.1cm}

\noindent \emph{Claim:} let $(j,\psi)\in W$ and $\xi$ be a subformula of $\psi$. Then:
\begin{compactenum}
\item for all $k\in [0,\depthX(\psi)]$
    such that
    $\depthX(\xi)\leq \depthX(\psi)-k$, $(\TModel,\TModel),j+k\models \xi$ \emph{iff} $(\HModel_W,\TModel),j+k\models \xi$.
  \item for all $k\in [0,j]$, $(\TModel,\TModel),k\models \xi$ \emph{iff} $(\HModel_W,\TModel),k\models \xi$.
\end{compactenum}
\end{IEEEproof}

 By well-known lower bounds for \LTL \cite{SistlaC85,DemriS02}, $\LTL$-satisfiability of
formulas of the form $\varphi \wedge \Eventually\Always \displaystyle{\bigwedge_{p\in P}\neg p}$, where $\varphi$ is a  $\THT^{1}(\Until)$ formula is \PSPACE-hard. Thus, since $\LTL$-satisfiability is \PSPACE-complete, by Lemmata~\ref{lemma:criteriumForRedToSAT} and~\ref{lemma:EquibibiumModelAlmostEmptyXUntil}, we obtain the following result.

\begin{corollary}\label{cor:MainCONXU} The \TEL consistency problems for $\THT^{\,1}(\Next,\Until)$ and $\THT^{\,1}(\Until)$ are \PSPACE-complete.
\end{corollary}

\subsection{\textbf{The fragment $\THT(\Next,\Eventually)$}}

It is well-known that $\LTL$-satisfiability for the \LTL fragment corresponding to $\THT{}{}(\Next,\Eventually)$ is already \PSPACE-complete \cite{SistlaC85}. By contrast and surprisingly,  we show that the \TEL consistency problem for $\THT{}{}(\Next,\Eventually)$ is just $\CompEL$-complete.

 The \emph{size} of an almost-empty total interpretation $(\TModel,\TModel)$ is $h+1$ where $h$ is the smallest position  such that
$\TModel(i)=\emptyset$ for all $i\geq h$. The main result for 
$\THT{}{}(\Next,\Eventually)$ is as follows.

\begin{proposition}\label{prop:MainResultXF} Let $\varphi$ be a $\THT{}{}(\Next,\Eventually)$ formula. If $\varphi$ has an equilibrium model, then
$\varphi$ has an almost-empty equilibrium model of size at most $|\varphi|^{3}$.
\end{proposition}

Given a $\THT{}{}(\Next,\Eventually)$ formula $\varphi$, nondeterministically guessing an almost-empty total interpretation $(\TModel,\TModel)$ of size at most $|\varphi|^{3}$ and checking that $(\TModel,\TModel)$ satisfies $\varphi$ can be done in polynomial time.
 Moreover, universally guessing $\HModel \sqsubset \TModel$ and checking that $(\HModel,\TModel)$  does not satisfy $\varphi$ can be done in polynomial time. Hence, since  $\CON(\HT)$ is $\CompEL$-complete, by Proposition~\ref{prop:MainResultXF}, we obtain the following.

\begin{corollary}\label{cor:MainResultXF} The \TEL consistency problems for $\THT{}{}(\Next,\Eventually)$, $\THT{}{}(\Next)$, and $\THT{}{}(\Eventually)$   are $\CompEL$-complete.
\end{corollary}

We now proceed with the proof of Proposition~\ref{prop:MainResultXF} which consists of the following two Lemmata~\ref{lemma:EquibibiumModelXFOne} and
\ref{lemma:EquibibiumModelXFTwo}.

\newcounter{lemma-EquibibiumModelXFOne}
\setcounter{lemma-EquibibiumModelXFOne}{\value{lemma}}
\newcounter{sec-EquibibiumModelXFOne}
\setcounter{sec-EquibibiumModelXFOne}{\value{section}}

\begin{lemma}\label{lemma:EquibibiumModelXFOne}
Let $\varphi$ be a $\THT{}{}(\Next,\Eventually)$ formula and $\Model=(\TModel,\TModel)$ be an equilibrium model of $\varphi$.
Then, $\Model$ has at most $\depthX(\varphi)\cdot(|\varphi|+1)$ non-empty positions.
\end{lemma}
\begin{IEEEproof} 
Let $W$ be a set of witnesses of $\Model$ for $\varphi$ according to Definition~\ref{Def:WitnessesForUX}. By Definition~\ref{Def:WitnessesForUX}, $W$ has cardinality at most $|\varphi|+1$.
Now, we define an $\LTL$ interpretation $\HModel_W \sqsubseteq \TModel$ as follows:
 for all $i\geq 0$, if there is $(j,\psi)\in W$ such that $j\leq i$ and $i-j\leq \depthX(\varphi)$, then
  $\HModel_W(i)=\TModel(i)$; otherwise, $\HModel_W(i)=\emptyset$.

 By construction, the set of non-empty positions of the interpretation $(\HModel_W,\TModel)$ has cardinality at most
 $\depthX(\varphi)\cdot(|\varphi|+1)$. We show that $\HModel_W=\TModel$, hence, the result follows.
 For this, since  $\Model=(\TModel,\TModel)$ is an equilibrium model of $\varphi$, it suffices to prove that
 $(\HModel_W,\TModel),0\models \varphi$. Since $(0,\varphi)\in W$, the result directly follows from the following claim, whose proof, based on Definition~\ref{Def:WitnessesForUX} and
 Proposition~\ref{prop:fundamentalsTHT}(1), is given in Appendix~\ref{APP:EquibibiumModelXFOne}: \vspace{0.1cm}

\noindent \emph{Claim:} for all $(i,\psi)\in W$, $k\in [0,\depthX(\psi)]$, and subformulas $\xi$ of $\psi$ such that
$\depthX(\xi)\leq \depthX(\psi)-k$, $(\TModel,\TModel),i+k\models \xi$ \emph{iff} $(\HModel_W,\TModel),i+k\models \xi$.
\end{IEEEproof}

\newcounter{lemma-EquibibiumModelXFTwo}
\setcounter{lemma-EquibibiumModelXFTwo}{\value{lemma}}
\newcounter{sec-EquibibiumModelXFTwo}
\setcounter{sec-EquibibiumModelXFTwo}{\value{section}}

The following result is straightforward (for details, see Appendix~\ref{APP:EquibibiumModelXFTwo}).

\begin{lemma}\label{lemma:EquibibiumModelXFTwo} Let $\varphi$ be a $\THT{}{}(\Next,\Eventually)$ formula, $n\geq 1$, and $\Model=(\TModel,\TModel)$ be an equilibrium model of $\varphi$ having $n$ non-empty positions. Then, there exists an \emph{almost-empty} equilibrium model of $\varphi$ of size at most $n\cdot  (\depthX(\varphi)+1)$.
\end{lemma}

\section{Minimal \LTL satisfiability}\label{sec:MinimalityLTL}

In this section we establish the complexity of the \emph{minimal \LTL satisfiability problem}, i.e., checking for a given \THT\ formula $\varphi$, whether $\varphi$ has a minimal \LTL model.

\begin{theorem}\label{theo:MinimalLTLSatifisfiability} Minimal \LTL satisfiability  is \EXPSPACE-complete even for the syntactical fragment $\THT^{\,1}_2(\Eventually,\Always)$.
\end{theorem}
\begin{IEEEproof}
For the lower bound, let $\Instance$ be an instance of the domino tiling problem considered in the proof of
Theorem~\ref{theorem:consistencyLowerBoundUntractable}, and $\varphi_\Instance$ be the $\THT^{\,1}_2(\Eventually,\Always)$ formula of Lemma~\ref{MainLemma:LowerBoundFGNested}. We show that $\varphi_\Instance$ has a minimal \LTL model \emph{iff} $\varphi_\Instance$ has an  equilibrium model.
Hence, by Lemma~\ref{MainLemma:LowerBoundFGNested}, the lower bound of Theorem~\ref{theo:MinimalLTLSatifisfiability} follows. Since $\varphi_\Instance$ is a $\THT^{\,1}_2(\Eventually,\Always)$ formula,  if $\varphi_\Instance$ has a minimal \LTL model $\TModel$, then  by Proposition~\ref{prop:fundamentalsTHT}(3--4), $(\TModel,\TModel)$ is an  equilibrium model of $\varphi_\Instance$. For the converse implication, let $(\TModel,\TModel)$ be an  equilibrium model of $\varphi_\Instance$. We assume that $\TModel$ is not a minimal \LTL model of $\varphi_\Instance$ and derive a contradiction. Hence, by Proposition~\ref{prop:fundamentalsTHT}(3), there is $\HModel \sqsubset \TModel$ such that $\HModel \models_\LTL \varphi_\Instance$ and $(\HModel,\HModel)\models \varphi_\Instance$. By the proof of
Lemma~\ref{MainLemma:LowerBoundFGNested},
\[
\varphi_\Instance =\varphi_\pseudo \wedge (\symSp \vee   \varphi_{\textit{bad}})
\]
where $\varphi_\pseudo$ and $\varphi_{\textit{bad}}$ are the $\THT_2^{1}(\Eventually,\Always)$ formulas of Propositions~\ref{Prop:FormulaPseudoFGNested} and~\ref{Prop:FormulaBadFGNested}, respectively. Moreover,
$(\TModel,\TModel)$ is a good pseudo-tiling code for $\THT_{2}^{1}(\Eventually,\Always)$.
Since $(\HModel,\HModel)\models \varphi_\Instance$ and $\HModel \sqsubset \TModel$, by Proposition~\ref{Prop:FormulaPseudoFGNested} and Definition~\ref{Def:PseudoTilingFGNested}, it follows that $(\HModel,\TModel)$ is a good pseudo-tiling code for $\THT_{2}^{1}(\Eventually,\Always)$ and $u\notin \HModel(0)$. Thus, since $\HModel \models_\LTL \varphi_\Instance$,
we have that $\HModel \models_\LTL \varphi_{\textit{bad}}$.
By Propositions~\ref{Prop:FormulaPseudoFGNested} and~\ref{Prop:FormulaBadFGNested}, we obtain that $(\HModel,\TModel)$ satisfies $\varphi_\Instance$. This contradicts the assumption that $(\TModel,\TModel)$ is an  equilibrium model of $\varphi_\Instance$, and we are done.

For the upper bound, we exploit an automata-theoretic approach. Let $\varphi$ be a \THT\ formula. It is well-known
 \cite{VardiW94} that one can construct in singly exponential time a B\"{u}chi nondeterministic finite-state automaton
 (B\"{u}chi \NFA) $\Au_\varphi$ over $2^{P}$ whose accepted language $\Lang(\Au)$ is the set of  \LTL interpretations which are \LTL models of $\varphi$. Moreover, given a B\"{u}chi \NFA $\Au$ over $2^{P}$, it is straightforward to construct in quadratic time  a B\"{u}chi \NFA $K(\Au)$ such that $\TModel\in\Lang(K(\Au))$ iff
 $\TModel\in\Lang(\Au)$ and there is $\HModel \sqsubset \TModel$ such that $\HModel\in \Lang(\Au)$. Hence,
 $K(\Au_\varphi)$ accepts the set of \LTL models of $\varphi$ which are not minimal. It follows that
 $\varphi$ has a minimal \LTL model \emph{iff}
 \begin{equation}\label{eq:MinimalLTLSatifisfiability}
   \Lang(\Au_\varphi)\cap [(2^{P})^{\omega}\setminus \Lang(K(\Au_\varphi))]\neq \emptyset
 \end{equation}
Now, checking non-emptiness of B\"{u}chi \NFA can be done in \NLOGSPACE. Moreover, the $\omega$-languages recognized by B\"{u}chi \NFA are closed under intersection and complementation, and complementation involves a singly exponential blow-up. Thus, by well-known results \cite{SistlaVW87,Safra89,VardiW94},  checking equation~(\ref{eq:MinimalLTLSatifisfiability}) can be done in single exponential space, which concludes the proof.
\end{IEEEproof}

It is well-known that for both the considered standard version of \LTL, whose interpretations are infinite words,
and \emph{finitary} \LTL (i.e. \LTL interpreted over \emph{finite} words), satisfiability is \PSPACE-complete~\cite{SistlaC85}.
On the other hand, Theorem~\ref{theo:MinimalLTLSatifisfiability}  highlights a meaningful difference arising from interpreting \LTL over finite words or infinite words. Indeed, while for finitary \LTL, minimal satisfiability evidently  coincides with satisfiability, for infinite words, minimal satisfiability turns out to be singly exponentially harder than satisfiability. 
\section{Conclusion}

We conclude with some observations and future research directions.
We have provided a systematic study of the computational complexity of the
\TEL consistency problem by considering natural syntactical fragments of
\THT. Our complexity results show that there is no difference in tractability
between implication height 2 and $k$ with $k>2$, and the same holds for the temporal
height. Moreover, unlike in the case of  \LTL, in \THT\  dual temporal modalities
need to be considered independently from one another, and they have quite different computational costs.
Additionally, we have shown that minimal \LTL satisfiability has, in the general case, the same complexity
as checking \TEL consistency. However, for some of the considered fragments, we have a different scenario.
An example is the fragment $\THT_1$ where there is no nesting of temporal modalities:
in this restricted case, the \TEL consistency problem is \NEXPTIME-complete, while, by Theorem~\ref{theorem:minimalityRXOneImplication},
mimimal \LTL satisfiabilty coincides with  \LTL satisfiability, the latter being just \NP-complete for the fragment $\THT_1$ \cite{DemriS02}.

Another subclass of \THT\ formulas, called temporal logic programs (\TLP) has been considered
in~\cite{Cabalar10,DBLP:journals/jancl/AguadoCDPV13}.
\TLP conforms to a logic programming style and corresponds to a fragment of $\THT_2^{2}(\Next,\Eventually,\Always)$.
As shown in \cite{Cabalar10}, for the \TEL consistency problem, the general case reduces in polynomial time to the case of \TLP formulas. Thus, our results imply that  checking \TEL consistency for \TLP is already
\EXPSPACE-complete.\footnote{In~\cite{DBLP:journals/jancl/AguadoCDPV13}  \TLP rules are divided into four different syntactical fragments.  {\em Initial rules} are in $\THT_{1}^{2} (\Next)$; {\em fulfillment rules} are of two types, either in $\THT_{2}^{1} (\Always)$ or in
$\THT_{2}^{1} (\Eventually,\Always)$; while so-called {\em dynamic rules} fall  in the  fragment $\THT_{2}^{2} (\Next,\Always)$.
}

As future research, we aim to address expressiveness issues for the \TEL framework. In particular, since we have individuated some non-trivial tractable fragments such as $\THT_1(\Next,\Always)$ and $\THT(\Next,\Eventually)$, it would be interesting to study what kind of temporal reasoning problems they can express. Moreover, an important question is to investigate from a semantical point of view the considered syntactical hierarchy of \THT\ fragments: is this hierarchy also semantically strict with respect to  \THT\ and/or \TEL semantics? Another relevant issue is to provide alternative characterizations of the class of \TEL\ languages (the $\omega$-languages of equilibrium models of
\THT\ formulas). It is known that this class is regular \cite{CabalarD11}. An intriguing open question is whether \TEL languages are \LTL-expressible.

\section*{Acknowledgments}
Work supported by the projects VIVAC (TIN2012-38137-C02, Bozzelli) and  MERLOT (TIN2013-42149-P, Pearce).



%



\bibliographystyle{IEEEtran}
\bibliography{IEEEabrv,biblio}

\newpage

\onecolumn

\newenvironment{changemargin}{%
  \begin{list}{}{%
    \setlength{\leftmargin}{40bp}%
    \setlength{\rightmargin}{40bp}%
    \setlength{\textheight}{610bp}%
   \setlength{\topmargin}{-30bp}
  }%
  \item[]}{\end{list}}

\makeatletter
\renewcommand{\normalsize}{%
\@setfontsize\normalsize {11pt}{14pt}
   \normalbaselineskip=13pt
   }

\makeatother

\makeatletter\makeatother

\normalsize

\begin{changemargin}

\newcounter{aux}
\newcounter{auxSec}

\begin{center}
\begin{LARGE}
  \noindent\textbf{Appendix}
\end{LARGE}
\end{center}

\section{Proofs from Section~\ref{sec:UntractableFragments}}

\subsection{\textbf{Full proof of Proposition~\ref{Prop:FormulaBadFGNested}}}\label{APP:FormulaBadFGNested}

\setcounter{aux}{\value{proposition}}
\setcounter{auxSec}{\value{section}}
\setcounter{section}{\value{sec-FormulaBadFGNested}}
\setcounter{proposition}{\value{prop-FormulaBadFGNested}}

\begin{proposition} One can construct in polynomial time a $\THT_{2}^{\,1}(\Eventually,\Always)$ formula $\varphi_{\textit{bad}}$ over $P\setminus \{u\}$
such that for all  total interpretations $(\TModel,\TModel)$ which are good pseudo-tiling codes for $\THT_{2}^{\,1}(\Eventually,\Always)$,
there exists a good pseudo-tiling code for $\THT_{2}^{\,1}(\Eventually,\Always)$ of the form $(\HModel,\TModel)$ with $\HModel\neq \TModel$ and satisfying $\varphi_{\textit{bad}}$ \emph{iff}
there is \emph{no} prefix of $\TModel$  whose projection over $P_\Main$  encodes a tiling. Moreover,
for all good pseudo-tiling codes $(\HModel,\TModel)$ for $\THT_{2}^{\,1}(\Eventually,\Always)$ with
 $\HModel\neq \TModel$, $(\HModel,\TModel)\models \varphi_{\textit{bad}}$ \emph{iff}
 $\HModel\models_\LTL \varphi_{\textit{bad}}$.
\end{proposition}
\setcounter{proposition}{\value{aux}}
\setcounter{section}{\value{auxSec}}
\begin{IEEEproof}
In the proof, we use  the following short-hands:
\[
\begin{array}{rcl}
  R_\Main & = &   P_\Main \setminus\{d_\Final\}\\
  P_\cell & = & P_\Main\setminus \{\$\} \\
  R_\cell & = & P_\cell \setminus\{d_\Final\}\\
  \Delta_R  & = & \Delta \setminus\{d_\Final\}
\end{array}
\]
Moreover, for an \LTL interpretation $\TModel$ over $P$, we say that a prefix of $\TModel$ is \emph{incomplete} if it has no position labeled by $d_\Final$.\vspace{0.2cm}

The $\THT_{2}^{\,1}(\Eventually,\Always)$ formula $\varphi_{\textit{bad}}$   is defined as follows:
\[
\begin{array}{rcl}
  \varphi_{\textit{bad}}& = &  \varphi_{\textit{bad\_in}}\vee  \varphi_{\textit{bad\_ord}} \vee \varphi_{\textit{bad\_acc}}\vee \varphi_{\textit{bad\_cell}}
   \vee \varphi_{\textit{bad\_first}} \vee \varphi_{\textit{bad\_last}} \vee\vspace{0.2cm}\\
   &&  \varphi_{\textit{bad\_inc}} \vee
  \varphi_{\textit{bad\_rr}} \vee   \varphi_{\textit{bad\_cr}}
\end{array}
\]
where for a total good pseudo-tiling code $(\TModel,\TModel)$, the different disjuncts in the definition of
$\varphi_{\textit{bad}}$ capture all the possible conditions
such that no prefix of $\TModel$ encodes a tiling iff some of these conditions is satisfied.
The construction of such disjuncts exploits the formulas $\theta(i_1 | P_1,\ldots ,i_k| P_k)$ of  Proposition~\ref{prop:FormualsForMarkFGNested}.

The disjunct $\varphi_{\textit{bad\_in}}$ checks that the content of the first cell is not $d_\Init$.
\[
\begin{array}{rcl}
  \varphi_{\textit{bad\_in}} & = &   \theta( 1 | \{\$\},  2|P_\num, 3 | \Delta\setminus\{d_\Init\},   4 | P_\Main)
\end{array}
\]

The disjunct $\varphi_{\textit{bad\_ord}}$ is used to check that \emph{either} some $\$$-position is preceded by an incomplete prefix and is followed by a $\Delta$-position,
\emph{or} some $P_\num$-position is preceded by an incomplete prefix and is followed by a $\$$-position.
\[
\begin{array}{rcl}
  \varphi_{\textit{bad\_ord}} & = &    \theta( 1| R_\Main, 2 | \{\$\},   3 | \Delta,   4 | P_\Main)\, \vee \, \theta(  2 | \{\$\},   3 | \Delta,   4 | P_\Main) \,\,\vee \vspace{0.3cm} \\
&   &  \theta( 1| R_\Main, 2 | P_\num, 3 | \{\$\}, 4 | P_\Main)
\end{array}
\]

The disjunct $\varphi_{\textit{bad\_acc}}$ asserts that there is \emph{no} cell $c$ preceded by an incomplete prefix such that $c$ has content $d_\Final$ and $c$
is the last cell of a row (recall that for a pseudo-tiling code some position is labeled by $d_\Final$).
\[
\begin{array}{rcl}
  \varphi_{\textit{bad\_acc}} & = &   \displaystyle{\bigvee_{i=1}^{i=n}}\Bigl(\theta( 1 | R_\Main,   2 |\{(i,0)\}, 3|P_\num, 4| \{d_\Final\},  5| P_\Main) \,\vee\,\\
  && \phantom{\displaystyle{\bigvee_{i=1}^{i=n}}}\,\, \theta( 1 | R_\Main,   2 |\{(i,0)\},  4| \{d_\Final\},  5| P_\Main)\Bigr)
\end{array}
\]

The disjunct $\varphi_{\textit{bad\_cell}}$ is used to individuate segments in $(\Delta_R\cup\{\$\}) P_\num^{+}\Delta^{+}$ which are preceded by incomplete prefixes and such that the suffix in
$P_\num^{+}\Delta^{+}$ is not a correct encoding of a cell.
\[
\begin{array}{rcl}
  \varphi_{\textit{bad\_cell}} & = &  \underbrace{\displaystyle{\bigvee_{(d,d')\in (\Delta\setminus\{d_\Final\})\times \Delta: d\neq d'}} \theta( 1 | R_\Main, 2| \{d\},  3 | \{d'\},  4 | P_\Main)}_{\text{a cell  contains two distinct elements in $\Delta$}} \,\,\,\vee \vspace{0.2cm} \\
 && \, \,\, \displaystyle{\bigvee_{i=1 }^{i=n}\bigvee_{(b,b')\in \{0,1\}: b\neq b'}}
 \Bigl(\theta(1 | R_\Main, 2| \{(i,b)\},  3 | \{(i,b')\},  4 | P_\Main)\,\vee  \vspace{0.2cm} \\
 && \underbrace{\phantom{\displaystyle{\bigvee_{i=1 }^{i=n}\bigvee_{(b,b')\in \{0,1\}}}}\theta(1 | R_\Main, 2| \{(i,b)\}, 3| P_\num,  4 | \{(i,b')\},  5 | P_\Main)\Bigr)}_{\text{two distinct bits in the encoding of a cell  have the same bit position}} \,\,\,\vee 
\end{array}
\]
\[
\begin{array}{rcl}
\phantom{\varphi_{\textit{bad\_cell}}} &  & \, \,\, \displaystyle{\bigvee_{((i,b),(j,b'))\in P_\num\times P_\num: i>j}}
 \Bigl(\theta(1 | R_\Main, 2| \{(i,b)\},  3 | \{(j,b')\},  4 | P_\Main)\,\vee  \vspace{0.2cm} \\
  && \underbrace{\phantom{\displaystyle{\bigvee_{((i,b),(j,b'))\in P_\num}}}\theta(1 | R_\Main, 2| \{(i,b)\}, 3| P_\num,  4 | \{(j,b')\},  5 | P_\Main)\Bigr)}_{\text{the bit positions in the encoding of a cell  are not ordered correctly}} \,\,\,\vee \vspace{0.2cm} \\
  && \, \,\,  \displaystyle{\bigvee_{i=1 }^{i=n}}\Bigl( \theta(1 | R_\Main, 2| \Delta_R\cup \{\$\}, 3| P_\num \setminus\{i\}\times\{0,1\},  4 | \Delta,  5 | P_\Main)  \,\,\vee  \vspace{0.2cm} \\
  && \, \,\, \underbrace{\phantom{1 | R_\Main, \{\$\},\{\$\},\,\,\,\,\,\,}  \theta(1 | \{\$\},   3| P_\num \setminus\{i\}\times\{0,1\},  4 | \Delta,  5 | P_\Main) \Bigr)}_{\text{some bit position in the encoding of a cell  is absent}} \,\,\,\Bigr]
\end{array}
\]
The disjunct $\varphi_{\textit{bad\_first}}$ (resp., $\varphi_{\textit{bad\_last}}$) checks the existence of rows which are preceded by incomplete prefixes and whose first (resp., last)
cell has column number distinct from $0$ (resp., $2^{n}-1$).
\[
\begin{array}{rcl}
  \varphi_{\textit{bad\_first}} & = &  \displaystyle{\bigvee_{i=1}^{i=n}} \Bigl(\theta( 1 | R_\Main, 2| \{\$\},  3 | P_\num,  4 | \{(i,1)\}, 5 | P_\Main)\,\vee\, \\
  && \phantom{\displaystyle{\bigvee_{i=1}^{i=n}}}\,\theta( 1 | R_\Main, 2| \{\$\},   4 | \{(i,1)\}, 5 | P_\Main)\,\vee\,\\
  && \theta(  2| \{\$\},  3 | P_\num,  4 | \{(i,1)\}, 5 | P_\Main)\,\vee\,
  \theta(  2| \{\$\},    4 | \{(i,1)\}, 5 | P_\Main)\Bigr)
\end{array}
\]

\[
\begin{array}{rcl}
  \varphi_{\textit{bad\_last}} & = & \theta( 1 | R_\Main,     2 | \{(n,0)\},  3| \Delta, 4 | \{\$\}, 5 | P_\Main)\, \vee \\
   &&\displaystyle{\bigvee_{i=1}^{i=n}} \theta( 1 | R_\Main,     2 | \{(i,0)\}, 3 | P_\num, 4| \Delta, 5 | \{\$\}, 6 | P_\Main)
\end{array}
\]

The disjunct $\varphi_{\textit{bad\_inc}}$  selects adjacent cells in a row  whose column numbers are not consecutive; moreover, the rightmost cell
is preceded by an incomplete prefix.
\[
\begin{array}{rcl}
 \varphi_{\textit{bad\_inc}} & = &   \Bigl(\theta( 1 | R_\Main, 2| \Delta_R\cup\{\$\},  3 | P_\num, 4| \Delta_R,  5 | P_\num, 6 | \Delta,  7 | P_\Main) \,\,\vee \vspace{0.2cm}\\
  &  & \underbrace{ \phantom{Ra, 2| \Delta_R\cup\{\$\}}   \theta( 1 | \{\$\},  3 | P_\num, 4| \Delta_R,  5 | P_\num, 6 | \Delta,  7 | P_\Main)\Bigr)}_{
  \text{mark by $t_3$ and $t_5$ the cell-numbers of two adjacent cells  in a row}} \,\,\wedge \vspace{0.2cm}\\
  && \,\,  \psi_{\textit{bad\_inc}}
  \end{array}
\]
where $\psi_{\textit{bad\_inc}}$   asserts that the cell numbers marked by $t_3$ and $t_5$, respectively, are not consecutive.
\[
\begin{array}{l}
 \psi_{\textit{bad\_inc}}  = \displaystyle{\bigwedge_{i=1}^{i=n}\Bigl(\Eventually((i,1)\wedge t_3) \,\,\wedge \,\, \Eventually((i,0)\wedge t_5)\Bigr)\,\vee\,
 \bigvee_{i=1}^{i=n}\Bigl(\Eventually((i,0)\wedge t_3) \,\,\wedge \,\, \Eventually((i,1)\wedge t_5)}\,\, \wedge \\
   \Bigl[\displaystyle{ \bigvee_{j=1}^{j=i-1} \bigl(\Eventually((j,0)\wedge t_3) \wedge  \Eventually((j,1)\wedge t_5)\bigr)\vee
 \bigvee_{j=i+1}^{j=n}\bigvee_{b,b'\in\{0,1\}:b\neq b'}\bigl(\Eventually((j,b)\wedge t_3) \wedge  \Eventually((j,b')\wedge t_5)\bigr) \Bigr]\Bigr)}
\end{array}
\]

The disjunct $\varphi_{\textit{bad\_rr}}$  checks that there are two adjacent cells in a row    which do \emph{not} have the same color on the shared edge; moreover, the rightmost cell
is preceded by an incomplete prefix.
\[
\begin{array}{rcl}
  \varphi_{\textit{bad\_rr}} & = & \displaystyle{\bigvee_{(d,d')\in (\Delta\setminus\{d_\Final\})\times \Delta: d_{\Right}\neq (d')_{\Left}}} \theta( 1 | R_\Main, 2| \{d\},  3 | P_\num,    4| \{d'\},     5 | P_\Main)
\end{array}
\]

Finally, the disjunct   $\varphi_{\textit{bad\_cr}}$  checks that there are two adjacent cells in a column  which do \emph{not} have the same color on the shared edge; moreover, the rightmost cell is preceded by an incomplete prefix. Formula $ \varphi_{\textit{bad\_cr}}$ is defined as follows.\vspace{0.2cm}

\noindent $ \Bigl( \theta( 1 | R_\Main, 2| P_\num , 3| \Delta_R,  4 | R_\cell, 5 | \{\$\},  6 | R_\cell, 7 | P_\num, 8|\Delta,  9 | P_\Main) \,\,\vee$\\
\text{\hspace{0.2cm}}$  \theta( 1 | R_\Main, 2| P_\num , 3| \Delta_R,  4 | R_\cell, 5 | \{\$\},   7 | P_\num, 8|\Delta,  9 | P_\Main)\,\,\vee$\\
\text{\hspace{0.2cm}}$  \underbrace{\theta( 1 | R_\Main, 2| P_\num , 3| \Delta_R,   5 | \{\$\},  6 | R_\cell, 7 | P_\num, 8|\Delta,  9 | P_\Main)\phantom{
  }\Bigr)}_{
  \text{mark with $t_2$ and $t_7$ the cell-numbers of two cells $c$ and $c'$ of two adjacent rows}} \,\,\wedge \,\, \psi_{\textit{bad\_cr}}$\vspace{0.2cm}

\noindent where $\psi_{\textit{bad\_cr}}$ asserts that the cells $c$ and $c'$ whose cell-numbers are marked by the propositions $t_2$ and $t_7$ and whose contents are marked by the propositions $t_3$ and $t_8$, respectively, have the same column number but distinct color on the shared edge. \[
\begin{array}{l}
 \psi_{\textit{bad\_cr}} = \underbrace{\displaystyle{\bigwedge_{i=1}^{i=n}\bigvee_{b\in \{0,1\}} \Bigl(\Eventually((i,b)\wedge t_2)\wedge \Eventually( (i,b)\wedge t_7)\Bigr)}}_{\text{the marked cells $c$ and $c'$ have the same column number}}\,\,\wedge\vspace{0.2cm}\\
  \underbrace{\displaystyle{\bigvee_{(d,d')\in \Delta\times \Delta: d_{\Up}\neq (d')_{\Down}}}\Bigl(\Eventually (d\wedge t_3)\wedge \Eventually ( d'\wedge t_8)\Bigr)}_{\text{the marked cells $c$ and $c'$ do \emph{not} have the same color on the shared edge}}
\end{array}
\]

By construction and  Proposition~\ref{prop:FormualsForMarkFGNested},
$\varphi_{\textit{bad}}$ is a $\THT_2^{1}(\Always)$  formula which can be constructed in polynomial time. Moreover,
  for all good pseudo-tiling codes $(\HModel,\TModel)$ such that
$\HModel\neq \TModel$,  if $(\HModel,\TModel)\models \varphi_{\textit{bad}}$, then there is no prefix of $\TModel$ whose projection over
 $P_\Main$ encodes a tiling.
On the other hand, by using Remark~\ref{remark:AssumptionTIling}, for each total good pseudo-tiling code $(\TModel,\TModel)$ such that no prefix of $\TModel$ encodes a tiling, there exists
a good pseudo-tiling code of the form $(\HModel,\TModel)$ such that $\HModel\neq \TModel$ and
$(\HModel,\TModel)$ satisfies $\varphi_{\textit{bad}}$.
Hence, the first part of Proposition~\ref{Prop:FormulaBadFGNested} follows. For the second part, notice that by construction, $\varphi_{\textit{bad}}$ is a positive boolean combinations of formulas $\psi$ such that either
$\psi$ is a $\THT^{0}$ formula, or $\psi$ is a formula of Proposition~\ref{prop:FormualsForMarkFGNested}.
By the semantics of \THT\,, for all $\THT^{0}$ formulas $\psi$ and interpretations $(\HModel,\TModel)$,
$(\HModel,\TModel)\models \psi$ iff $\HModel\models_\LTL \psi$.
Thus, by   Proposition~\ref{prop:FormualsForMarkFGNested}, it follows that
for all good pseudo-tiling codes $(\HModel,\TModel)$ for $\THT_{2}^{\,1}(\Eventually,\Always)$ with
 $\HModel\neq \TModel$, $(\HModel,\TModel)\models \varphi_{\textit{bad}}$ \emph{iff}
 $\HModel\models_\LTL \varphi_{\textit{bad}}$. Hence, the result follows.
\end{IEEEproof}

\bigskip
\bigskip

\subsection{\textbf{Proof of Theorem~\ref{theorem:consistencyLowerBoundUntractable}: reductions  for the fragments $\THT_2^{\,2}(\Always)$ and   $\THT_2^{\,2}(\Until)$}}\label{APP:LowerBoundsUntilRelease}

For the  fragments $\THT_2^{2}(\Always)$ and   $\THT_2^{2}(\Until)$,  we give distinct notions of pseudo-tiling code which in turn are different from the one adopted for the fragment
$\THT_2^{1}(\Eventually,\Always)$. Then, we give corresponding versions of Propositions~\ref{Prop:FormulaPseudoFGNested}, \ref{prop:FormualsForMarkFGNested} and~\ref{Prop:FormulaBadFGNested}.

For an $\LTL$ interpretation $\TModel $ over $P$ and $i\geq 0$, we say that $i$ is an \emph{empty position} of $\TModel$ if $\TModel(i)=\emptyset$.\vspace{0.2cm}

\begin{definition}[Pseudo-tiling codes for $\THT_2^{2}(\Always)$ and   $\THT_2^{2}(\Until)$]\label{def:PseudoTilingUGNested} Let $\Lang \in \{\THT_2^{\,2}(\Always),\THT_2^{\,2}(\Until)\}$.
An interpretation  $\Model=(\HModel,\TModel)$  is a \emph{pseudo-tiling code} for $\Lang$  if there is  $L\in \Nat\cup \{\infty\}$, with
 $L$ being an empty position of
  $\TModel$ if $\Lang=\THT_2(\Until) $, and $L$ being $\infty$ otherwise, such that the following holds:
   \begin{compactitem}
  \item \emph{Pseudo-tiling $\TModel$-requirement}: $\$\in \TModel(0)$ and the following holds:
  \begin{compactitem}
  \item  for all $0\leq i< L$, $\TModel(i)\cap P_\Main$ is a singleton and $\TModel(i)\cap P_\Main=\HModel(i)\cap P_\Main$;
  \item there is $0\leq i< L$ such that $d_\Final\in \TModel(i)$.
  \end{compactitem}
  \item \emph{Full  $\TModel$-requirement}: for all $0\leq i< L$, $\TModel(i)\cap P_\Tag=P_\Tag$ and $\symSp\in \TModel(i)$.
  \item \emph{$\HModel$-requirement}: if $\HModel\neq \TModel$, then
  \begin{compactitem}
  \item \emph{Case $\Lang = \THT_2^{\,2}(\Always)$:}  there is $k_\infty \in \Nat \cup \{\infty\}$ such that (i) for all $i\leq k_\infty$, $\HModel(i)\cap P_\Tag$ is a singleton and $u\notin \HModel(i)$, and
  (ii) for all $i>k_\infty$, $\HModel(i)\cap P_\Tag=P_\Tag$ and $\symSp\in \HModel(i)$.
  \item  \emph{Case $\Lang = \THT_2^{\,2}(\Until)$:}   for all $0\leq i< L$,   $\HModel(i)\cap P_\Tag\neq \emptyset$. Moreover,
      if there is $0\leq i< L$ such that either $u\in \HModel(i)$ or  $|\HModel(i)\cap P_\Tag|\geq 2$, then
      for all $0\leq j< L$, $\HModel(j)\cap P_\Tag=P_\Tag$ and $\symSp\in \HModel(j)$.
  \end{compactitem}
\end{compactitem}
\end{definition}\vspace{0.2cm}

\begin{definition}[Slices and good pseudo-tiling codes for $\THT_2^{\,2}(\Always)$ and   $\THT_2(\Until)$]\label{def:SlicesUGNested} Let $\Lang \in \{\THT_2^{\,2}(\Always),\THT_2^{\,2}(\Until)\}$.
  For every pseudo-tiling code $\Model=(\HModel,\TModel)$ for $\Lang$ such that $\HModel\neq \TModel$, the \emph{slice of $(\HModel,\TModel)$} is defined as follows:
   \begin{compactitem}
  \item \emph{Case $\Lang = \THT_2^{\,2}(\Always)$:} the slice of $\Model$
is $\HModel$ if $u\notin \HModel(i)$ for all $i\geq 0$; otherwise, the slice of $\Model$ is the maximal prefix of $\HModel$ whose positions are not labeled by $u$ (note that such a prefix is non-empty, otherwise $\HModel=\TModel$). Observe that for every position $i$ of the slice of $\Model$, there is exactly one proposition $t$ in $P_\Tag$ such that $t\in \HModel(i)$.
  \item  \emph{Case $\Lang = \THT_2^{\,2}(\Until)$:} the slice of $\Model$ is the maximal prefix of $\HModel$ consisting of non-empty positions.
  \end{compactitem} \vspace{0.2cm}

A pseudo-tiling code    $\Model=(\HModel,\TModel)$ for $\Lang $ is  \emph{good} if whenever $\HModel\neq \TModel$ and $\Lang =\THT_2(\Until)$, then for all positions $i$ of the slice of $\Model$,
$u\notin \HModel(i)$ and $\HModel(i)\cap P_\Tag$ is a singleton.
\end{definition}\vspace{0.2cm}

An interpretation $\Model=(\HModel,\TModel)$ satisfies the \emph{empty suffix requirement} if there is an empty position $L$ of $\TModel$ such that for all $i>L$ (resp., $i<L$),
$i$ is an empty position (resp., $i$ is not an empty position) of $\TModel$.
Evidently, by Definitions~\ref{def:PseudoTilingUGNested} and~\ref{def:SlicesUGNested}, the following holds.\vspace{0.2cm}

\begin{remark}\label{Remark:LowerBoundU} If $\Model$ is a pseudo-tiling code for $\THT_2^{\,2}(\Until)$  which satisfies the empty suffix requirement, then $\Model$ is \emph{good}.
\end{remark}

The  notion of pseudo-tiling code for  $\Lang \in \{\THT_2^{\,2}(\Always),\THT_2^{\,2}(\Until)\}$ can be captured by an $\Lang$-formula.\vspace{0.2cm}

\begin{proposition}\label{Prop:FormulaPseudoU} Let $\Lang \in \{\THT_2^{\,2}(\Always),\THT_2^{\,2}(\Until)\}$. Then, one can construct in polynomial time an $\Lang$-formula   $\varphi_\pseudo$  such that
 $(\HModel,\TModel)\models \varphi_\pseudo$ \emph{iff}
$(\HModel,\TModel)$ is a pseudo-tiling code for $\Lang$.
\end{proposition}
\begin{IEEEproof}\emph{Case $\Lang=\THT_{2}^{\,\,2}(\Always)$:} we use the fact that $(\HModel,\TModel)\models (\neg \symSp \rightarrow \symSp)$ iff $\symSp\in \TModel(0)$.
\[
\begin{array}{l}
\varphi_{\pseudo} = \underbrace{\$ \,\wedge \, \displaystyle{\Always(\bigvee_{p\in P_\Main}(\,\, p\,\wedge \, \bigwedge_{p'\in P_\Main\setminus\{p\}}\neg p')) \,\wedge \,  (\neg\,\,\Always \bigvee_{p\in P_\Main\setminus\{d_\Final\}} p)}}_{\text{pseudo-tiling $\TModel$-requirement}} \,\,\wedge \vspace{0.4cm}\\
 \underbrace{\bigl(\neg \symSp \rightarrow \symSp\bigr) \wedge  \Always(\bigvee_{p\in P_\Tag}p)\,\wedge\, \displaystyle{\Always\bigl([\bigvee_{(p,p')\in P_\Tag: p\neq p'}(p\wedge p')] \rightarrow \symSp\bigr)
               \wedge   \Always\bigl(\symSp \rightarrow \Always(\symSp\wedge\bigwedge_{p\in P_\Tag}p)\bigr)}
}_{\text{ Full $\TModel$-requirement and $\HModel$-requirement}}
\end{array}
\]
The first three conjuncts in the definition of $\varphi_\pseudo$ evidently capture the pseudo-tiling $\TModel$-requirement.
Moreover, since $(\HModel,\TModel)\models (\neg \symSp \rightarrow \symSp)$ iff $\symSp\in \TModel(0)$, the last four conjuncts ensure the full $\TModel$-requirement and  the $\HModel$-requirement.\vspace{0.2cm}

\noindent \emph{Case $\Lang=\THT_2^{\,\,2}(\Until)$:}
 let $\eta_\emptyset=   \displaystyle{\bigwedge_{p\in P}\,\neg p}$ (characterizing the empty posititions).

 \[
\begin{array}{rcl}
\varphi_{\pseudo} &=& \underbrace{\$ \,\wedge \,   \displaystyle{\Bigl( \bigvee_{p\in P_\Main}(\,\, p\,\wedge \, \bigwedge_{p'\in P_\Main\setminus\{p\}}\neg p')\,\,\Until\, \eta_\emptyset\Bigr)}  \,\wedge\,
\displaystyle{\neg\bigl((\bigvee_{p\in P_\Main\setminus\{d_\Final\}}\, p )\,\, \Until \, \eta_\emptyset\bigr) }}_{\text{pseudo-tiling $\TModel$-requirement}}\,\,\wedge\vspace{0.2cm}\\
&& \displaystyle{\Bigl((\bigvee_{p\in P_\Tag}p)\,\,\Until\, \eta_\emptyset\Bigr)\,\wedge\, \Bigl([(\bigvee_{p\in P}\,p)\,\,\Until\,\bigvee_{(p,p')\in P_\Tag: p\neq p'}\Eventually(p\wedge p')] \rightarrow \symSp\Bigr)}\,\,\wedge\vspace{0.2cm}\\
&& \underbrace{\bigl(\neg \symSp \rightarrow \symSp\bigr)\,\,\wedge \,\, \displaystyle{\Bigl[\Bigl((\bigvee_{p\in P}\,p)\,\,\Until\,\symSp\Bigr)  \rightarrow \Bigl((\symSp\wedge\bigwedge_{p\in P_\Tag}p)\,\,\Until\, \eta_\emptyset\Bigr)\Bigr]\text{\hspace{1.5cm}}}}_{\text{ Full $\TModel$-requirement and $\HModel$-requirement}}
\end{array}
\]

\end{IEEEproof}\vspace{0.2cm}

The following Propositions~\ref{prop:FormualsForMarkU} and~\ref{Prop:FormulaBadU} represent the versions  of  Propositions~\ref{prop:FormualsForMarkFGNested} and~\ref{Prop:FormulaBadFGNested}
for the considered fragments $\THT_2^{2}(\Always)$ and   $\THT_2^{2}(\Until)$.\vspace{0.2cm}

\begin{proposition}\label{prop:FormualsForMarkU} Let $\Lang \in \{\THT_2^{\,2}(\Always),\THT_2^{\,2}(\Until)\}$,   $t_{i_1},\ldots, t_{i_k}$ be  distinct propositions in $P_\Tag$, and
 $P_1,\ldots,P_k$ be non-empty subsets of $P_\Main$. Then,  one can construct in polynomial time an  $\Lang$-formula
$\theta(i_1|P_1,\ldots,i_k|P_k)$ satisfying the following: for all \emph{good pseudo-tiling codes} $(\HModel,\TModel)$ for $\Lang$ such that $\HModel\neq \TModel$,
$$(\HModel,\TModel)\models\theta(i_1|P_1,\ldots,i_k|P_k) \text{ \emph{ iff}}$$
the projection of the slice of $(\HModel,\TModel)$ over $P_\Tag$ is
\[
\text{{either in }} \{t_{i_1}\}^{+}\ldots \{t_{i_{k-1}}\}^{+}\{t_{i_k}\}^{\omega} \text{ or in }
\{t_{i_1}\}^{+}\ldots \{t_{i_{k}}\}^{+},
 \]
 and for all $1\leq j\leq k$, all the main propositions which label the segment of $\HModel$ marked by $t_{i_j}$ are in $P_j$.
\end{proposition}
\begin{IEEEproof} \emph{Case $\Lang=\THT_{2}^{\,\,2}(\Always)$:} we use the fact that for a  pseudo-tiling code  $(\HModel,\TModel)$ for
 $\THT_{2}^{2}(\Always)$ such that  $\HModel\neq \TModel$,
 a position $i\geq 0$, and  $t\in P_\Tag$,  formula $t \rightarrow u$ holds at position $i$ iff either $t\notin \HModel(i)$ or $i$ is not a position of the slice of $(\HModel,\TModel)$; moreover, if $i$ is not a position of the slice of $(\HModel,\TModel)$, then $\HModel(i)\cap P_\Tag = P_\Tag$. Furthermore, for a
pseudo-tiling code  $(\HModel,\TModel)$  of $\THT_{2}^{2}(\Always)$, $\HModel\neq \TModel$ iff $u\notin \HModel(0)$.
 \[
\begin{array}{rcl}
  \theta(i_1 | P_1,\ldots ,i_k| P_k)   =   \underbrace{
  \displaystyle{\bigwedge_{t\in P_\Tag\setminus\{t_{i_1},\ldots,t_{i_k}\}} \Always(t \rightarrow u)}}_{
  \text{the slice is only marked by tag propositions in $\{t_{i_1},\ldots,t_{i_k}\}$ }}\,\,\,\wedge \vspace{0.2cm}\\
   \underbrace{
  \displaystyle{\bigwedge_{j=1}^{j=k} \Bigl([\Always(t_{i_j} \rightarrow u)] \longrightarrow \, u \Bigr)}}_{
  \text{every tag $t_{i_j}$ ($j=1,\ldots,k$) marks some position of the   slice }}\,\,\,\wedge\vspace{0.2cm}\\
   \underbrace{
  \displaystyle{\bigwedge_{j=1}^{j=k} \Always\Bigl((t_{i_j} \rightarrow u) \vee   \bigvee_{p\in P_j}\, p  \Bigr)}}_{
  \text{the positions of the slice marked by $t_{i_j}$  are labeled by main propositions in $P_j$ }}\,\,\,\wedge \vspace{0.2cm}\\
   \underbrace{
  \displaystyle{\bigwedge_{j=1}^{j=k} \Always\Bigl(t_{i_j} \rightarrow \Always(\bigvee_{r=j}^{r=k}t_{i_r})  \Bigr)}}_{
  \text{the tags $t_{i_j}$ mark the slice according to the order $t_{i_1},\ldots,t_{i_k}$}}
\end{array}
\]
\emph{Case $\Lang=\THT_{2}^{\,\,2}(\Until)$:} we use the fact that for a good pseudo-tiling code $(\HModel,\TModel)$ for $\THT_{2}^{2}(\Until)$ such that $\HModel\neq \TModel$,   $u\in\TModel(0)$, $u\notin\HModel(0)$ and for all the positions $i$ of the slice of $(\HModel,\TModel)$,   $\HModel(i)\cap P_\Tag$ is a singleton.
 In order to define the $\THT_{2}^{2}(\Until)$-formula
$\theta(i_1|P_1,\ldots,i_k|P_k)$, we use for all $t,t'\in P_\Tag$ and for all implication-free propositional formulas $\xi$, the following auxiliary $\THT_{2}^{2}(\Until)$-formulas $\psi(t,t')$ and $\phi(\xi)$
    \[
    \psi(t,t') = \displaystyle{(\bigvee_{p\in P}\,p)\,\,\Until\,\,\Bigl(t\,\,\wedge \,\,[(\bigvee_{p\in P}\,p)\,\,\Until \,\, t']\Bigr)}
    \]
    \[
    \phi(\xi) = \displaystyle{(\bigvee_{p\in P}\,p)\,\,\Until \,\, \xi}
    \]

Formula $\psi(t,t')$ asserts that along the slice of the given good pseudo-tiling code for $\THT_{2}^{2}(\Until)$, there is a position marked by $t$ followed by a position marked by $t'$. Formula $\phi(\xi)$ requires that there is a position along the slice, where $\xi$ holds.
The $\THT_{2}^{2}(\Until)$-formula
$\theta(i_1|P_1,\ldots,i_k|P_k)$ is defined as follows.

\[
\begin{array}{rcl}
  \theta(i_1 | P_1,\ldots ,i_k| P_k) = \underbrace{\Bigl(\displaystyle{\bigwedge_{j=1}^{j=k-1}\psi(t_{i_j},t_{i_{j+1}})\Bigr)}}_{\text{partial order requirement}}\,\wedge\,\Bigl(\psi(i_1 | P_1,\ldots ,i_k| P_k) \longrightarrow\, u\Bigr)
\end{array}
\]
\[
\begin{array}{rcl}
  \psi(i_1 | P_1,\ldots ,i_k| P_k) = \underbrace{\displaystyle{\Bigl(\bigvee_{t\in P_\Tag  \setminus\{t_{i_1},\ldots,t_{i_k}\}}\phi(t)\Bigr)}}_{\text{some position in the slice is marked by some $t\in P_\Tag  \setminus\{t_{i_1},\ldots,t_{i_k}\}$}}\,\,\vee\vspace{0.2cm}\\
   \phantom{}\underbrace{\displaystyle{\Bigl(\bigvee_{(t,t')\in P_\Tag\times P_\Tag : t\neq t'}(\psi(t,t')\wedge \psi(t',t)) \Bigr)}}_{\text{in the slice,  a $t'$-marked position occurs  between two $t$-marked positions with $t\neq t'$}}\,\,\vee\vspace{0.2cm}\\
   \phantom{ } \underbrace{\displaystyle{\Bigl(\bigvee_{j=1}^{j=k}\bigvee_{p\in P_\Main \setminus P_j}\phi(t_{i_j}\wedge p) \Bigr)}}_{\text{for some $ \,1\leq j\leq k$,  a $t_{i_j}$-marked position in the slice  is labeled by a $(P_\Main \setminus P_j)$-proposition}}
\end{array}
\]

\end{IEEEproof}\vspace{0.2cm}

\begin{proposition}\label{Prop:FormulaBadU} Let $\Lang \in \{\THT_2^{\,2}(\Always),\THT_2^{\,2}(\Until)\}$. Then, one can construct in polynomial time an $\Lang$-formula   $\varphi_{\textit{bad}}$
such that for all  total interpretations $(\TModel,\TModel)$ which are  pseudo-tiling codes for $\Lang$,
there exists a \emph{good} pseudo-tiling code for $\Lang$   of the form $(\HModel,\TModel)$ with $\HModel\neq \TModel$ and satisfying $\varphi_{\textit{bad}}$ \emph{iff}
there is \emph{no} prefix of $\TModel$  whose projection over $P_\Main$  encodes a tiling.
\end{proposition}
\begin{IEEEproof}
The $\Lang$-formula   $\varphi_{\textit{bad}}$    is defined as follows
\[
\begin{array}{rcl}
  \varphi_{\textit{bad}}& = &  \varphi_{\textit{bad\_in}}\vee  \varphi_{\textit{bad\_ord}} \vee \varphi_{\textit{bad\_acc}}\vee \varphi_{\textit{bad\_cell}}
   \vee \varphi_{\textit{bad\_first}} \vee \varphi_{\textit{bad\_last}} \vee\vspace{0.2cm}\\
   &&  \varphi_{\textit{bad\_inc}} \vee
  \varphi_{\textit{bad\_rr}} \vee   \varphi_{\textit{bad\_cr}}
\end{array}
\]
where for a total  pseudo-tiling code $(\TModel,\TModel)$ for $\Lang$, the different disjuncts in the definition of $\varphi_{\textit{bad}}$ have the same intended meaning as the homonym disjuncts in the proof of Proposition~\ref{Prop:FormulaBadFGNested}. In particular, they capture all the possible conditions
such that no prefix of $\TModel$ encodes a tiling iff some of these conditions is satisfied.
The construction of such disjuncts exploits the formulas $\theta(i_1 | P_1,\ldots ,i_k| P_k)$ of  Proposition~\ref{prop:FormualsForMarkU}.
In particular, all the above disjuncts -- except $\varphi_{\textit{bad\_inc}}$ and  $\varphi_{\textit{bad\_cr}}$ -- are defined as the
homonym disjuncts in the proof of Proposition~\ref{Prop:FormulaBadFGNested}, but we use the formulas  $\theta(i_1 | P_1,\ldots ,i_k| P_k)$ of  Proposition~~\ref{prop:FormualsForMarkU} instead of the formulas of Proposition~\ref{prop:FormualsForMarkFGNested}.

The construction of $\varphi_{\textit{bad\_inc}}$ and  $\varphi_{\textit{bad\_cr}}$ is as follows.
Recall from the proof of Proposition~\ref{Prop:FormulaBadFGNested} that
for an \LTL interpretation $\TModel$ over $P$, a prefix of $\TModel$ is \emph{incomplete} if it has no position labeled by $d_\Final$.

We use  the following short-hands:
\[
\begin{array}{rcl}
  R_\Main & = &   P_\Main \setminus\{d_\Final\}\\
  P_\cell & = & P_\Main\setminus \{\$\} \\
  R_\cell & = & P_\cell \setminus\{d_\Final\}\\
  \Delta_R  & = & \Delta \setminus\{d_\Final\}
\end{array}
\]

For a good pseudo-tiling code $(\HModel,\TModel)$ for $\Lang$ such that $\HModel\neq \TModel$, the disjunct $\varphi_{\textit{bad\_inc}}$  selects along the slice  adjacent cells in a row  whose column numbers are not consecutive; moreover, the rightmost cell
is preceded by an incomplete prefix. In order to define $\varphi_{\textit{bad\_inc}}$, we use the following auxiliary $\Lang$-formulas $\phi(p,t)$ where $p\in P_\Main$ and $t\in P_\Tag$:
\begin{itemize}
  \item Case $\Lang= \THT_{2}^{2}(\Always)$:
  \[
\phi(p,t) =\displaystyle{\Bigl(\Always[(t\rightarrow u)\vee \bigvee_{p'\in P_\Main \setminus\{p\}}\,p']\Bigr)\rightarrow \, u}
\]
  \item Case $\Lang= \THT_{2}^{2}(\Until)$: $\phi(p,t)= \displaystyle{(\bigvee_{p'\in P}\,p')\,\,\Until\,(p\wedge t)}$
\end{itemize}

It is easy to check that for a good pseudo-tiling code $(\HModel,\TModel)$ for $\Lang$ such that $\HModel\neq \TModel$,
$(\HModel,\TModel)\models \phi(p,t)$ iff there is a position $i$ of the slice of $(\HModel,\TModel)$ marked by $t$ and where $p$ holds.
The formula  $\varphi_{\textit{bad\_inc}}$ is defined as follows:
\[
\begin{array}{rcl}
 \varphi_{\textit{bad\_inc}} & = &   \Bigl(\theta( 1 | R_\Main, 2| \Delta_R\cup\{\$\},  3 | P_\num, 4| \Delta_R,  5 | P_\num, 6 | \Delta,  7 | P_\Main) \,\,\vee \vspace{0.2cm}\\
  &  & \underbrace{ \phantom{Ra, 2| \Delta_R\cup\{\$\}}   \theta( 1 | \{\$\},  3 | P_\num, 4| \Delta_R,  5 | P_\num, 6 | \Delta,  7 | P_\Main)\Bigr)}_{
  \text{mark by $t_3$ and $t_5$ the cell-numbers of two adjacent cells  in a row}} \,\,\wedge \vspace{0.2cm}\\
  && \,\,  \psi_{\textit{bad\_inc}}
  \end{array}
\]
where $\psi_{\textit{bad\_inc}}$ uses the above formulas $\phi(p,t)$ and asserts that the cell numbers marked by $t_3$ and $t_5$, respectively, are not consecutive.
\[
\begin{array}{l}
 \psi_{\textit{bad\_inc}}  = \displaystyle{\bigwedge_{i=1}^{i=n}\Bigl(\phi((i,1),t_3) \,\,\wedge \,\, \phi((i,0), t_5)\Bigr)\,\vee\,
 \bigvee_{i=1}^{i=n}\Bigl(\phi((i,0),t_3) \,\,\wedge \,\, \phi((i,1), t_5)}\,\, \wedge \\
   \Bigl[\displaystyle{ \bigvee_{j=1}^{j=i-1} \bigl(\phi((j,0),t_3) \wedge  \phi((j,1),t_5)\bigr)\vee
 \bigvee_{j=i+1}^{j=n}\bigvee_{b,b'\in\{0,1\}:b\neq b'}\bigl(\phi((j,b),t_3) \wedge  \phi((j,b'), t_5)\bigr) \Bigr]\Bigr)}
\end{array}
\]

Finally, the disjunct   $\varphi_{\textit{bad\_cr}}$  checks that, along the slice, there are two adjacent cells in a column  which do \emph{not} have the same color on the shared edge; moreover, the rightmost cell is preceded by an incomplete prefix. Formula $ \varphi_{\textit{bad\_cr}}$ is defined as follows.\vspace{0.2cm}

\noindent $ \Bigl( \theta( 1 | R_\Main, 2| P_\num , 3| \Delta_R,  4 | R_\cell, 5 | \{\$\},  6 | R_\cell, 7 | P_\num, 8|\Delta,  9 | P_\Main) \,\,\vee$\\
\text{\hspace{0.2cm}}$  \theta( 1 | R_\Main, 2| P_\num , 3| \Delta_R,  4 | R_\cell, 5 | \{\$\},   7 | P_\num, 8|\Delta,  9 | P_\Main)\,\,\vee$\\
\text{\hspace{0.2cm}}$  \underbrace{\theta( 1 | R_\Main, 2| P_\num , 3| \Delta_R,   5 | \{\$\},  6 | R_\cell, 7 | P_\num, 8|\Delta,  9 | P_\Main)\phantom{
  }\Bigr)}_{
  \text{mark with $t_2$ and $t_7$ the cell-numbers of two cells $c$ and $c'$ of two adjacent rows}} \,\,\wedge \,\, \psi_{\textit{bad\_cr}}$\vspace{0.2cm}

\noindent where $\psi_{\textit{bad\_cr}}$ asserts that the   cells $c$ and $c'$ whose cell-numbers are marked by the propositions $t_2$ and $t_7$ and whose contents are marked by the propositions $t_3$ and $t_8$, respectively, have the same column number but distinct color on the shared edge. For the construction of $\psi_{\textit{bad\_cr}}$, we use the $\Lang$-formulas $\phi(p,t)$ (where $p\in P_\Main$ and $t\in P_\Tag$) exploited in the construction of  $\varphi_{\textit{bad\_inc}}$.
 \[
\begin{array}{l}
 \psi_{\textit{bad\_cr}} = \underbrace{\displaystyle{\bigwedge_{i=1}^{i=n}\bigvee_{b\in \{0,1\}} \Bigl(\phi((i,b),t_2)\wedge \phi( (i,b),t_7)\Bigr)}}_{\text{the marked cells $c$ and $c'$ have the same column number}}\,\,\wedge\vspace{0.2cm}\\
  \underbrace{\displaystyle{\bigvee_{(d,d')\in \Delta\times \Delta: d_{\Up}\neq (d')_{\Down}}}\Bigl(\phi (d,t_3)\wedge \phi ( d',t_8)\Bigr)}_{\text{the marked   cells $c$ and $c'$ do \emph{not} have the same color on the shared edge}}
\end{array}
\]

\end{IEEEproof}\vspace{0.2cm}

By using Propositions~\ref{Prop:FormulaPseudoU} and~\ref{Prop:FormulaBadU}, we prove the following result from which Theorem~\ref{theorem:consistencyLowerBoundUntractable} for the fragments $\THT_2^{2}(\Always)$ and  $\THT_2^{2}(\Until)$   directly follows.\vspace{0.2cm}

\begin{lemma}\label{MainLemma:LowerBoundU} Let $\Lang \in \{\THT_2^{\,\,2}(\Always),\THT_2^{\,\,2}(\Until)\}$.
Then, one can construct in polynomial time an $\Lang$-formula $\varphi_\Instance$ such that there is a temporal equilibrium model of $\varphi_\Instance$ \emph{iff} there is a tiling of $\Instance$.
\end{lemma}
\begin{IEEEproof} Let $\varphi_\pseudo$ be the $\Lang$-formula of Proposition~\ref{Prop:FormulaPseudoU} and
$\varphi_{\textit{bad}}$ be the $\Lang$-formula of Proposition~\ref{Prop:FormulaBadU}. Then:
\[
\varphi_\Instance =\varphi_\pseudo \wedge (\symSp \vee   \varphi_{\textit{bad}})
\]
Now, we prove that the construction is correct.
First, assume that there exists a temporal equilibrium model $(\TModel,\TModel)$   of $\varphi_\Instance$. By construction of $\varphi_\Instance$ and Proposition~\ref{Prop:FormulaPseudoU}, $(\TModel,\TModel)$ is a pseudo-tiling code for $\Lang$. If no prefix of $\TModel$  encodes a tiling, by Proposition~\ref{Prop:FormulaBadU}, there exists $\HModel \sqsubset \TModel$ such that $(\HModel,\TModel)\models \varphi_{\textit{bad}}$
and $(\HModel,\TModel)$ is a good pseudo-tiling code for $\Lang$; hence, by  Proposition~\ref{Prop:FormulaPseudoU},
$(\HModel,\TModel)$ satisfies $\varphi_\Instance$, which contradicts the assumption that $(\TModel,\TModel)$ is a temporal equilibrium model. Thus,
some prefix of $\TModel$ encodes a tiling, and the result follows.

Now, assume that there exists a tiling $f$ of $\Instance$. Assume that $\Lang =\THT_2^{\,2}(\Until)$ (the other case being simpler). Let $(\TModel,\TModel)$ be any pseudo-tiling code for $\THT_2^{\,2}(\Until)$ satisfying the \emph{empty suffix requirement}
such that   the projection of some prefix of $\TModel$ over $P_\Main$ is an encoding of $f$. Note that such a $(\TModel,\TModel)$ exists.
Since $\symSp\in\TModel(0)$, by construction and Proposition~\ref{Prop:FormulaPseudoU},   $(\TModel,\TModel)$
satisfies $\varphi_\Instance$.
We assume that $(\TModel,\TModel)$ is not an equilibrium model and derive a contradiction, hence, the result follows. Thus, there is
$\HModel \sqsubset \TModel$ such that $(\HModel,\TModel)\models \varphi_\Instance$. By construction and  Proposition~\ref{Prop:FormulaPseudoU},
$(\HModel,\TModel)$ is a pseudo-tiling code for $\THT_2^{\,2}(\Until)$. Since $(\TModel,\TModel)$ satisfies the \emph{empty suffix requirement},
$(\HModel,\TModel)$ satisfies the empty suffix requirement as well. Thus, by Remark~\ref{Remark:LowerBoundU},
 $(\HModel,\TModel)$ is a good pseudo-tiling code, and in particular, $u\notin \HModel(0)$. Since
$(\HModel,\TModel)\models \varphi_\Instance$, by construction,  $(\HModel,\TModel)\models  \varphi_{\textit{bad}}$. Thus, by
Proposition~\ref{Prop:FormulaBadU}, there is no prefix of $\TModel$ which encodes a tiling. This contradicts the hypothesis, and we are done.

\end{IEEEproof}

\newpage

\subsection{\textbf{Proof of Proposition~\ref{Prop:FormulaPseudoForFGOne}}}\label{APP:FormulaPseudoForFGOne}

\setcounter{aux}{\value{proposition}}
\setcounter{auxSec}{\value{section}}
\setcounter{section}{\value{sec-FormulaPseudoForFGOne}}
\setcounter{proposition}{\value{prop-FormulaPseudoForFGOne}}

\begin{proposition} One can construct in polynomial time a $\THT_1^{\,2}(\Eventually,\Always)$ formula $\varphi_\pseudo$
 such that
 $(\HModel,\TModel)\models \varphi_\pseudo$ \emph{iff}
$(\HModel,\TModel)$ is a pseudo-tiling code for $\THT_1^{\,2}(\Eventually,\Always)$.
\end{proposition}
\setcounter{proposition}{\value{aux}}
\setcounter{section}{\value{auxSec}}
\begin{IEEEproof} We use the fact that $(\HModel,\TModel)\models (\neg \symSp \rightarrow \symSp)$ iff $\symSp\in \TModel(0)$.
The $\THT_1^{2}(\Eventually,\Always)$ formula  $\varphi_{\pseudo}$ is defined as follows:
\[
\begin{array}{l}
\varphi_{\pseudo} = \Bigl(\neg \symSp \rightarrow \symSp\Bigr) \wedge \varphi_{\TModel} \wedge \varphi_{\textit{full}}\wedge \varphi_{\HModel}
\end{array}
\]
where $\varphi_\TModel$, $\varphi_{\textit{full}}$, and $\varphi_\HModel$ are $\THT_1^{1}(\Eventually,\Always)$ formulas, and:   $\varphi_\TModel$ ensures the pseudo-tiling $\TModel$-requirement, $\varphi_{\textit{full}}$ together with the conjunct
$\neg \symSp \rightarrow \symSp$ ensures the full $\TModel$-requirement, and $\varphi_\HModel$  together with
the conjuncts
$\neg \symSp \rightarrow \symSp$ and $\varphi_{\textit{full}}$ guarantees the $\HModel$-requirement.
\[
\begin{array}{rlc}
\varphi_{\TModel} & = &\displaystyle{\Always\,\bigvee_{d\in \Delta}\Bigl(\,d\,\wedge \, \bigwedge_{d'\in \Delta\setminus\{d\}} \neg d'\Bigr) \,\wedge\, \bigwedge_{i=1}^{i=n}\bigwedge_{\tau\in\{r,c\}}\Always \bigvee_{b\in\{0,1\}}\Bigl((\tau,i,b)\wedge \neg (\tau,i,1-b)\Bigr)}
 \,\,\wedge \vspace{0.2cm}\\
 && \underbrace{\displaystyle{\Eventually \Bigl( d_\Init\wedge \bigwedge_{i=1}^{i=n}\bigwedge_{\tau\in\{r,c\}}(\tau,i,0)\Bigr)}}_{\text{initialization}} \,\,\wedge \,\, \underbrace{\displaystyle{\Eventually \Bigl( d_\Final \wedge \bigwedge_{i=1}^{i=n}\bigwedge_{\tau\in\{r,c\}}(\tau,i,1)\Bigr)}}_{\text{acceptance}}
\end{array}
\]
\[
\begin{array}{rlc}
\varphi_{\textit{full}} & = &\Eventually\symSp \rightarrow \Always(\symSp\wedge\displaystyle{\bigwedge_{p\in P_\Tag}p)}
\end{array}
\]
\[
\begin{array}{l}
\varphi_{\HModel} = \Always\Bigl(t_1\vee t_2 \vee t_3 \vee \displaystyle{\bigwedge_{i=1}^{i=n}\bigwedge_{\tau\in\{r,c\}}\bigvee_{b\in \{0,1\}}\overline{(\tau,i,b)}}\Bigr) \,\,\wedge \,\, \Bigl(\varphi_{\textit{bad}\_\HModel}\rightarrow \symSp\Bigr)
\end{array}
\]
\[
\begin{array}{rlc}
\varphi_{\textit{bad}\_\HModel} & = & \displaystyle{\Bigl[\Eventually(t_1\vee t_2\vee t_3)\wedge \Eventually(\bigvee_{t\in P_\Tag\setminus\{t_1,t_2,t_3\}} \,t) \Bigr] \, \vee \, \Eventually[\bigvee_{t,t'\in \{t_1,t_2,t_3\}: t\neq t'} (\,t \wedge t')]} \,\, \vee \vspace{0.2cm}\\
&& \displaystyle{\bigvee_{i=1}^{i=n}\bigvee_{\tau\in\{r,c\}}[(\Eventually\, \overline{(\tau,i,0)})\wedge (\Eventually\, \overline{(\tau,i,1)})]}
\end{array}
\]
\end{IEEEproof}

\bigskip
\bigskip

\subsection{\textbf{Proof of Theorem~\ref{theorem:consistencyLowerBoundTHTFGOneTemporalNesting} for the fragments $\THT_1^{\,\,2}(\Until)$ and $\THT_1^{\,\,2}(\Release)$}}\label{APP:consistencyLowerBoundTHTFGOneTemporalNesting}

\myparagraph{Encoding of  tilings} The notions of cell-codes and cell-number codes (over $P_\Tag$) are defined as for the reduction given for the fragment
 $\THT_1^{2}(\Eventually,\Always)$.  However, a tiling $f:[0,2^{n}-1]\times [0,2^{n}-1] \rightarrow \Delta$ (of the given instance $\Instance$)
 is   encoded by finite words (and not infinite words) $w$ over $2^{P_\Main}$ satisfying the following, where
 $|w|$ denotes the length of $w$:
 \begin{itemize}
   \item for all $i,j\in [0,2^{n}-1]$, there is $0\leq h< |w|$ such that $w(h)$ is the cell-code of the $(i,j)^{th}$ cell of $f$;
   \item for all $0\leq h< |w|$, $w(h)$ encodes the $(i,j)^{th}$ cell of $f$ for some $i,j\in [0,2^{n}-1]$.
 \end{itemize}\vspace{0.2cm}

\myparagraph{Reductions for $\THT_1^{\,2}(\Until)$ and $\THT_1^{\,2}(\Release)$}
for  these two fragments, we give two slightly different notions of pseudo-tiling code which in turn are different from the one adopted for the fragment
$\THT_1^{2}(\Eventually,\Always)$. Then, we provide
corresponding versions of Propositions~\ref{Prop:FormulaPseudoForFGOne} and~\ref{Prop:FormulaBadFGOne}.

Recall from Appendix~\ref{APP:LowerBoundsUntilRelease} that for an $\LTL$ interpretation $\TModel$ over $P$ and for $i\geq 0$, $i$ is an \emph{empty position} of $\TModel$ if $\TModel(i)=\emptyset$. An interpretation $\Model=(\HModel,\TModel)$ satisfies the \emph{empty suffix requirement} if there is an empty position $L$ of $\TModel$ such that for all $i>L$ (resp., $i<L$),
$i$ is an empty position (resp., $i$ is not an empty position) of $\TModel$.\vspace{0.2cm}

\begin{definition}[Pseudo-tiling codes for $\THT_1^{2}(\Until)$]\label{def:pseudoTilingForUOne}
An interpretation  $\Model=(\HModel,\TModel)$  is a \emph{pseudo-tiling code} for $\THT_1^{\,2}(\Until)$  if there is an empty position $L$ of $\TModel$ such that the following holds:
   \begin{compactitem}
     \item \emph{Pseudo-tiling $\TModel$-requirement}: for all $0\leq i< L$, $\TModel(i)\cap P_\Main$ is a cell-code and $\HModel(i)\cap P_\Main=\TModel(i)\cap P_\Main$. Moreover,
      \begin{compactitem}
      \item there is $0\leq i<L$ such that $\TModel(i)\cap P_\Main$ has row-number $0$, column-number $0$ and $d_\Init\in \TModel(i)$ (\emph{initialization});
        \item there is $0\leq i<L$ such that $\TModel(i)\cap P_\Main$ has row-number $2^{n}-1$, column-number $2^{n}-1$, and $ d_\Final\in \TModel(i)$ (\emph{acceptance}).
      \end{compactitem}
  \item \emph{Full  $\TModel$-requirement}: for all $0\leq i< L$, $\TModel(i)\cap P_\Tag=P_\Tag$ and $\symSp\in \TModel(i)$.
  \item \emph{$\HModel$-requirement}: for all $0\leq i< L$,  either  $\HModel(i)\cap \{t_1,t_2,t_3\}\neq \emptyset$, or there is a cell-number code $P'\subseteq P_\Tag$ such that $P'\subseteq\HModel(i)$. Moreover, if the following \emph{goodness condition} is \emph{not} satisfied, then for all $0\leq i< L$, $\HModel(i)=\TModel(i)$:
      \\ \emph{Goodness condition:} for all $1\leq i< L$, $u\notin \HModel(i)$ and
      \begin{compactitem}
        \item \emph{either} there is a cell-number code $P'\subseteq P_\Tag$  such that for all $0\leq i<L$, $\HModel(i)\cap P_\Tag=P'$;
        \item \emph{or} for all $0\leq i<L$, $\HModel(i)\cap P_\Tag$ is a singleton contained in $\{t_1,t_2,t_3\}$.
      \end{compactitem}
\end{compactitem}\vspace{0.1cm}
The \emph{slice} of $(\HModel,\TModel)$ is the prefix of $\HModel$ of length $L$ (i.e., the maximal prefix of $\HModel$ consisting of non-empty positions of $\TModel$).
\end{definition}\vspace{0.2cm}

\begin{definition}[Pseudo-tiling codes for $\THT_1^{2}(\Release)$]
An interpretation  $\Model=(\HModel,\TModel)$  is a \emph{pseudo-tiling code} for $\THT_1^{\,2}(\Release)$  if there is an empty position $L$ of $\TModel$ such that the following holds:
   \begin{compactitem}
     \item \emph{Pseudo-tiling $\TModel$-requirement}: for all $i\geq 0$, \emph{either} $i$ is an empty position of $\TModel$, \emph{or} $\TModel(i)\cap P_\Main$ is a cell-code and $\HModel(i)\cap P_\Main=\TModel(i)\cap P_\Main$. Moreover,
      \begin{compactitem}
      \item there is $0\leq i<L$ such that $\TModel(i)\cap P_\Main$ has row-number $0$, column-number $0$ and $d_\Init\in \TModel(i)$ (\emph{initialization});
        \item there is $0\leq i<L$ such that $\TModel(i)\cap P_\Main$ has row-number $2^{n}-1$, column-number $2^{n}-1$, and $ d_\Final\in \TModel(i)$ (\emph{acceptance}).
      \end{compactitem}
  \item \emph{Full  $\TModel$-requirement}: for all non-empty positions $i$ of $\TModel$, $\TModel(i)\cap P_\Tag=P_\Tag$ and $\symSp\in \TModel(i)$.
  \item \emph{$\HModel$-requirement}: for all non-empty positions $i$ of $\TModel$,  either  $\HModel(i)\cap \{t_1,t_2,t_3\}\neq \emptyset$, or there is a cell-number code $P'\subseteq P_\Tag$ such that $P'\subseteq\HModel(i)$. Moreover, if the goodness condition is \emph{not} satisfied, then  $\HModel =\TModel$, where the goodness condition is defined as in Definition~\ref{def:pseudoTilingForUOne}.
\end{compactitem}\vspace{0.1cm}
The \emph{slice} of $(\HModel,\TModel)$ is defined as in Definition~\ref{def:pseudoTilingForUOne}.
\end{definition}\vspace{0.2cm}

A pseudo-tiling code for $\THT_1^{2}(\Until)$ (resp., $\THT_1^{2}(\Release)$)  $\Model=(\HModel,\TModel)$ is  \emph{good} if whenever $\HModel\neq \TModel$, then $\Model$ satisfies the goodness condition.
Evidently, the following holds.\vspace{0.2cm}

\begin{remark}\label{Remark:LowerBoundUROne}
If $\Model$ is a pseudo-tiling code for $\THT_1^{\,\,2}(\Release)$, then $\Model$ is \emph{good}. Moreover,
if $\Model$ is a pseudo-tiling code for $\THT_1^{\,\,2}(\Until)$  which satisfies the empty suffix requirement, then $\Model$ is \emph{good}.
\end{remark}\vspace{0.2cm}

The following Propositions~\ref{Prop:FormulaPseudoForUROne} and~\ref{Prop:FormulaBadUROne} represent the variants for the fragments
$\THT_1^{2}(\Release)$ and $\THT_1^{2}(\Until)$ of Propositions~\ref{Prop:FormulaPseudoForFGOne} and~\ref{Prop:FormulaBadFGOne}.\vspace{0.2cm}

\begin{proposition}\label{Prop:FormulaPseudoForUROne} Let $\Lang\in\{\THT_1^{\,2}(\Until),\THT_1^{\,2}(\Release)\}$. Then, one can construct in polynomial time an $\Lang$-formula $\varphi_\pseudo$  such that
 $(\HModel,\TModel)\models \varphi_\pseudo$ \emph{iff}
$(\HModel,\TModel)$ is a pseudo-tiling code for $\Lang$.
\end{proposition}
\begin{IEEEproof}
The $\Lang$-formula  $\varphi_{\pseudo}$ is defined as follows:
\[
\begin{array}{l}
\varphi_{\pseudo} = \Bigl(\neg \symSp \rightarrow \symSp\Bigr) \wedge \varphi_{\TModel} \wedge \varphi_{\textit{full}}\wedge \varphi_{\HModel}
\end{array}
\]
where $\varphi_\TModel$, $\varphi_{\textit{full}}$, and $\varphi_\HModel$ are $\Lang$ formulas, and:   $\varphi_\TModel$ ensures the pseudo-tiling $\TModel$-requirement for $\Lang$, $\varphi_{\textit{full}}$ together with the conjunct
$\neg \symSp \rightarrow \symSp$ ensures the full $\TModel$-requirement for $\Lang$, and $\varphi_\HModel$  together with
the conjuncts
$\neg \symSp \rightarrow \symSp$ and $\varphi_{\textit{full}}$ guarantees the $\HModel$-requirement for $\Lang$.

We use the propositional formula $\eta_0=\displaystyle{\bigwedge_{p\in P}\,\neg p}$ (which characterizes the empty positions).\vspace{0.2cm}

\noindent \textbf{Case $\Lang= \THT_1^{\,2}(\Until)$}: for each propositional formula $\xi$, let $\psi(\xi)$ be the
$\THT_1^{2}(\Until)$-formula  given by
\[
\psi(\xi)= (\displaystyle{\bigvee_{p\in P}\,p})\,\, \Until \,\,\xi
\]
 Then:
\[
\begin{array}{rlc}
\varphi_{\TModel} & = & \displaystyle{\,\Bigl[\,\bigl(\bigvee_{d\in \Delta}\,(d\,\wedge \, \bigwedge_{d'\in \Delta\setminus\{d\}} \neg d')\bigr)\,\,\Until \,\,\eta_0 \Bigr]} \,\wedge\, \vspace{0.2cm}\\
&& \displaystyle{\Bigl[\,\Bigl(\bigwedge_{i=1}^{i=n}\bigwedge_{\tau\in\{r,c\}} \bigvee_{b\in\{0,1\}}((\tau,i,b)\wedge \neg (\tau,i,1-b))\Bigr)
\,\,\Until \,\,\eta_0 \Bigr]} \,\,\wedge \vspace{0.2cm}\\
 && \underbrace{\displaystyle{\psi \Bigl( d_\Init\wedge \bigwedge_{i=1}^{i=n}\bigwedge_{\tau\in\{r,c\}}(\tau,i,0)\Bigr)}}_{\text{initialization}} \,\,\wedge \,\, \underbrace{\displaystyle{\psi \Bigl( d_\Final \wedge \bigwedge_{i=1}^{i=n}\bigwedge_{\tau\in\{r,c\}}(\tau,i,1)\Bigr)}}_{\text{acceptance}}
\end{array}
\]
\[
\begin{array}{rlc}
\varphi_{\textit{full}} & = &\psi(\symSp) \rightarrow \Bigl( (\symSp\wedge\displaystyle{\bigwedge_{p\in P_\Tag}\,p)}\,\,\Until\,\,\eta_0\Bigr)
\end{array}
\]
\[
\begin{array}{l}
\varphi_{\HModel} = \Bigl[\Bigl(t_1\vee t_2 \vee t_3 \vee \displaystyle{\bigwedge_{i=1}^{i=n}\bigwedge_{\tau\in\{r,c\}}\bigvee_{b\in \{0,1\}}\overline{(\tau,i,b)}}\Bigr)\,\,\Until\,\,\eta_0\Bigr] \,\,\wedge \,\, \Bigl(\varphi_{\textit{bad}\_\HModel}\rightarrow \symSp\Bigr)
\end{array}
\]
\[
\begin{array}{rlc}
\varphi_{\textit{bad}\_\HModel} & = & \displaystyle{\Bigl[\psi(t_1\vee t_2\vee t_3)\wedge \psi(\bigvee_{t\in P_\Tag\setminus\{t_1,t_2,t_3\}} \,t) \Bigr] \, \vee \, \psi[\bigvee_{t,t'\in \{t_1,t_2,t_3\}: t\neq t'} (\,t \wedge t')]} \,\, \vee \vspace{0.2cm}\\
&& \displaystyle{\bigvee_{i=1}^{i=n}\bigvee_{\tau\in\{r,c\}}[\psi(\overline{(\tau,i,0)})\wedge \psi(\overline{(\tau,i,1)})]}
\end{array}
\]
\noindent \textbf{Case $\Lang= \THT_1^{\,2}(\Release)$}: for each propositional formula $\xi$, let $\psi(\xi)$ be the
$\THT_1^{2}(\Release)$-formula  given by
\[
\psi(\xi)=  \displaystyle{\xi\,\, \Release \,\,\bigvee_{p\in P}\,p}
\]
 Then:
\[
\begin{array}{rlc}
\varphi_{\TModel} & = &\displaystyle{(\neg \Always \bigvee_{p\in P}\,p)\,\,\wedge\,\Always\Bigr(\eta_0\,\vee\,\,\bigvee_{d\in \Delta}\bigl(\,d\,\wedge \, \bigwedge_{d'\in \Delta\setminus\{d\}} \neg d'\bigr)\Bigr) \,\wedge\,}
 \vspace{0.2cm}\\
 && \Always\Bigl(\eta_0\,\vee\,\displaystyle{\bigwedge_{i=1}^{i=n}\bigwedge_{\tau\in\{r,c\}} \bigvee_{b\in\{0,1\}}\bigl((\tau,i,b)\wedge \neg (\tau,i,1-b)\bigr)\Bigr)}
 \,\,\wedge \vspace{0.2cm}\\
 && \underbrace{\displaystyle{\psi \Bigl( d_\Init\wedge \bigwedge_{i=1}^{i=n}\bigwedge_{\tau\in\{r,c\}}(\tau,i,0)\Bigr)}}_{\text{initialization}} \,\,\wedge \,\, \underbrace{\displaystyle{\psi \Bigl( d_\Final \wedge \bigwedge_{i=1}^{i=n}\bigwedge_{\tau\in\{r,c\}}(\tau,i,1)\Bigr)}}_{\text{acceptance}}
\end{array}
\]
\[
\begin{array}{rlc}
\varphi_{\textit{full}} & = &\psi(\symSp) \rightarrow \Always\Bigl(\eta_0\vee (\symSp\wedge\displaystyle{\bigwedge_{p\in P_\Tag}p)\Bigr)}
\end{array}
\]
\[
\begin{array}{l}
\varphi_{\HModel} = \Always\Bigl(\eta_0\vee t_1\vee t_2 \vee t_3 \vee \displaystyle{\bigwedge_{i=1}^{i=n}\bigwedge_{\tau\in\{r,c\}}\bigvee_{b\in \{0,1\}}\overline{(\tau,i,b)}}\Bigr) \,\,\wedge \,\, \Bigl(\varphi_{\textit{bad}\_\HModel}\rightarrow \symSp\Bigr)
\end{array}
\]
\[
\begin{array}{rlc}
\varphi_{\textit{bad}\_\HModel} & = & \displaystyle{\Bigl[\psi(t_1\vee t_2\vee t_3)\wedge \psi(\bigvee_{t\in P_\Tag\setminus\{t_1,t_2,t_3\}} \,t) \Bigr] \, \vee \, \psi[\bigvee_{t,t'\in \{t_1,t_2,t_3\}: t\neq t'} (\,t \wedge t')]} \,\, \vee \vspace{0.2cm}\\
&& \displaystyle{\bigvee_{i=1}^{i=n}\bigvee_{\tau\in\{r,c\}}[\psi(\overline{(\tau,i,0)})\wedge \psi( \overline{(\tau,i,1)})]}
\end{array}
\]

\end{IEEEproof}\vspace{0.2cm}

\begin{proposition}\label{Prop:FormulaBadUROne} Let $\Lang\in\{\THT_1^{\,2}(\Until),\THT_1^{\,2}(\Release)\}$. Then, one can construct in polynomial time an $\Lang$-formula $\varphi_{\textit{bad}}$  such that for all  total interpretations $\Model=(\TModel,\TModel)$ which are  pseudo-tiling codes for $\Lang$,
there exists a \emph{good} pseudo-tiling code for $\Lang$ of the form $(\HModel,\TModel)$ with $\HModel\neq \TModel$ and satisfying $\varphi_{\textit{bad}}$ \emph{iff}
the projection of the \emph{slice} of $\Model$ over $P_\Main$ does \emph{not} encode a tiling.
\end{proposition}
\begin{IEEEproof}
First, we define some auxiliary formulas. As in the proof of Proposition~\ref{Prop:FormulaPseudoForUROne}, for each implication-free propositional formula $\xi$, we consider the following $\Lang$-formula $\psi(\xi)$.
  \begin{itemize}
    \item Case $\Lang=\THT_1^{2}(\Until)$:\,\, $\psi(\xi)= (\displaystyle{\bigvee_{p\in P}\,p})\,\, \Until \,\,\xi$
    \item Case $\Lang=\THT_1^{2}(\Release)$:\,\, $\psi(\xi)= \xi\,\,\Release \,\,\displaystyle{(\bigvee_{p\in P}\,p}) $
  \end{itemize}
For a  pseudo-tiling code $\Model$ for $\Lang$, the $\Lang$-formula $\psi(\xi)$ asserts that there is a position of the slice of $\Model$, where
$\xi$ holds.

Moreover, for
 all $t,t'\in \{t_1,t_2,t_3\}$ and $\tau\in\{r,c\}$, we construct an  $\Lang$-formula $\phi(t,t',\tau)$ such that for each \emph{good} pseudo-tiling code $(\HModel,\TModel)$ for $\Lang$ with $\HModel\neq \TModel$, $(\HModel,\TModel)\models\phi(t,t',r)$ (resp., $(\HModel,\TModel)\models\phi(t,t',c)$) \emph{iff} for all the positions of the slice of $(\HModel,\TModel)$ which are  marked by the propositions $t$ and $t'$, the  associated cell-codes have the same row-number (resp., column-number).
\[
\phi(t,t',\tau) = \Bigl(\bigvee_{i=1}^{i=n} \Bigl[\psi\bigl( (t\vee t')\wedge (\tau,i,0)\bigr)\wedge \psi\bigl( (t\vee t') \wedge (\tau,i,1)\bigr)\Bigr]\Bigr) \rightarrow \symSp
\]

Then, the $\Lang$-formula $\varphi_{\textit{bad}}$  consists of four disjuncts
which are defined similarly to the disjuncts in the proof of   Proposition~\ref{Prop:FormulaBadFGOne} but for their construction, we use the above formulas $\psi(\xi)$ and $\phi(t,t',\tau)$.
\end{IEEEproof}\vspace{0.2cm}

Fix $\Lang\in\{\THT_1(\Until),\THT_1(\Release)\}$. Let  $\varphi_\Instance$ be the $\Lang$-formula defined as follows:
\[
\varphi_\Instance =\varphi_\pseudo \wedge (\symSp \vee   \varphi_{\textit{bad}})
\]
where $\varphi_\pseudo$ is the $\Lang$-formula of Proposition~\ref{Prop:FormulaPseudoForUROne} and
$\varphi_{\textit{bad}}$ is the $\Lang$-formula of Proposition~\ref{Prop:FormulaBadUROne}. By Propositions~\ref{Prop:FormulaPseudoForUROne} and~\ref{Prop:FormulaBadUROne},
$\varphi_\Instance$ can be constructed in polynomial time. Moreover, by Propositions~\ref{Prop:FormulaPseudoForUROne} and~\ref{Prop:FormulaBadUROne}, we easily deduce the following result, hence,
 Theorem~\ref{theorem:consistencyLowerBoundTHTFGOneTemporalNesting} for the fragment $\Lang\in\{\THT_1^{2}(\Until),\THT_1^{2}(\Release)\}$ directly follows.\vspace{0.2cm}

\begin{lemma}[Correctness of the construction] There exists a temporal equilibrium model of  $\varphi_\Instance$ \emph{iff} there exists a tiling of $\Instance$.
\end{lemma}
\begin{IEEEproof}
The proof is similar to the one of Lemma~\ref{MainLemma:LowerBoundU} in Appendix~\ref{APP:LowerBoundsUntilRelease}, and we omit the details here.
\end{IEEEproof}

\newpage

\subsection{\textbf{Proof of Lemma~\ref{lemma:SingleExpontialEMforTemporalOne}}}\label{APP:SingleExpontialEMforTemporalOne}

In order to prove Lemma~\ref{lemma:SingleExpontialEMforTemporalOne}, we exploit a notion of
 similarity and contraction for interpretations.

\begin{definition}[Similarity and contraction] \emph{Let $\Model$ and $\Model'$ be two interpretations. We say that $\Model'$ is a \emph{simulation} of $\Model$ if:
\begin{compactitem}
  \item $\Model(0)=\Model'(0)$ and $\Model(1)=\Model'(1)$;
  \item for all $i\geq 0$, there is $i'\geq 0$ such that $\Model'(i')=\Model(i)$ and for all $k'\in [0,i'-1]$, there is $k\in [0,i-1]$ such that
  $\Model'(k')=\Model(k)$.
\end{compactitem}
$\Model$ and $\Model'$ are \emph{bisimilar} if $\Model$ is a simulation of $\Model'$ and vice versa.}

\emph{$\Model'$ \emph{is a contraction of} $\Model$ if $\Model'$ is of the form $\Model(n_0),\Model(n_1),\ldots$, where
$n_0<n_1<\ldots$  is an infinite sequence  of increasing natural numbers
such that there is $k\geq 0$ so that $\Model(n_i)=\Model(n_{k+1})$ for all $i\geq k+1$, and the finite set of positions
$W=\{n_0,\ldots,n_k\}$ \emph{minimally} satisfies the following conditions:
\begin{compactitem}
  \item $0,1\in W$;
  \item for all $i\geq 0$, let $i_m$ be the smallest position such that $\Model(i_m)=\Model(i)$. Then, $i_m\in W$.
\end{compactitem}
We also say that $\Model'$ is a contraction of $\Model$ with respect to the sequence $n_0<n_1<\ldots$.}
\end{definition}

Note that a contraction of a total interpretation over $P$ is a strongly ultimately periodic total interpretation of size at most $2+ 2^{|P|}$. Now, we observe the following.

\begin{lemma}\label{lemma:SimilarityAndContraction} Let $\varphi\in \THT_1$ and $\Model$ and $\Model\,'$ be two interpretations. Then:
\begin{compactitem}
  \item if $\Model$ and $\Model\,'$ are bisimilar, then $\Model\models \varphi$ iff $\Model\,'\models \varphi$;
  \item if $\Model\,'$ is a contraction of $\Model$, then $\Model$ and $\Model\,'$ are bisimilar.
\end{compactitem}
\end{lemma}
\begin{IEEEproof}\emph{Property 1:} let $\Model$ and $\Model'$ be bisimilar. We show that $\Model\models \varphi$ iff $\Model'\models \varphi$ by induction on the structure of $\varphi$. Since $\Model(0)=\Model'(0)$ and $\Model(1)=\Model'(1)$, the unique non-trivial cases are when $\varphi$   is either of the form $\varphi_1\,\Until\,\varphi_2$ or of the form $\varphi_1\,\Release\,\varphi_2$. For these two cases, we consider the implication
$\Model\models \varphi$ $\Rightarrow$ $\Model'\models \varphi$ (the converse implication is symmetric). We crucially use the fact that since
$\varphi\in \THT_1$,  the subformulas $\varphi_1$ and $\varphi_2$ have no temporal modalities.
\begin{compactitem}
  \item Case $\varphi=\varphi_1\,\Until\,\varphi_2$: let $\Model\models \varphi$. Hence, there exists $i\geq 0$ such that
  $\Model,i\models \varphi_2$ and $\Model,k\models \varphi_1$ for all $k\in [0,i-1]$. Since $\Model'$ is a simulation of $\Model$, there exists
  $i'\geq 0$ such that $\Model'(i')=\Model(i)$ and for all $k'\in [0,i'-1]$, there is $k\in [0,i-1]$ such that
  $\Model'(k')=\Model(k)$. Since $\varphi_1$ and $\varphi_2$ have no temporal modalities, we obtain that $\Model',i'\models \varphi_2$ and $\Model',k'\models \varphi_1$ for all $k'\in [0,i'-1]$. Hence, $\Model'\models \varphi$, and the result follows.
    \item Case $\varphi=\varphi_1\,\Release\,\varphi_2$: let $\Model\models \varphi$. By the semantics of $\Release$, there are two cases:
     \begin{compactitem}
       \item $\Model,i\models \varphi_2$ for all $i\geq0$: since $\Model$ is a simulation of $\Model'$, for all $i'\geq 0$, there is $i\geq 0$ such that
       $\Model'(i')=\Model(i)$. Thus, since $\varphi_2$ has no temporal modalities, we obtain that $\Model',i\models \varphi_2$ for all $i\geq0$, hence,
       $\Model'\models \varphi$.
       \item There is $i\geq 0$ such that $\Model,i\models \varphi_1\wedge \varphi_2$ and $\Model,k\models \varphi_2$ for all $k\in [0,i-1]$. We proceed as for the case of the until modality.
     \end{compactitem}
\end{compactitem}
\vspace{0.2cm}

\noindent \emph{Property 2:} let $\Model'=\Model(n_0),\Model(n_1),\ldots$ be a contraction of $\Model$ with respect to the sequence
 $n_0<n_1<\ldots$. We need to show that  $\Model$ is a simulation of $\Model'$ and vice versa.
 Let $i\geq 0$. By construction, there is $j\geq 0$ such that $\Model(n_j)=\Model(i)$ and $n_j\leq i$. Thus, since $n_0=0$, $n_1=1$, and
  $\Model'=\Model(n_0),\Model(n_1),\ldots$, we obtain that $\Model'$ is a simulation of $\Model$.

  Now, we prove that
  $\Model$ is a simulation of $\Model'$. Let $i'\geq 0$. We need to show that there is $i\geq 0$ such that $\Model(i)=\Model'(i')$ and for all
  $k\in [0,i-1]$, there is $k'\in [0,i'-1]$ such that $\Model(k)=\Model'(k')$. By construction
  $\Model'(i')=\Model(n_{i'})$ and one of the following holds:
  \begin{compactitem}
    \item for all $h\geq 0$, there is $k'\in [0,i'-1]$ such that $\Model(h)=\Model'(k')$ (in particular, $i'$ is a position of the periodic part of $\Model'$). In this case, by setting $i=n_{i'}$, the result follows.
    \item $n_{i'}$ is the smallest position $h$ such that $\Model(h)=\Model'(i')$. We set $i=n_{i'}$. Let $k\in [0,n_{i'}-1]$ and $k_m$ be the smallest position
    such that $\Model(k_m)=\Model(k)$. Since $k<n_{i'}$, by construction, $k_m=n_h$ for some $h<i'$ and $\Model'(h)=\Model(n_h)$. Hence, the result follows.
  \end{compactitem}
\end{IEEEproof}

Now, we prove Lemma~\ref{lemma:SingleExpontialEMforTemporalOne}.

\setcounter{aux}{\value{lemma}}
\setcounter{auxSec}{\value{section}}
\setcounter{section}{\value{sec-SingleExpontialEMforTemporalOne}}
\setcounter{lemma}{\value{lemma-SingleExpontialEMforTemporalOne}}

\begin{lemma} Let $\varphi$ be a  $\THT_1$ formula having some equilibrium model.
Then, there exists a strongly ultimately periodic equilibrium model of $\varphi$ of size at most $2+2^{|\varphi|}$.
\end{lemma}
\setcounter{lemma}{\value{aux}}
\setcounter{section}{\value{auxSec}}
\begin{IEEEproof} We assume without loss of generality that all the propositions in $P$ occur in $\varphi$. Let $(\TModel,\TModel)$ be an equilibrium model of $\varphi$ and $(\TModel',\TModel')$ be any contraction of $(\TModel,\TModel)$. By construction, $(\TModel',\TModel')$ is a
strongly ultimately periodic interpretation of size at most $2+2^{|\varphi|}$. We show that $(\TModel',\TModel')$ is an equilibrium model of $\varphi$, hence, the result follows. Since
$(\TModel,\TModel)$ is an equilibrium model of $\varphi$, by Lemma~\ref{lemma:SimilarityAndContraction}, $(\TModel',\TModel')\models \varphi$. Now, let $\HModel'\sqsubset \TModel'$ and $\Model'=(\HModel',\TModel')$. It remains to show that $\Model'\not\models\varphi$.
Let $n_0<n_1<\ldots$ be the infinite sequence of increasing natural numbers  such that $(\TModel',\TModel')$ is a contraction of $(\TModel,\TModel)$ with respect to
$n_0<n_1<\ldots$. In particular, $\TModel'=\TModel(n_0),\TModel(n_1),\ldots$. Let $\Model$ be the interpretation defined as follows: for each position $n_j$ along the sequence $n_0<n_1<\ldots$,
 $\Model(n_j)= (\HModel'(j),\TModel'(j))$, and for each position $i$ which does not occur along the sequence $n_0<n_1<\ldots$,
 $\Model(i)=(\HModel'(h),\TModel(i))$, where $h$ is the smallest position such that $\TModel(i)=\TModel'(h)$ (since $(\TModel',\TModel')$ is a contraction of $(\TModel,\TModel)$ with respect to
$n_0<n_1<\ldots$ such a $h$ exists). Evidently, $\Model$ is of the form $(\HModel,\TModel)$ with $\HModel \sqsubset \TModel$, and
$\Model' =\Model(n_0),\Model(n_1),\ldots$. Since  $(\TModel',\TModel')$ is a contraction of $(\TModel,\TModel)$ with respect to
$n_0<n_1<\ldots$, one can easily show that $\Model$ and $\Model'$ are bisimilar. Thus, since $\Model\not\models \varphi$ ($(\TModel,\TModel)$ is an equilibrium model of $\varphi$), by Lemma~\ref{lemma:SimilarityAndContraction}, the result follows.

\end{IEEEproof}

\bigskip
\bigskip

\subsection{\textbf{Proof of Lemma~\ref{lemma:WitnessExtractionTemporalOne}}}\label{APP:WitnessExtractionTemporalOne}

\setcounter{aux}{\value{lemma}}
\setcounter{auxSec}{\value{section}}
\setcounter{section}{\value{sec-WitnessExtractionTemporalOne}}
\setcounter{lemma}{\value{lemma-WitnessExtractionTemporalOne}}

\begin{lemma} Given  $\varphi\in \THT_1$, the following holds.
\begin{compactenum}
  \item Let $\Model$ and $\Model\,'$ be two interpretations such that $\Model\,'=\Model(n_0),\Model(n_1),\ldots$ where $n_0<n_1<\ldots$ is an infinite sequence of increasing natural numbers containing all the positions of some witness pattern of $\Model$ for $\varphi$. Then, for each subformula $\psi$ of $\varphi$, $\Model\models \psi$ \emph{iff} $\Model\,'\models \psi$.
  \item Let $\Model=(\TModel,\TModel)$ be a total strongly ultimately periodic interpretation satisfying $\varphi$ of size $m$. Then
   $\Model$ is an equilibrium model of $\varphi$ \emph{iff} for each $\HModel \sqsubset \TModel$ such that
   $(\HModel, \TModel)$ is a strongly ultimately periodic interpretation of size at most $m+|\varphi|+3$, $(\HModel, \TModel)\not\models\varphi$.
\end{compactenum}
\end{lemma}
\setcounter{lemma}{\value{aux}}
\setcounter{section}{\value{auxSec}}
\begin{IEEEproof} \emph{Property 1:} The proof is by induction on the structure of $\psi$.
The non-trivial cases is when
$\psi$ has
an until or release modality as root operator.  Hence, either $\psi=\varphi_1\Until\varphi_2$ or $\psi=\varphi_1\Release\varphi_2$ for some formulas $\varphi_1$ and $\varphi_2$ which have no temporal  modalities. Here, we focus on the case $\psi=\varphi_1\Until\varphi_2$ (the case $\psi=\varphi_1\Release\varphi_2$ being similar).
First, assume that $\Model\models \varphi_1\Until\varphi_2$. Let $i$ be the smallest position
  such that $\Model,i\models \varphi_2$. We have that $\Model,k\models \varphi_1$ for all $k\in [0,i-1]$, and by Definition~\ref{Def:WitnessExtraction} $i=n_j$ for some $j\geq 0$.
Thus, since $\Model'=\Model(n_0),\Model(n_1),\ldots$, $n_0<n_1<\ldots$, and $\varphi_1$ and $\varphi_2$  have no temporal  modalities, we obtain that
 $\Model',j\models \varphi_2$ and $\Model',h\models \varphi_1$ for all $h\in [0,j-1]$. Hence, $\Model'\models \varphi_1\Until\varphi_2$.

 Now, assume that $\Model\not\models \varphi_1\Until\varphi_2$. If $\Model\not\models \Eventually\varphi_2$, then since
 $\Model'=\Model(n_0),\Model(n_1),\ldots$ and $\varphi_2$ has no temporal modalities, we obtain that $\Model'\not\models \Eventually\varphi_2$, hence,
   $\Model'\not\models \varphi_1\Until\varphi_2$. Now, assume that $\Model\models \Eventually\varphi_2$.
   Let $i$ be the smallest position
  such that $\Model,i\not\models \varphi_1$. Note that $\Model,k\not\models \varphi_2$ for all $k\in [0,i]$. By Definition~\ref{Def:WitnessExtraction}, $i=n_j$ for some $j\geq 0$.
Thus, since $\Model'=\Model(n_0),\Model(n_1),\ldots$, $n_0<n_1<\ldots$, and $\varphi_1$ and $\varphi_2$  have no temporal  modalities, we obtain that
 $\Model',j\not\models \varphi_1$ and $\Model',h\not\models \varphi_2$ for all $h\in [0,j]$. Hence, $\Model'\not\models \varphi_1\Until\varphi_2$, and we are done.
\vspace{0.2cm}

\noindent \emph{Property 2:} let $(\TModel,\TModel)$ be a strongly ultimately periodic interpretation of size $m$ and $\HModel\sqsubset \TModel$.
 We prove that there is $\HModel_m \sqsubset \TModel$ such that $(\HModel_m,\TModel)$ is a strongly ultimately periodic interpretation of size at most
 $m+|\varphi|+3$ and for each subformula $\psi$ of $\varphi$, $(\HModel,\TModel)\models \psi$ iff $(\HModel_m,\TModel)\models \psi$. Hence, Property~2 follows.

Let $\Model =(\HModel,\TModel)$  and $\Model_W =(\HModel_W,\TModel_W)$ be a witness extraction of $\Model$ for $\varphi$.
 Recall that $\Model_W=\Model(n_0), \Model(n_1),\ldots$, where $n_0<n_1<\ldots$ is a witness pattern of $\Model$ for $\varphi$. Since
 $\HModel\sqsubset \TModel$, by Definition~\ref{Def:WitnessExtraction}, $\HModel_W\sqsubset \TModel_W$. Let $j$ be the smallest position such that $n_j> m$. Define $\Model_m=\Model(0),\ldots, \Model(m),\Model(n_{j}),\Model(n_{j+1}),\ldots$. Since $\Model_W$ is strongly  ultimately periodic of size at most $|\varphi|+3$ and $(\TModel,\TModel)$ is strongly ultimately periodic of size $m$, it holds that $\Model_m$ is a strongly ultimately periodic interpretation of the form $\Model_m=(\HModel_m,\TModel)$ having size at most $m+|\varphi|+3$ and such that $\HModel_m\sqsubset \TModel$. It remains to show that for each subformula $\psi$ of $\varphi$, $\Model\models \psi$ iff $\Model_m\models \psi$.
  Since $0<\ldots <m<n_j <n_{j+1}$ contains all the positions of a witness pattern of $\Model$ for $\varphi$, the result directly follows from
  Property~1.

\end{IEEEproof}

\newpage

\section{Proofs from Section~\ref{sec:TractableFragments}}

\subsection{\textbf{Proof of Theorem~\ref{theorem:LowerBoundPositiveTHTMain}}}\label{APP:LowerBoundPositiveTHT}

By the semantics of \THT\ and \LTL, the following holds.

\begin{proposition}\label{prop:fundamentalsTHTzero} Let $(\HModel,\TModel)$ be an interpretation and $\varphi$ be a $\THT^{\,0}$ formula. Then, $(\HModel,\TModel)\models\varphi$ \emph{iff} $\HModel \models_\LTL \varphi$.
\end{proposition}

By Proposition~\ref{prop:fundamentalsTHTzero}, for a $\THT^{0}$ formula $\varphi$ and
a total interpretation $(\TModel,\TModel)$, $(\TModel,\TModel)$ is an equilibrium model of
$\varphi$ \emph{iff} $\TModel$ is a minimal \LTL model of $\varphi$. Hence,
Theorem~\ref{theorem:LowerBoundPositiveTHTMain} directly follows from the following result.

\begin{theorem}\label{LowerBoundPositiveTHT} For $\THT^{\,0}$ formulas, checking the existence of minimal \LTL models is \PSPACE-hard.
\end{theorem}

Theorem~\ref{LowerBoundPositiveTHT} is proved by
a  polynomial-time reduction  from a domino-tiling problem for grids with rows of linear length~\cite{Harel92}.
  An instance $\Instance =\tupleof{C,\Delta,n,d_\Init,d_\Final}$ of
  this problem is  as in the proof of
   Theorem~\ref{theorem:consistencyLowerBoundUntractable}.
  However, here, a  tiling of $\Instance$ is defined as a mapping $f:[0,k]\times [0,n-1] \rightarrow \Delta$,  i.e.,  the number of  columns is $n$.
 It is well-known that checking the existence of a tiling for
  $\Instance$ is \PSPACE-complete~\cite{Harel92}. We construct in polynomial time a $\THT^{0}(\Next,\Eventually,\Always)$ formula $\varphi_\Instance$ which admits a minimal \LTL model iff there exists a tiling of $\Instance$.  Hence, Theorem~\ref{LowerBoundPositiveTHT} follows.

\myparagraph{Encoding of tilings} We use the set $P$ of atomic propositions given by
$P = \{1,\ldots,n\}\times \Delta$.
Rows of tilings are encoded by finite  words of the form $\{(1,d_1)\}\ldots \{(n,d_n)\}$, and a
  tiling $f$ is encoded by the finite word   $w$  over $2^{P}$
 corresponding to the sequence of row encodings of $f$, starting from the first row of $f$.

\myparagraph{Construction of $\varphi_\Instance$} fix an \LTL interpretation $\TModel$ over $P$.  The \LTL interpretation $\TModel$ is \emph{well-formed} if for every position $i\geq 0$, $\TModel(i)$ is a singleton. $\TModel$ is \emph{almost well-formed} if there exists a suffix of $\TModel$ which is well-formed.
First, we observe the following.

\begin{lemma}\label{lemma:firstLemmaLowerBoundPositiveTHT} One can construct in polynomial time a
$\THT^{\,0}(\Next,\Eventually,\Always)$ formula $\psi_\Instance$ such that for all \LTL interpretations $\TModel$ which are almost well-formed, the following holds:
\begin{compactitem}
  \item $\TModel\models \psi_\Instance$ \emph{iff} some suffix of  $\TModel$ is of the form $w_0\cdot w_1 \cdot$ such that
  $w_i$ encodes a tiling for all $i\geq 0$ (i.e., some suffix of $\TModel$ is the $\omega$-concatenation of tiling encodings).
\end{compactitem}
\end{lemma}
\begin{IEEEproof} The $\THT^{\,0}(\Next,\Eventually,\Always)$ formula $\psi_\Instance$ is defined as follows.
 \[
\begin{array}{rcl}
  \psi_\Instance   =   \underbrace{
 \Always\Eventually((n,d_\Final)\wedge \Next \, (1,d_\Init)) }_{
  \text{initialization and acceptance}}\,\,\wedge \,\,\Eventually\Always\Bigl\{ \vspace{0.2cm}\\
   \underbrace{
  \displaystyle{ \Bigl(\bigvee_{d,d'\in \Delta} \Bigl[ ((n,d) \wedge \Next\,(1,d')) \vee
  \bigvee_{i=1}^{i=n-1}((i,d)\wedge \Next\,(i+1,d')) \Bigr]\Bigr)}}_{
  \text{there is a suffix which is a sequence of row encodings }}\,\,\,\wedge\vspace{0.2cm}\\
     \underbrace{
  \displaystyle{ \Bigl(\bigvee_{d\in \Delta} \Bigl[ (n,d)  \vee \bigvee_{i=1}^{i=n-1}\bigvee_{d'\in \Delta:\, (d')_\Left = d_\Right}((i,d)\wedge \Next\,(i+1,d')) \Bigr]\Bigr)}}_{
  \text{adjacent-row requirement}}\,\,\,\wedge\vspace{0.2cm}\\
   \underbrace{
  \displaystyle{ \Bigl(\bigvee_{j=0}^{n-1}\Next^{j}\bigl[ (n,d_\Final)  \wedge \Next\, (1,d_\Init)\bigr] \,\vee
  \, \bigvee_{d,d'\in \Delta:\, (d')_\Down = d_\Up}\bigvee_{i=1}^{i=n}\bigl[(i,d)\wedge \Next^{n}\,(i,d') \bigr]\Bigr)}}_{
  \text{adjacent-column requirement}}\,\,\Bigr\}
\end{array}
\]

\end{IEEEproof}

Let $\psi_\Instance$ be the $\THT^{0}(\Next,\Eventually,\Always)$ formula of Lemma~\ref{lemma:firstLemmaLowerBoundPositiveTHT}
and $\psi_{\textit{no\_cell}}$ be the propositional $\THT^{0}$ formula given
by $\displaystyle{\bigvee_{p,p'\in P:\, p\neq p'}}(p\wedge p')$. Then,
the $\THT^{0}(\Next,\Eventually,\Always)$ formula $\varphi_\Instance$ is defined as follows:
\[
\varphi_\Instance = \displaystyle{\Always(\bigvee_{p\in P}\, p)\,\wedge\, \Bigl(\Always\Eventually(\psi_{\textit{no\_cell}})\,\,\vee\,\, \psi_\Instance\Bigr)}
\]
Correctness of the construction directly follows from the following lemma, which concludes the proof of Theorem~\ref{LowerBoundPositiveTHT}.

\begin{lemma} There is a tiling of $\Instance$ iff there is a minimal \LTL model of $\varphi_\Instance$.
\end{lemma}
\begin{IEEEproof}
First, assume that there exists a minimal \LTL model $\TModel$ of $\varphi_\Instance$. By construction of
$\varphi_\Instance$, for all positions $i\geq 0$, $\TModel(i)\neq \emptyset$. Hence,
 $\TModel$ is almost well-formed iff $\TModel \not\models_\LTL \Always\Eventually(\psi_{\textit{no\_cell}})$.
 We claim that  $\TModel$ is almost well-formed. We assume the contrary and derive a contradiction.
Hence, $\TModel \models_\LTL \Always\Eventually(\psi_{\textit{no\_cell}})$. This implies that there exists
$\HModel \sqsubset \TModel $ such that $\HModel \models \Always\Eventually(\psi_{\textit{no\_cell}})$ and for all positions $i\geq 0$, $\HModel(i)\neq \emptyset$. By construction of $\varphi_\Instance$, we obtain that
$\HModel\models_\LTL \varphi_\Instance$ which contradicts the minimality of $\TModel$. Thus, the claim holds, and
 by construction of $\varphi_\Instance$,
$\TModel$ is almost well-formed and $\TModel\models \psi_\Instance$. By Lemma~\ref{lemma:firstLemmaLowerBoundPositiveTHT}, some suffix of $\TModel$ is the $\omega$-concatenation of tiling encodings. Hence, there exists a tiling of $\Instance$.

For the converse implication, assume that there exists a tiling $f$ of $\Instance$. Let $\TModel$ be the \LTL interpretation given by $(w_f)^{\omega}$ where $w_f$ is the encoding of $f$. Note that $\TModel$ is well-formed and by Lemma~\ref{lemma:firstLemmaLowerBoundPositiveTHT}, $\TModel$ is an \LTL model of $\varphi_\Instance$. Moreover, since $\TModel$ is well-formed, for all $\HModel \sqsubset \TModel$, there exists a position $i$ such that $\HModel(i)=\emptyset$. Hence, by construction of $\varphi_\Instance$,
$\HModel \not\models_\LTL \varphi_\Instance$. Thus, $\TModel$ is a minimal \LTL model of $\varphi_\Instance$ and we are done.

\end{IEEEproof}

\bigskip
\bigskip

\subsection{\textbf{Proof of Theorem~\ref{theorem:minimalityRXOneImplication} for the fragment $\THT_1$}}\label{APP:minimalityRXOneImplication}

\begin{theorem} Let  $\varphi$ be a $\THT_1$ formula which is $\LTL$ satisfiable. Then, there exists a minimal \LTL  model of $\varphi$.
\end{theorem}
\begin{IEEEproof}
First, we need additional definitions. A $\THT_1$ formula $\varphi$ is in disjunctive normal form if $\varphi$ is of the form $D_1\vee \ldots \vee D_k$, where for all $i\in [1,k]$,
$D_i$, \emph{called main disjunct of $\varphi$}, is of the form
\[
\eta \wedge (\Next \chi) \wedge (\Always \psi) \wedge (\xi_1\Until \phi_1) \wedge \ldots \wedge (\xi_m\Until \phi_m)
\]
where $\eta$ has no temporal modalities. Since $\varphi_1\Release \varphi_2$ can be seen as a shorthand for $(\varphi_2 \Until (\varphi_1\wedge \varphi_2)) \vee \Always\varphi_2$, given
a  $\THT_1$ formula $\varphi$, one can construct a $\THT_1$ formula $\psi$ in disjunctive normal form such that for all \LTL interpretations $\TModel$ and positions $i\geq 0$, $\TModel,i\models_\LTL \varphi$ iff $\TModel,i\models_\LTL \psi$. Thus, without loss of generality, we can assume that the given
\LTL satisfiable formula $\varphi$ in $\THT_1$  is in disjunctive normal form.

Let $D_1,\ldots,D_k$ be the main disjuncts of $\varphi$ and $\TModel$ be an \LTL model of $\varphi$. Hence, there exists $i\in [1,k]$ such that
$\TModel\models_\LTL D_i$. We claim that there exists $\TModel_i\sqsubseteq \TModel$ such that $\TModel_i$ is a minimal  \LTL model of $D_i$. Before proving this, we first observe that the claim implies the existence of a minimal \LTL model of $\varphi$. Indeed, if there exists
$j\neq i$ and $\TModel_j \sqsubset \TModel_i$ such that $\TModel_j\models_\LTL D_j$, by applying the claim, there must exist  $\TModel'_j\sqsubseteq\TModel_j$
such that $\TModel'_j$ is a minimal \LTL model of $D_j$ and for all $\TModel'' \sqsubseteq \TModel'_j$, $\TModel''$ is not an \LTL model of $D_i$. Thus, by iterating the reasoning to the remaining set $\{D_1,\ldots,D_k\}\setminus \{D_i,D_j\}$ of main disjuncts, the existence of a minimal \LTL model of $\varphi$ follows.

Now, we prove the claim. The main disjunct $D_i$ is of the form
\[
\eta \wedge (\Next \chi) \wedge (\Always \psi) \wedge (\xi_1\Until \phi_1) \wedge \ldots \wedge (\xi_m\Until \phi_m)
\]
where $\eta$ has no temporal modalities. Moreover, since $D_i\in\THT_1$, the subformulas $\chi,\psi,\xi_1,\phi_1,\ldots,\xi_m,\phi_m$
have no temporal modalities. Since $\TModel\models_\LTL D_i$,  for all $j\in [1,m]$, there exists the smallest position $\ell_j$
such that $\TModel,\ell_j\models_\LTL \phi_j$ (note that $\TModel,h\models_\LTL \xi_j$ for all $h\in [0,\ell_j-1]$).
Let $\ell=\max(\{\ell_1,\ldots,\ell_m,1\})$ and $\TModel'$ be the \LTL interpretation defined as follows:
\begin{itemize}
  \item for all $n\geq 0$, $\TModel'(n)=\TModel(n)$ if $n\leq \ell$; otherwise, $\TModel'(n)$ is a minimal subset of $\TModel(n)$ such that
  $\TModel'(n)$ satisfies the propositional formula $\psi$.
\end{itemize}
 By construction $\TModel'\sqsubseteq \TModel$ and since $\eta,\chi,\psi,\xi_1,\phi_1,\ldots,\xi_m,\phi_m$ are propositional formulas,
 $\TModel'$ is an \LTL model of $D_i$. Moreover, for all \LTL interpretations $\TModel''$ such that $\TModel''(n)\subset \TModel'(n)$ for some
 $n> \ell$, $\TModel''\not\models_\LTL D_i$. Hence, the set of \LTL interpretations $\TModel''$ such that
 $\TModel''\sqsubset \TModel'$ and $\TModel''$ is an \LTL model of $D_i$ is finite. Thus, since $\TModel'\sqsubseteq \TModel$, there exists a minimal \LTL model $\TModel_i$ of $D_i$ such that $\TModel_i\sqsubseteq\TModel$, and we are done.

\end{IEEEproof}

\bigskip
\bigskip

\subsection{\textbf{Full proof of Lemma~\ref{lemma:EquibibiumModelAlmostEmptyXUntil}}}\label{APP:EquibibiumModelAlmostEmptyXUntil}

\setcounter{aux}{\value{lemma}}
\setcounter{auxSec}{\value{section}}
\setcounter{section}{\value{sec-EquibibiumModelAlmostEmptyXUntil}}
\setcounter{lemma}{\value{lemma-EquibibiumModelAlmostEmptyXUntil}}

\begin{lemma}
Let $\varphi$ be a $\THT(\Next,\Until)$ formula and $\Model=(\TModel,\TModel)$ be an equilibrium model of $\varphi$.
Then, $\Model$ is almost-empty.
\end{lemma}
\setcounter{proposition}{\value{aux}}
\setcounter{section}{\value{auxSec}}
\begin{IEEEproof} Let $\varphi$ and $\Model=(\TModel,\TModel)$ be as in the statement of the lemma. We assume without loss of generality that $\varphi$ is \emph{not} of the form $\psi_1\,\Until\,\psi_2$ (otherwise, we consider the formula
$(\psi_1\,\Until\,\psi_2) \wedge \top$).

Fix a set of witnesses $W$ of $\Model$ for $\varphi$. Let $\ell$ be the greatest position occurring in $W$.
We define an $\LTL$ interpretation $\HModel_W \sqsubseteq \TModel$ as follows:
\begin{itemize}
  \item for all $i\geq 0$, $\HModel_W(i)=\TModel(i)$ if $i\leq \ell + \depthX(\varphi)$, and
  $\HModel_W(i)=\emptyset$ otherwise.
\end{itemize}

 We show that $\HModel_W=\TModel$, hence, $\Model=(\TModel,\TModel)$ is almost empty, and  the result follows.
 For this, since  $\Model=(\TModel,\TModel)$ is an equilibrium model of $\varphi$, it suffices to prove that
 $(\HModel_W,\TModel),0\models \varphi$. Since $(0,\varphi)\in W$, the result directly follows from the following claim: \vspace{0.2cm}

\noindent \emph{Claim:} for all $(j,\psi)\in W$ and subformulas $\xi$ of $\psi$, the following holds:
\begin{compactenum}
\item for all $k\in [0,\depthX(\psi)]$
    such that
    $\depthX(\xi)\leq \depthX(\psi)-k$, $(\TModel,\TModel),j+k\models \xi$ \emph{iff} $(\HModel_W,\TModel),j+k\models \xi$.
  \item for all $k\in [0,j]$, $(\TModel,\TModel),k\models \xi$ \emph{iff} $(\HModel_W,\TModel),k\models \xi$;
\end{compactenum}
\vspace{0.2cm}

\noindent \emph{Proof of the claim:} Let $(j,\psi)\in W$ and  $\xi$ be a subformula of $\psi$. We prove Properties~1 and~2
by  induction on the structure of $\xi$. We only consider the cases where $\xi$ has a temporal modality as root operator (the other cases easily follow from the construction and induction hypothesis). Thus, since $\xi$ is a $\THT{}{}(\Next,\Until)$ formula, either $\xi=\Next\xi_1$ or $\xi=\xi_1\,\Until\xi_2$ for some
$\THT{}{}(\Next,\Until)$ formulas $\xi_1$ and $\xi_2$. We prove the implication  $(\TModel,\TModel),j+k\models \xi$ $\Rightarrow$ $(\HModel_W,\TModel),j+k\models \xi$ of Property~1, and the implication $(\TModel,\TModel),k\models \xi$ $\Rightarrow$ $(\HModel_W,\TModel),k\models \xi$ of Property~2
(since the converse implications directly follow  from Proposition~\ref{prop:fundamentalsTHT}(1)).\vspace{0.2cm}

\noindent \emph{Property~1:} let $(\TModel,\TModel),j+k\models \xi$, where $k\in [0,\depthX(\psi)]$ and
    $\depthX(\xi)\leq \depthX(\psi)-k$.  If $\xi=\Next\xi_1$, then $(\TModel,\TModel),j+(k+1)\models \xi_1$, $k<\depthX(\psi)$, and $\depthX(\xi_1)\leq \depthX(\psi)-(k+1)$. Hence, by applying the induction hypothesis,  Property~1 follows.

    Now, assume that $\xi=\xi_1\,\Until \xi_2$.   First, we consider the case when $\xi=\psi$.  Since $\depthX(\xi)\leq \depthX(\psi)-k$, it follows that $k=0$. Since $(j,\xi_1\,\Until \xi_2)\in W$ and  $\varphi\neq \xi_1\,\Until \xi_2$, by Definition~\ref{Def:WitnessesForUX}, we have that $(\TModel,\TModel),j\models \xi_2$. Hence, by applying the induction hypothesis for Property~1, the result follows.

     Now, assume that $\xi_1\,\Until \xi_2$ is a strict subformula of $\psi$.
Since  $(\TModel,\TModel),j+k\models \xi_1\,\Until\xi_2$ and $(j,\psi)\in W$, by  Definition~\ref{Def:WitnessesForUX}, for some position $j'$,
 $(j',\xi_1\,\Until \xi_2)\in W$ and $(\TModel,\TModel),j'\models  \xi_2$. Moreover, either
$\xi_1\,\Until\xi_2\in\Fin(\varphi,\Model)$ and $j'$ is the greatest position such that
$(\TModel,\TModel),j'\models\xi_2$, or $\xi_1\,\Until\xi_2\in\Inf(\varphi,\Model)$ and $j'>j+\depthX(\varphi)\geq j+k$. Hence, $j'\geq j+k$.
Thus, since $(\TModel,\TModel),j+k\models \xi_1\,\Until\xi_2$ and $(\TModel,\TModel),j'\models  \xi_2$, there must be $\ell\in [j+k,j']$ such that
$(\TModel,\TModel),\ell\models  \xi_2$ and $(\TModel,\TModel),m\models  \xi_1$ for all $m\in [j+k,\ell-1]$.
Since $(j',\xi_1\,\Until \xi_2)\in W$, by applying
 the induction hypothesis on Property~2 for the subformulas $\xi_1$ and $\xi_2$ of $\xi_1\,\Until \xi_2$, the result follows.

\noindent \emph{Property~2:} for the case $\xi=\Next\xi_1$, Property~2 directly follows from
  Property~1 and the induction hypothesis. Now, let us consider the case $\xi=\xi_1\,\Until\xi_2$.
  Let $(\TModel,\TModel),k\models \xi$ with $k\in [0,j]$. First, assume that $\xi_1\,\Until\xi_2=\psi$.  Since $(j,\xi_1\,\Until \xi_2)\in W$ and  $\varphi\neq \xi_1\,\Until \xi_2$, by Definition~\ref{Def:WitnessesForUX}, we have that $(\TModel,\TModel),j\models \xi_2$.
   Thus, since $(\TModel,\TModel),k\models \xi$ and $k\in [0,j]$, there must be $\ell\in [k,j]$ such that
$(\TModel,\TModel),\ell\models  \xi_2$ and $(\TModel,\TModel),m\models  \xi_1$ for all $m\in [k,\ell-1]$.
Since $(j,\xi_1\,\Until \xi_2)\in W$, by applying
 the induction hypothesis on Property~2 for the subformulas $\xi_1$ and $\xi_2$ of $\xi_1\,\Until \xi_2$, the result follows.

 Now,    assume that $\xi_1\,\Until \xi_2$ is a strict subformula of $\psi$.
Since  $(\TModel,\TModel),k\models \xi_1\,\Until\xi_2$ and $(j,\psi)\in W$, by  Definition~\ref{Def:WitnessesForUX}, for some position $j'$,
 $(j',\xi_1\,\Until \xi_2)\in W$ and $(\TModel,\TModel),j'\models  \xi_2$. Moreover, either
$\xi_1\,\Until\xi_2\in\Fin(\varphi,\Model)$ and $j'$ is the greatest position such that
$(\TModel,\TModel),j'\models\xi_2$, or $\xi_1\,\Until\xi_2\in\Inf(\varphi,\Model)$ and $j'>j+\depthX(\varphi)\geq k$. Hence, $j'\geq k$.
Thus, since $(\TModel,\TModel), k\models \xi_1\,\Until\xi_2$ and $(\TModel,\TModel),j'\models  \xi_2$, there must be $\ell\in [ k,j']$ such that
$(\TModel,\TModel),\ell\models  \xi_2$ and $(\TModel,\TModel),m\models  \xi_1$ for all $m\in [k,\ell-1]$.
Since $(j',\xi_1\,\Until \xi_2)\in W$, by applying
 the induction hypothesis on Property~2 for the subformulas $\xi_1$ and $\xi_2$ of $\xi_1\,\Until \xi_2$, the result follows.
\end{IEEEproof}

\bigskip
\bigskip

\subsection{\textbf{Full proof of Lemma~\ref{lemma:EquibibiumModelXFOne}}}\label{APP:EquibibiumModelXFOne}

\setcounter{aux}{\value{lemma}}
\setcounter{auxSec}{\value{section}}
\setcounter{section}{\value{sec-EquibibiumModelXFOne}}
\setcounter{lemma}{\value{lemma-EquibibiumModelXFOne}}

\begin{lemma}
Let $\varphi$ be a $\THT{}{}(\Next,\Eventually)$ formula and $\Model=(\TModel,\TModel)$ be an equilibrium model of $\varphi$.
Then, $\Model$ has at most $\depthX(\varphi)\cdot(|\varphi|+1)$ non-empty positions.
\end{lemma}
\setcounter{proposition}{\value{aux}}
\setcounter{section}{\value{auxSec}}
\begin{IEEEproof} We assume without loss of generality that $\varphi$ is \emph{not} of the form $\Eventually\,\psi$ (otherwise, we consider the formula
$(\Eventually\,\psi) \wedge \top$).
Let $W$ be a set of witnesses of $\Model$ for $\varphi$ according to Definition~\ref{Def:WitnessesForUX}. By Definition~\ref{Def:WitnessesForUX}, $W$ has cardinality at most $|\varphi|+1$.
Now, we define an $\LTL$ interpretation $\HModel_W \sqsubseteq \TModel$ as follows:
\begin{compactitem}
  \item for all $i\geq 0$, if there is $(j,\psi)\in W$ such that $j\leq i$ and $i-j\leq \depthX(\varphi)$, then
  $\HModel_W(i)=\TModel(i)$; otherwise, $\HModel_W(i)=\emptyset$.
\end{compactitem}
 By construction, the set of non-empty positions of the interpretation $(\HModel_W,\TModel)$ has cardinality at most
 $\depthX(\varphi)\cdot(|\varphi|+1)$. We show that $\HModel_W=\TModel$, hence, the result follows.
 For this, since  $\Model=(\TModel,\TModel)$ is an equilibrium model of $\varphi$, it suffices to prove that
 $(\HModel_W,\TModel),0\models \varphi$. Since $(0,\varphi)\in W$, the result directly follows from the following claim: \vspace{0.2cm}

\noindent \emph{Claim:} for all $(i,\psi)\in W$, $k\in [0,\depthX(\psi)]$, and subformulas $\xi$ of $\psi$ such that
$\depthX(\xi)\leq \depthX(\psi)-k$, $(\TModel,\TModel),i+k\models \xi$ \emph{iff} $(\HModel_W,\TModel),i+k\models \xi$.\vspace{0.2cm}

\noindent \emph{Proof of the claim:} Let $(i,\psi)\in W$, $k\in [0,\depthX(\psi)]$, and  $\xi$ be a subformula of $\psi$ such that
$\depthX(\xi)\leq \depthX(\psi)-k$. The implication  $(\HModel_W,\TModel),i+k\models \xi$ $\Rightarrow$
$(\TModel,\TModel),i+k\models \xi$ directly follows from Proposition~\ref{prop:fundamentalsTHT}(1). For the converse implication,
 assume that $(\TModel,\TModel),i+k\models \xi$. We show that $(\HModel_W,\TModel),i+k\models \xi$ by
  induction on the structure of $\xi$. We only consider the cases where $\xi$ has a temporal modality as root operator (the other cases easily follow from the induction hypothesis and the fact that by construction $\HModel_W(i+k)=\TModel(i+k)$). Thus, since $\xi$ is a $\THT{}{}(\Next,\Eventually)$ formula, either $\xi=\Next\xi'$ or $\xi=\Eventually\xi'$.
  First, assume that $\xi=\Next\xi'$. Hence, $(\TModel,\TModel),i+(k+1)\models \xi'$.  Since $\xi$ is a subformula of $\psi$ such that
$\depthX(\xi)\leq \depthX(\psi)-k$ and $k\in [0,\depthX(\psi)]$, we have that $k<\depthX(\psi)$ and $\depthX(\xi')\leq \depthX(\psi)-(k+1)$.
By the induction hypothesis, $(\HModel_W,\TModel),i+(k+1)\models \xi'$, hence, $(\HModel_W,\TModel),i+k\models \xi$, and the result follows.

Now, assume that $\xi=\Eventually\xi'$.   First, assume that $\xi=\psi$. Since $\depthX(\xi)\leq \depthX(\psi)-k$, it follows that $k=0$. Since $(i,\Eventually\xi')\in W$ and  $\varphi\neq \Eventually\xi'$, by Definition~\ref{Def:WitnessesForUX}, we have that $(\TModel,\TModel),i\models \xi'$. Hence, by applying the induction hypothesis, the result follows. Now, assume that $\Eventually\xi'$ is a strict subformula of $\psi$.
Since  $(\TModel,\TModel),i+k\models \Eventually\xi'$, there exists $j\geq i+k$ such that
$(\TModel,\TModel),j\models \xi'$. We need to show that $(\HModel_W,\TModel),i+k\models \Eventually\xi'$.
By construction, $\Eventually\xi'\in \Fin(\varphi,\Model)\cup \Inf(\varphi,\Model)$. We distinguish two cases:
\begin{itemize}
  \item $\Eventually\xi'\in \Fin(\varphi,\Model)$: since $(\TModel,\TModel),j\models \xi'$, by Definition~\ref{Def:WitnessesForUX}, there exists the greatest position $j'$ such that  $(\TModel,\TModel),j'\models \xi'$ and $(j',\Eventually\xi')\in W$. Hence,  $j'\geq j$.
  Moreover, by applying the induction hypothesis, we have that $(\HModel_W,\TModel),j'\models \xi'$. Thus, since $j\geq i+k$ and $j'\geq j$, we obtain
that $(\HModel_W,\TModel),i+k\models \Eventually\xi'$, and the result holds.
  \item $\Eventually\xi'\in \Inf(\varphi,\Model)$: by Definition~\ref{Def:WitnessesForUX}, there exists a position $m$ such that $(m,\Eventually\xi')\in W$
   and $(\TModel,\TModel),m\models \xi'$. Hence, by applying the induction hypothesis,
   $(\HModel_W,\TModel),m\models \xi'$.
  Moreover,  since $(i,\psi)\in W$, $\Eventually\xi'$ is a strict subformula of $\psi$, and $k\leq \depthX(\varphi)$, by Definition~\ref{Def:WitnessesForUX}, it follows that $m>i+k$. Hence, $(\HModel_W,\TModel),i+k\models \Eventually\xi'$, and the result follows, which concludes.
\end{itemize}
\end{IEEEproof}

\bigskip
\bigskip

\subsection{\textbf{Proof of Lemma~\ref{lemma:EquibibiumModelXFTwo}}}\label{APP:EquibibiumModelXFTwo}

\setcounter{aux}{\value{lemma}}
\setcounter{auxSec}{\value{section}}
\setcounter{section}{\value{sec-EquibibiumModelXFTwo}}
\setcounter{lemma}{\value{lemma-EquibibiumModelXFTwo}}

\begin{lemma} Let $\varphi$ be a $\THT{}{}(\Next,\Eventually)$ formula, $n\geq 1$, and $\Model=(\TModel,\TModel)$ be an equilibrium model of $\varphi$ having $n$ non-empty positions. Then, there exists an \emph{almost-empty} equilibrium model of $\varphi$ of size at most $n\cdot  (\depthX(\varphi)+1)$.
\end{lemma}
\setcounter{proposition}{\value{aux}}
\setcounter{section}{\value{auxSec}}
\begin{IEEEproof} By hypothesis $\Model$ is an almost-empty equilibrium model of $\varphi$ having $n$  non-empty positions. Let $\ell$ be the size of $\Model$. If $\ell\leq n\cdot  (\depthX(\varphi)+1) $, we are done. Otherwise, we show that there exists an almost-empty equilibrium model of $\varphi$ of size $\ell-1$ and having $n$ non-empty positions. Hence, by iterating the reasoning, the result follows.
Since $\ell >n\cdot(\depthX(\varphi)+1)$, there must be a set of empty positions of $\Model$ of the form $[h,k]$ such that $k\leq \ell$ and $k-h>\depthX(\varphi)+1$. Let $\Model'$ be the total interpretation defined as follows: for all $i\geq 0$, $\Model'(i)=\Model(i)$ if $i<k$, and
$\Model'(i)=\Model(i+1)$ otherwise. Intuitively, $\Model'$ is obtained from $\Model$ by contracting the interval $[h,k]$ of one position. Note that
 $\Model'$ is an almost-empty total interpretation of size $\ell-1$ and having $n$ non-empty positions. One can easily show that $\Model'$ is still an equilibrium model of $\varphi$, which concludes.
\end{IEEEproof}

\end{changemargin} 

\end{document}